\documentclass[11pt, a4paper, notitlepage]{article}

\usepackage[utf8]{inputenc}
\usepackage[T1]{fontenc}
\usepackage{lmodern}
\usepackage[english]{babel}

\usepackage{amsmath}
\usepackage{amsfonts}
\usepackage{amssymb}

\usepackage{graphicx}
\usepackage{caption}
\usepackage{float}

\usepackage[left=2cm,right=2cm,top=2cm,bottom=2cm]{geometry}
\usepackage{multicol}
\usepackage{fullpage}

\usepackage[square,sort,comma,numbers, sort&compress]{natbib}

\usepackage{authblk}
\usepackage[symbol]{footmisc}

\usepackage{hyperref}
\hypersetup{
	colorlinks=true,
	urlcolor=blue,
    citecolor=blue,
    linkcolor=black,
    }

\DeclareMathOperator*{\argmin}{arg\,min}

\title{Uncertainty Quantification and Propagation for ACORN, a geometric deep learning tracking pipeline for HEP experiments}

\makeatletter

\begin{document}

\author[1, 2]{Lukas Péron\footnote[1]{Email:
\href{mailto:lukas.peron@ens.psl.eu}{lukas.peron@ens.psl.eu}}}
\author[2]{Paolo Calafiura}
\author[2]{Xiangyang Ju}
\author[2]{Jay Chan}
\affil[1]{École Normale Supérieure, Physics Department, Paris, France}
\affil[2]{Lawrence Berkeley National Laboratory, Berkeley, CA, USA}

\newcommand{\HRule}{\rule{\linewidth}{0.65mm}}
\begin{titlepage}
\begin{center}
\HRule
\vspace*{0.2cm}

\textbf{\Large{\@title}}

\HRule

\vspace*{1cm}

\@author

\vspace*{1cm}

\begin{abstract}
Geometric learning pipelines have achieved state-of-the-art performance in High-Energy and Nuclear Physics reconstruction tasks like flavor tagging and particle tracking. Starting from a point cloud of detector or particle-level measurements, a graph can be built where the measurements are nodes, and where the edges represent all possible physics relationships between the nodes. Depending on the size of the resulting input graph, a filtering stage may be needed to sparsify the graph connections. A Graph Neural Network will then build a latent representation of the input graph that can be used to predict, for example, whether two nodes (measurements) belong to the same particle or to classify a node as noise. The graph may then be partitioned into particle-level subgraphs, and a regression task used to infer the particle properties. Evaluating the uncertainty of the overall pipeline is important to measure and increase the statistical significance of the final result. How do we measure the uncertainty of the predictions of a multistep pattern recognition pipeline? How do we know which step of the pipeline contributes the most to the prediction uncertainty, and how do we distinguish between irreducible uncertainties arising from the aleatoric nature of our input data (detector noise, multiple scattering, etc) and epistemic uncertainties that we could reduce by using, for example, a larger model, or more training data?
 
We have developed an Uncertainty Quantification process for multistep pipelines to study these questions and applied it to the ACORN particle tracking pipeline. All our experiments are made using the TrackML open dataset. Using the Monte Carlo Dropout method, we measure the data and model uncertainties of the pipeline steps, study how they propagate down the pipeline, and how they are impacted by the training dataset's size, the input data's geometry and physical properties. We will show that for our case study, as the training dataset grows, the overall uncertainty becomes dominated by aleatoric uncertainty, indicating that we had sufficient data to train the ACORN model we chose to its full potential. We show that the ACORN pipeline yields high confidence in the track reconstruction and does not suffer from the miscalibration of the GNN model.
\end{abstract}
\vspace*{1cm}

\end{center}

\vspace*{1cm}

\end{titlepage}

\begin{multicols*}{2}

\section{Introduction}
\label{sec:Introduction}

\begin{figure}[H]
    \centering
    \includegraphics[width=.5\textwidth]{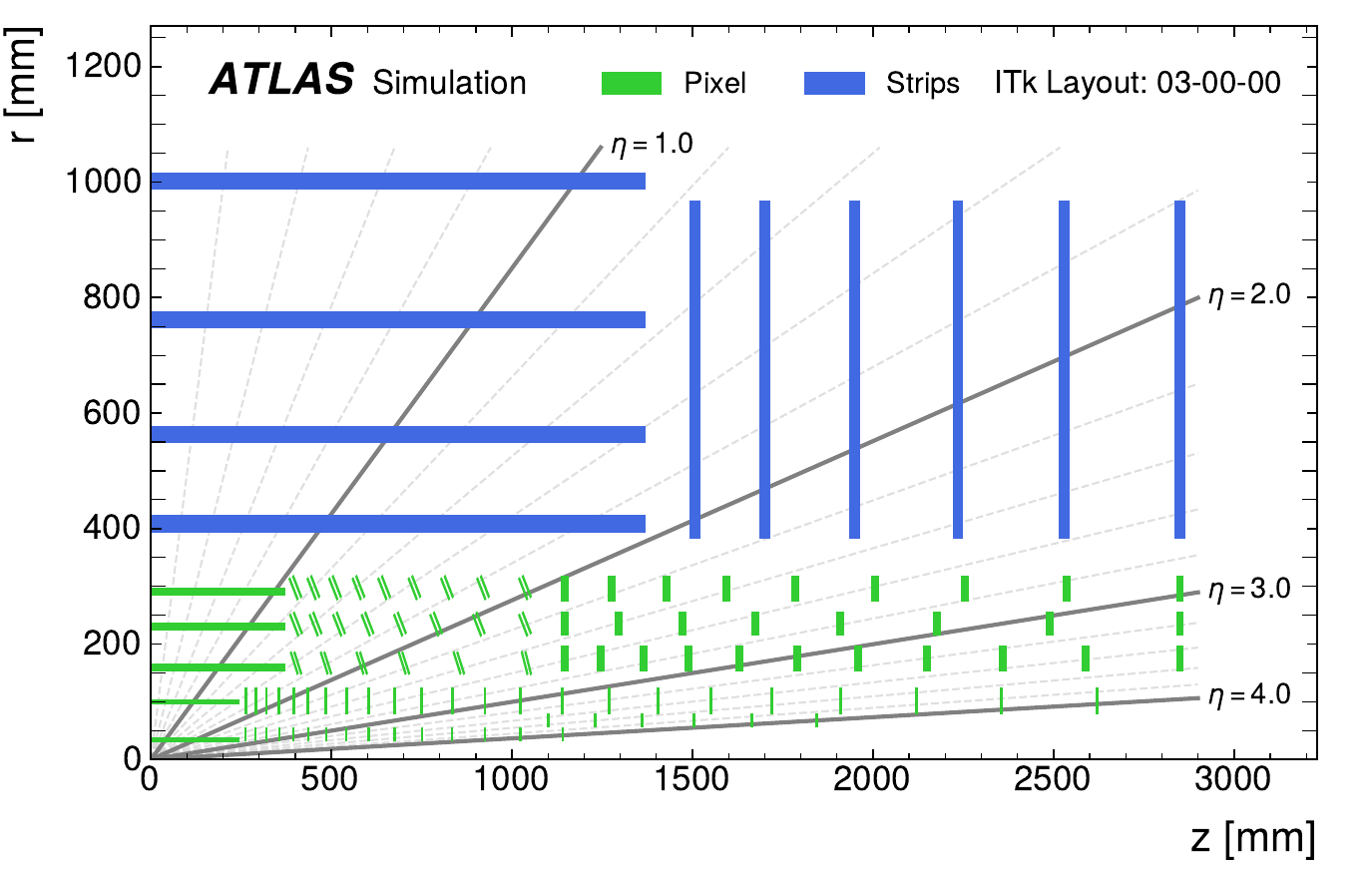}
    \caption{\centering ITk layout. Extracted from \cite{Aad_2025}}
    \label{fig:ITk_layout}
\end{figure}

\begin{figure}[H]
    \centering
    \includegraphics[width=.5\textwidth]{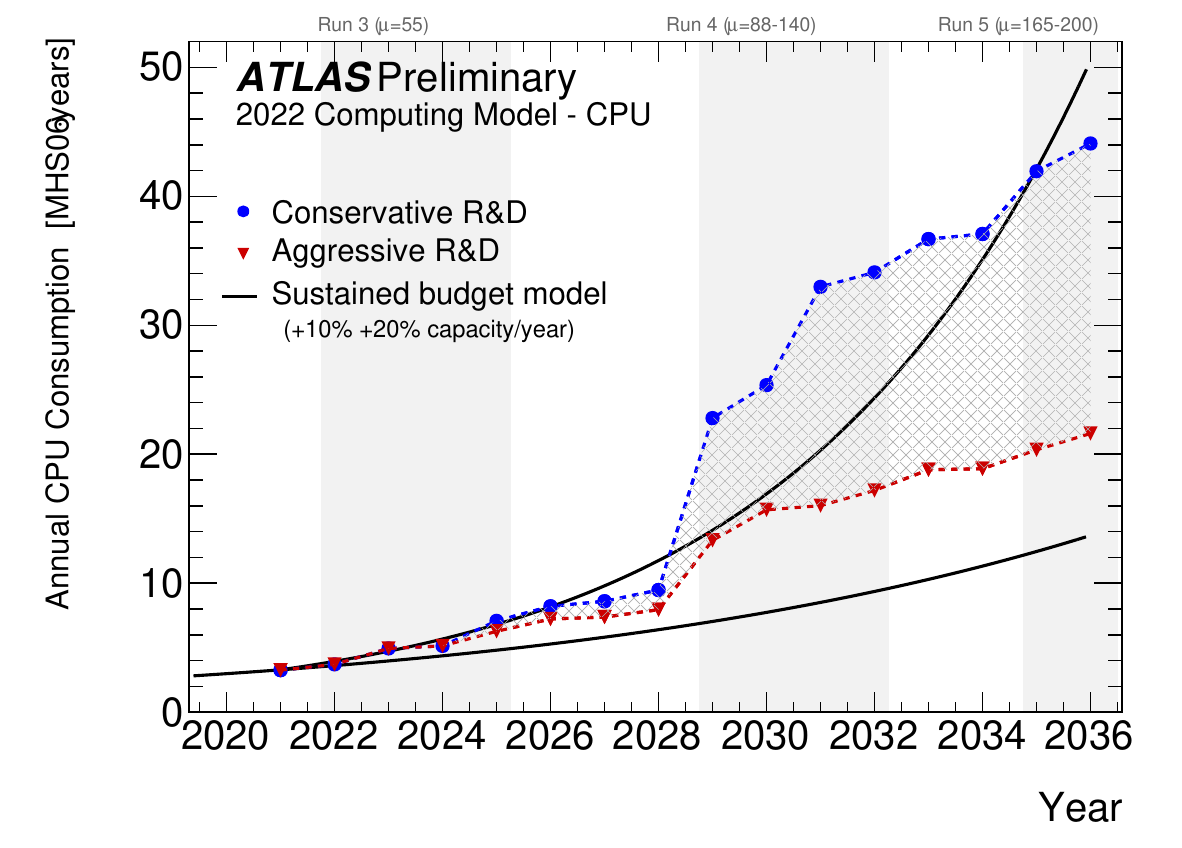}
    \caption{\centering ATLAS CPU consumption prevision. Extracted from \cite{CERN-LHCC-2022-005}}
    \label{fig:cpu_conso}
\end{figure}

The Large Hadron Collider (LHC) has, since its inauguration, provided outstanding results that have peaked with the discovery of the Higgs boson in 2012 \cite{20121, 201230}. In the upcoming years, the LHC will undergo important upgrades to begin its high luminosity phase. During this phase, the mean pile-up $\langle \mu \rangle$ (i.e. number of collisions per event) is expected to be multiplied by almost 3, reaching $\langle \mu \rangle = 200$ and the data volume to be multiplied by 10. Upgrading the hardware and software infrastructures of the experiments is an important technological challenge. Within the ATLAS collaboration, a new inner tracker (ITk) \cite{GONELLA2023167597} is developed to fulfill the requirements of radiation resilience, granularity, resolution... It is composed by two parts respectively consisting of pixels and strips detectors organized on barrels and end caps. The layout of ITk is depicted on Figure \ref{fig:ITk_layout}.
If the hardware challenge is important, the software challenge is no less so. The amount and complexity of the data that will be produced by the HL-LHC, along with the expected significant growth of the ATLAS CPU consumption, shown on Figure \ref{fig:cpu_conso}, lead to the necessity of having better algorithms for the data processing and especially the tracking, i.e. the reconstruction of particle trajectories and properties from the hits left in the tracker. We refer to these trajectories as particle tracks. In this context, the ExaTrkX project has been formed in order to develop a deep learning-based tracking algorithm. In the last decade, deep learning has been successfully applied to a very wide range of topics, leading to significant advances. Geometric Deep Learning \cite{bronstein2021geometricdeeplearninggrids} has been proposed as a solution to the tracking challenge, with a trained pipeline that could be used to infer track candidates from the hits information of an event in short amount of time. This solution, named ACORN (A Charge Object Reconstruction Network) \cite{ACORN1, ACORN2, ACORN3, ACORN4}, consist of a succession of potentially deep learning algorithms, one of which is a Graph Neural Network. The knowledge of the systematic uncertainties downstream is of great importance to confidently claim discoveries. 
Measuring the uncertainty of the outputs of a deep learning model is the purpose of Uncertainty Quantification (UQ). When working with a model pipeline, uncertainty of a certain model may affect, in non-trivial ways, the uncertainties of the subsequent models. In this report, we construct a method, based on Monte Carlo Dropout (MCD) \cite{gal2016dropoutbayesianapproximationrepresenting}, to study the model-wise uncertainties of the ACORN pipeline and their propagation through the pipeline. We apply this scheme on the TrackML dataset \cite{trackml-particle-identification} and show that the ACORN pipeline has high confidence and reliability in its track predictions. We also show that the use of binary classifier calibration method does not significantly affect the final tracking reconstruction efficiency.

\section{Prior works}
\label{sec:prior_works_review}
UQ is a long-standing and active field of research as some applications like medical diagnosis \cite{sundararajan2017axiomaticattributiondeepnetworks} or autonomous driving \cite{Poland28122018} have very strict reliability requirements. Several methods have emerged over time such as Bayesian Neural Networks \cite{charnock2020bayesianneuralnetworks}, Deep Ensembles \cite{jain2020maximizingoveralldiversityimproved}, Evidential Deep Learning \cite{sensoy2018evidentialdeeplearningquantify}, Bootstrapping \cite{bootstrap} and MCD. We focus here on the latter which has been proven to be resource efficient and equivalent to Bayesian Neural Networks under mild assumptions \cite{gal2016dropoutbayesianapproximationrepresenting, foong2020expressivenessapproximateinferencebayesian}. MCD aims to sample from the network output's posterior $p(\mathbf{y}|\mathbf{x},\mathcal D^{\text{train}}, \mathbf{\theta})$ with $\mathcal D^{\text{train}}$ the training dataset and $\mathbf{\theta}$ the model weights.

In HEP, UQ has been applied to various tasks such as parameter estimation \cite{atlas_uq}, LArTPC event reconstruction \cite{koh2023deepneuralnetworkuncertainty}, out of distribution detection \cite{OOD_detection_with_EDL_HEP} and many others. HEP experiments typically rely on maximizing of a likelihood function in order to achieve these tasks. However, with significant growth and increased complexity of the available data, deep learning has become a key part of HEP analysis. As the standard significance and uncertainty thresholds required to claim a discovery are stringent in HEP, UQ is essential to obtain a complete description of systematic uncertainties \cite{atlas_uq}.

For some particular tasks, model pipelines (i.e., succession of deep learning models) have demonstrated state-of-the-art abilities and easier implementation than single end-to-end model \cite{https://doi.org/10.1111/mice.13164, WAHID2023212403, 10466328}. In the case of tasks subject to critical reliability requirements, such as those previously mentioned, understanding the propagation of the uncertainty through the chained models is important. Such developments have been made outside HEP context \cite{pmlr-v28-wang13a, petersen2024uncertaintyquantificationstabledistribution}. In the context of particle physics we found one study conducted on neutrino event reconstruction in LArTPC \cite{douglas2025uncertaintypropagationchainedmodels}. This reference shows that an uncertainty-aware chain of models achieves better performance than an uncertainty-blind chain. Layer-wise uncertainty propagation (UP) has been studied with stochastic differential equations \cite{kong2020sdenetequippingdeepneural}. 

\section{Uncertainty propagation in model pipeline}
\label{sec:uncer_prop}
Using model pipeline naturally yields the question of the impact of one model's uncertainty on the subsequent models' uncertainties. To the best of our knowledge, there is no theoretical analysis of the UP for deep learning model pipeline in the literature. We will not give a detailed nor exhaustive treatment of this question, but rather give hints on the difficulties it yields for realistic cases. First, let us state in clear terms what the question we ask is. Given a model pipeline composed of $N$ models, we denote by $(X_n)_{n\leq N}$ the sequence of model random variables such that for all $n\leq N$, $X_{n+1}|X_n$ has the law distribution $p(\mathbf{x_{n+1}}|\mathbf{x_n}, \mathcal D^{\text{train}}_{n+1}, \mathbf{\theta_{n+1}}) = p_{n+1}$ with $\mathcal D^{\text{train}}_{n+1}$ the training dataset of the $(n +1)$th model and $\mathbf{\theta_{n+1}}$ the weights of the $(n+1)$th model. In usual UQ settings, it is common to work directly with the distribution of the weights. We prefer to work with the distribution of the results of the models.  In what follows, we assume that all the $X_n$ belong to the same probability space. We would like to know if an evolution equation of the form
\begin{equation}
    \mathbb V[X_{n+1}] = f(\mathbb V[X_n])
    \label{eq:evolution_models_uncertainty}
\end{equation}
can be written. We first start by writing the law of total variance
\begin{align}
    &\mathbb V[X_{n+1}]=\mathbb V\left[\mu_{n}(X_n)\right]+\mathbb E\left[\varsigma^2_{n}(X_n)\right]\label{eq:law_total_var}\\
    &\text{with}\;\;\left \lbrace \begin{matrix}
        \mu_{n}(X_n) = \mathbb E_{p_{n+1}}[X_{n+1}|X_n]\\
        \varsigma^2_{n}(X_n) = \mathbb V_{p_{n+1}}[X_{n+1}|X_n]\\
    \end{matrix}\right. .
\end{align}
The functions $\mu_n$ and $\varsigma_n^2$ cannot be expressed analytically and therefore an equation such as \eqref{eq:evolution_models_uncertainty} is intractable. Hypothesis could be assumed in order to obtain a tractable relation. For instance having $\mu_n(x)\sim x^k$ and $\varsigma_n^2(x)\sim x^m$ leads to $\mathbb V[X_{n+1}] \sim \mathbb V[X_n]^{\max(m/2, k)}$. One could also ask for the laws $p_n$ to be Gaussian, homoscedastic or with other simple behavior. Unfortunately, these behaviors are not suited for real life situations. However these assumptions could be helpful when applied to simple models, like single hidden layer neural networks.

Another remark can be done. In equation \eqref{eq:law_total_var} the term $\mathbb V[\mu_n(X_n)]$ can be understand as an intrinsic uncertainty of the $(n+1)$th model. Each input from $X_n$ yield an average result $\mu_n(X_n)$ that is subject to variation. On the other hand, the second term, $\mathbb E\left[\varsigma_n^2(X_n)\right]$ can be understood as a propagated uncertainty from the $n$th model to the $(n+1)$th model.

\section{ACORN}
\label{sec:ACORN}
\subsection{Pipeline description}
\begin{figure}[H]
    \centering
    \includegraphics[width=.5\textwidth]{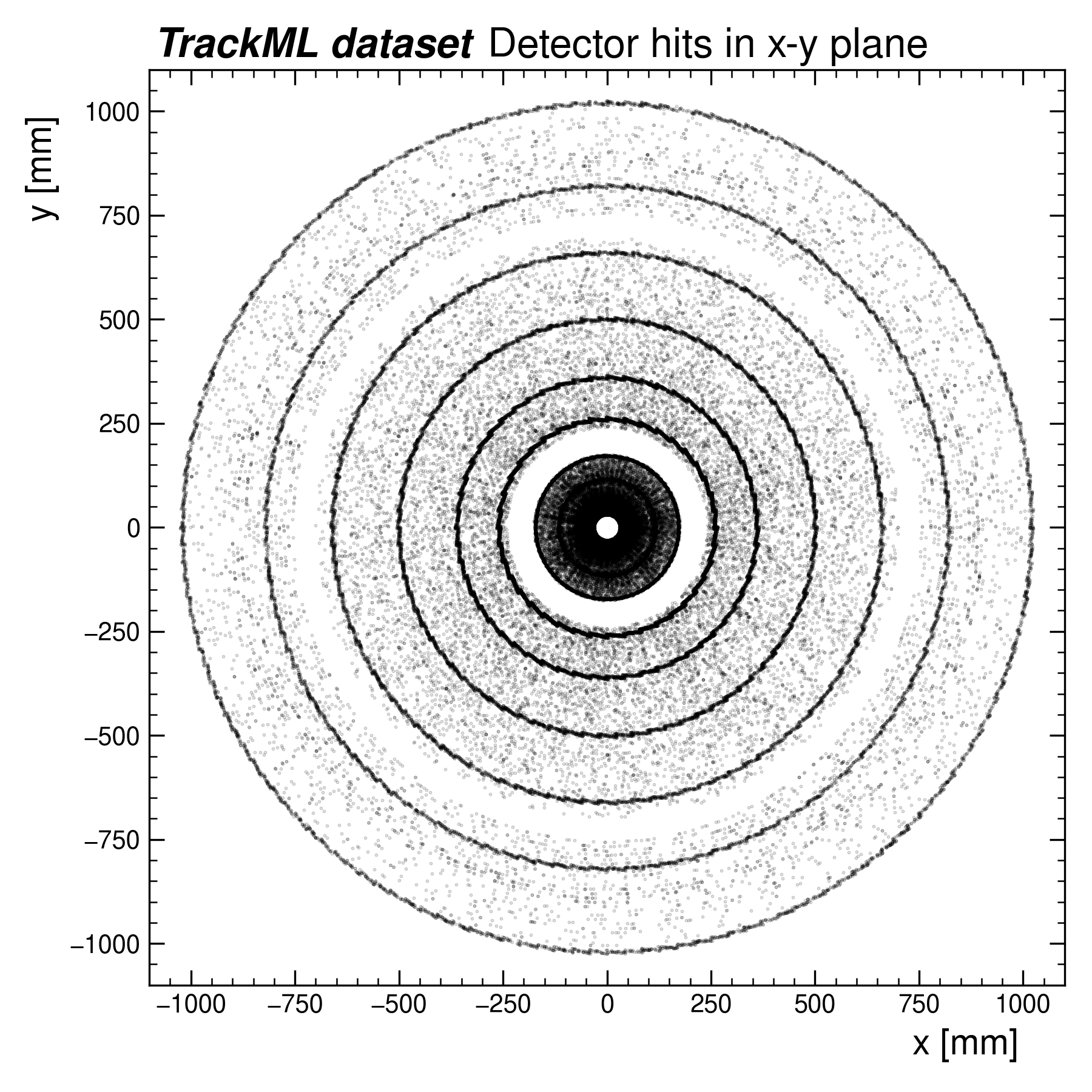}
    \caption{\centering Example of detector hits from the TrackML dataset in the $(xy)$-plane.}
    \label{fig:detector_hits}
\end{figure}

In the rest of the report, we use the ATLAS coordinate system: $z$ is the axis of the beamline, $r$ the radial distance, $(xy)$ is the transverse plane. When particle go through the tracker, they leave energy in the pixels and strips. These deposits (named hits) form the data output by the detector. Figure \ref{fig:detector_hits} shows an example of such hits in the TrackML detector. The goal of tracking is to cluster the deposits corresponding to the same particle into a full track that can then be used to infer physical parameters ($d_0$, $z_0$, $\phi$, $\theta$, $q/p$). and be utilized for physics analysis downstream.
\begin{figure*}
\centering
    \includegraphics[width=\textwidth]{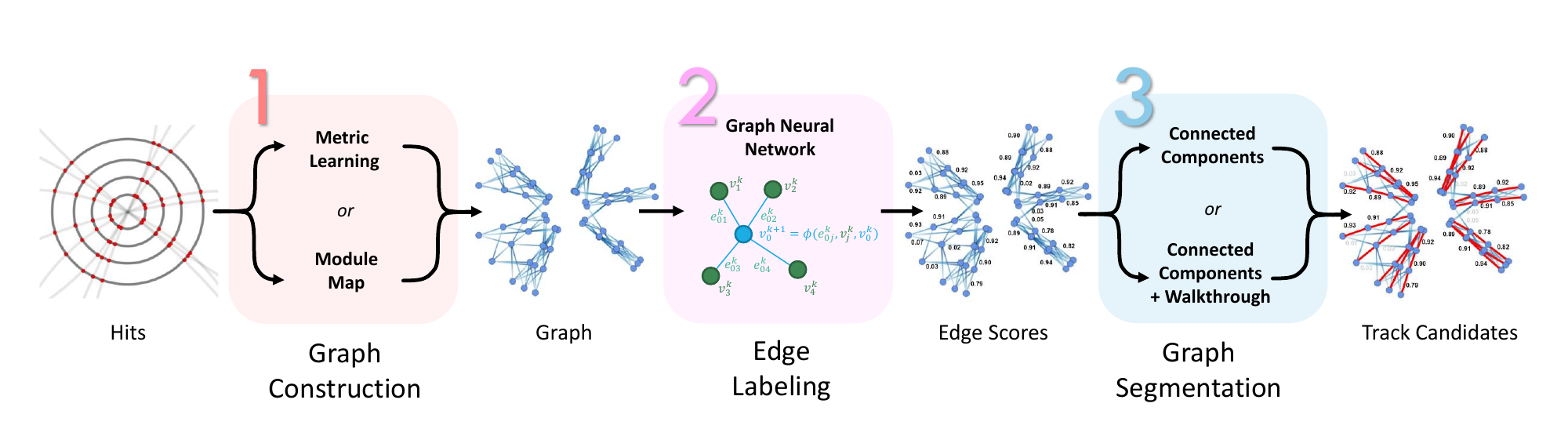}
    \caption{\centering The ACORN pipeline. Extracted from \cite{ACORN3}.}
    \label{fig:acorn_pipeline}
\end{figure*}
ACORN is a tracking pipeline developed by the ExaTrkX project. It is an alternative to the Combinatorial Kalman Filter (CKF) \cite{ai2022common} currently in use in the ATLAS software framework, Athena \cite{athena}. The core idea of ACORN is to use a Graph Neural Network (GNN) acting on a graph generated from the point cloud information received from the inner tracker. The pipeline consists on three successive tasks : graph construction, edge scoring and track construction. There are different algorithms for each task. In our analysis we chose to use metric learning as the graph constructor followed by a Filter used to sparsify the graph edges. An edge scoring GNN is then applied, and the track candidates are constructed with the CC\&Walk algorithm. A sketch of the pipeline is shown on Figure \ref{fig:acorn_pipeline}.

The metric learning algorithm is a Multilayer Perceptron (MLP) that embeds the hits in a high dimension space in which the hits belonging to the same track are close to each other. A fixed radius nearest neighbors (FRNN) clustering is applied, such that a hit is connected to all the hits neighboring it within a self-centered hypersphere of radius $r_{embedding}$.

The obtained graphs contain $O(10^5)$ edges on average. To sparsify this graph we use an MLP acting on the graph nodes (hits) features to give each edge a score between 0 and 1. All the edges with a score below a certain threshold are dropped before the graphs are passed to the GNN. This Filter reduces the graph size down to $O(10^4)$ edges.

The GNN consists on a three steps model. First, an MLP (encoder) is used to embed the nodes (resp. edges) features in a high dimension space. Then a message passing MLP is applied to the graph embedded nodes (resp. edges) features. This message passing is done by passing the features of each node (resp. edge) through an MLP and iteratively aggregating the results with the nodes (resp. edges) neighbors. The aggregation function used is a sum over the message passing MLP outputs. After the message passing, a decoder is used to infer the edge scores. Neither the nodes nor edges features are changed by the GNN, only the edge scores is changed.

\begin{figure*}
    \centering
    \includegraphics[width=\textwidth]{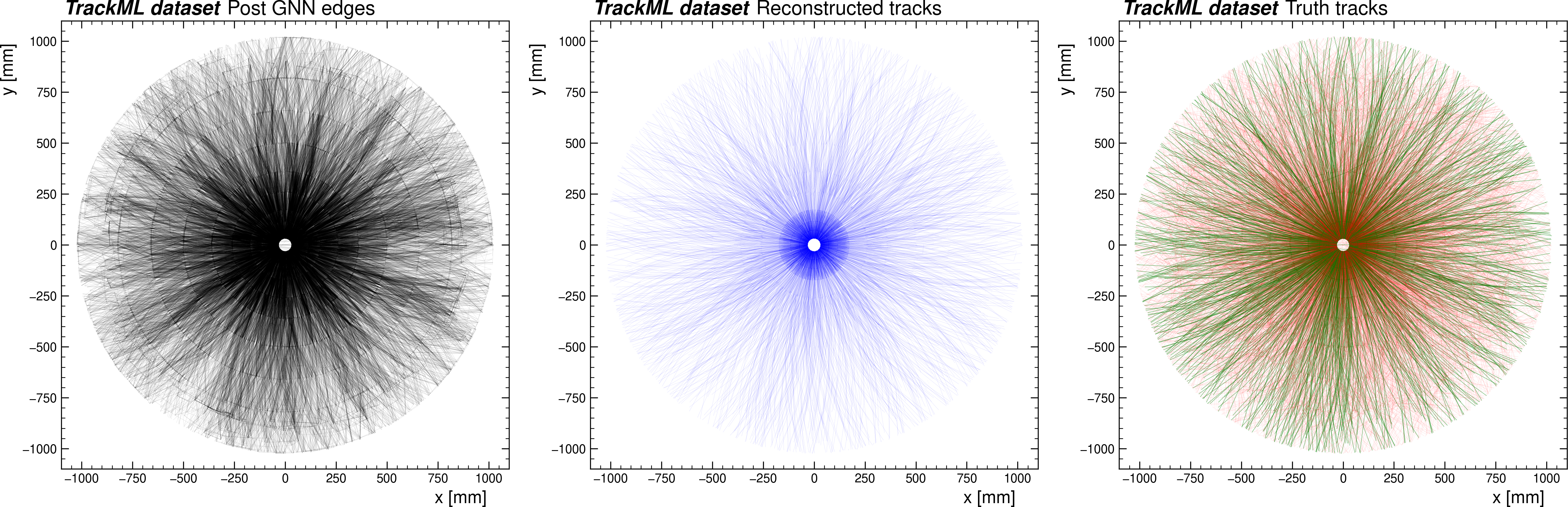}
    \caption{\centering Example of GNN output graph, true tracks objective and reconstructed tracks by CC\&Walk in the $(xy)$-plane. Green (red) tracks are tracks with  $p_T\geq(<)1$GeV.}
    \label{fig:example_graph_show}
\end{figure*}

Finally, the track candidates are formed using the Connected Component \& Walkthrough algorithm. The CC\&Walk method involves first removing all edges with a score below a certain threshold. Then, all sequences of edges in which a node is never connected to more than two edges, i.e., there is no branching, are proposed as track candidates. This initial selection is called the Connected Components part. Sequences with branching are addressed subsequently. First, a stricter score cutoff is applied to the remaining edges. If the branching is no longer present, connected components is applied and the edge sequence is proposed as a track candidate. If the branching remains, two situations are possible. First, the sum of the edge scores right after the branching is lower than a third score cutoff, the edge with the highest score is chosen to continue the sequence. Otherwise, all the edges are retained and tracks are reconstructed by selecting the longest sequences possible. As a matter of visualization, Figure \ref{fig:example_graph_show} displays an example of CC\&Walk input graph, the true tracks intended for reconstruction, and the actual track candidates proposed by CC\&Walk. A particle track is deemed \textbf{reconstructed} in the ATLAS standards if there exist a track candidate that contains at least half the hits of the particle track.

\subsection{Sources of uncertainty}
\label{subsec:source_of_uncertainty}
UQ usually defines two sources of uncertainty. First the aleatoric, or data, uncertainty which arises from the data used for training and is, therefore, irreducible. Then, the epistemic uncertainty which originates from the choices made in the architecture of the deep learning model, the training scheme and the random initialization of the model weights. Epistemic uncertainty is reducible and approaches zero in the limit of infinite training and infinite training dataset size.

\subsection{Uncertainty Propagation}
\label{subsec:uncer_prop_in_acorn}
Because ACORN is a model pipeline, it is subject to UP. As discussed in Section \ref{sec:uncer_prop}, the law of total variance allow us to separate the uncertainty of a model in two terms. The first is intrinsic and the second is propagated from the preceding model. In this study we are interested in the uncertainty quantification of the GNN and we consider the uncertainty propagated from the Filter stage to the GNN. As described in Section \ref{sec:ACORN} the Filter outputs a scored graph. If an edge has a score below a defined threshold, it is dropped, leading to a new graph that is then used as an input to the GNN. However, the Filter, being a deep learning model, possesses its own uncertainty. Thus, two slightly different instances of the Filter model (for instance not trained with the exact same dataset, of with different weight initialization) may not predict the same score for all edges. The edges that pass the score cut are then likely to be different. For this reason, we call the Filter uncertainty \textit{topological uncertainty}. As it leads to different graphs, it is not trivial to quantify the propagation of this uncertainty to the GNN. In Section \ref{sec:methods} we describe how we chose to proceed in order to have a consistent way of measuring this UP.

We also study the uncertainty propagated from the GNN and the Filter to the track building stage. This last stage is a rule-based algorithm and therefore its uncertainty is only due to the upstream uncertainty. We describe in Section \ref{sec:methods} how we chose to quantify this UP.

\section{Methods}
\label{sec:methods}
\subsection{Edges selection}
Our analysis was conducted on the TrackML dataset, consisting of 1,400 training events, 50 validation events and 50 test events. No hard cuts were applied on the $p_T$ of the particles, i.e. all edges are kept no matter their $p_T$. Edges fall into three groups: false (unrelated hits), true non-target ($<3$ hits or $p_T<1$ GeV), and true target ($\geq3$ hits \& $p_T\geq1$ GeV). The true non target edges have a null contribution to the calculation of the loss in each model. This choice reflects the need for three hits for curvature and topology differences that aid performance. In the approximation of a stationary magnetic field $B$ aligned with the beamline axis $z$, we have
\begin{equation}
    p_T[\text{GeV}] = 0.3 R[\text{m}]B[\text{T}],
\end{equation}
with $R$ the radius of curvature. Thus, the high $p_T$ tracks are less curved than the low $p_T$ tracks. The high $p_T$ tracks are of primary interest since they carry more valuable information as they are related to hard-scatter interactions.

\subsection{Uncertainty Quantification}
The MCD implementation we conducted is the same for the Filter and for the GNN. We first added dropout layers to the models. For the GNN, we chose to add the dropout layers only in the message passing MLPs. This choice is motivated by the will to not altering the high dimensional embedding of the edges and nodes features in order to keep realistic inputs without pruned information. The Filter has dropout before every layer of the MLP except the input one. 

Once the models are trained, we measure their uncertainties in three ways, using MCD. For each input graph, we generate $T=100$ predictions of score for each edge $\mathcal E_n$. The edge score $s_n$ is then considered like a random variable valued in $(0,1)$, and we compute the empirical mean predicted scores $\mathbb E_{p(s_n|\mathcal G_n, \mathcal D^{\text{train}}, \theta)}\left[s_n\right]=\langle s_n \rangle$ and the empirical standard deviation of the predicted scores $\mathbb V_{p(s_n|\mathcal G_n, \mathcal D^{\text{train}}, \theta)}\left[s_n\right] = \sigma^2_n$ on the $T$ stochastic samples. This is the first measurement of uncertainty. This measurement has the advantage of being easily interpreted: an edge has, on average of the model weights, a score of $\langle s_n \rangle \pm \sigma_n$. However, this standard deviation uncertainty provides no information about the relative proportion of aleatoric and epistemic uncertainty. To address this issue, we utilize the measurement proposed in Ref. \cite{Gal2016Uncertainty}. It is important to note though that Ref. \cite{wimmer2023quantifyingaleatoricepistemicuncertainty} has highlighted some issues with this method of measuring the aleatoric and epistemic uncertainties, therefore, the results we present should be considered as indications rather than absolute truths. We will assume the additivity of the aleatoric and epistemic uncertainty as we work only with fully trained model and reasonable sizes of dataset. The total uncertainty of the model (aleatoric + epistemic) for an edge $\mathcal E_n$ is measured as the Shannon's entropy of the average score prediction on the $T$ stochastic samples: 
\begin{align}
    \mathbb H\left[\langle s_n\rangle\right] &= - \langle s_n\rangle \ln\left(\langle s_n\rangle\right)\\\nonumber
    &- (1-\langle s_n\rangle)\ln\left(1-\langle s_n\rangle\right).
\end{align}
The epistemic uncertainty for an edge is measured as the mutual information
\begin{equation}
    \mathbb I[s_n] = \mathbb H\left[\langle s_n\rangle\right] - \mathbb E_{p(s_n|\mathcal G_n, \mathcal D^{\text{train}}, \theta)}\left[\mathbb H[s_n]\right].
    \label{eq:def_I}
\end{equation}
One should notice here that the second term corresponds to the empirical average of the Shannon's entropy of each stochastic sample of the score prediction. Seeing $\mathbb H$ and $\mathbb E_{p(s_n|\mathcal G_n, \mathcal D^{\text{train}}, \theta)}$ as operators acting on the random variable $s_n$, we can rewrite \eqref{eq:def_I} as
\begin{equation}
    \mathbb I[s_n] = \left[\mathbb H, \mathbb E_{p(s_n|\mathcal G_n, \mathcal D^{\text{train}}, \theta)}\right][s_n]
\end{equation}
with $[\cdot,\cdot]$ the commutator. The aleatoric uncertainty for an edge $\mathcal E_n$ is then defined as the conditional entropy $\mathbb H[\langle s_n\rangle ] - \mathbb I[s_n] = \mathbb E_{p(s_n|\mathcal G_n, \mathcal D^{\text{train}}, \theta)}\left[\mathbb H[s_n]\right]$. In Section \ref{sec:results} we present the results for the UQ ($\sigma_n$, $\mathbb I[s_n]$ and $\mathbb H[s_n]$) of the GNN only, the results for the UQ of the Filter are presented in Appendix \ref{app:Filter_uq}. In the rest of this report, we will refer to $\mathbb H[\langle s_n\rangle]$, the sum of aleatoric and epistemic uncertainties, as the \textbf{total} uncertainty.

\subsection{Uncertainty Propagation}
\label{subsec:methods_uncer_prop}
To obtain the uncertainty propagated from the Filter to the GNN we proceed as follows. For each input graph of the Filter, we infer $T$ stochastic scored graphs with the Filter. We apply a cut $s_{\text{cut}}^{\text{GNN}}=0.05$ such that all the edges with score $s_n < s_{\text{cut}}^{\text{GNN}}$ after the Filter are pruned. Then, all the graphs are passed through the same GNN which has not dropout layer activated. To have reasonable comparison between the $T$ graphs, we record how many times each edge has been kept, and we compute $s_n^{\text{GNN}}$, $\sigma_n^{\text{GNN}}$, $\mathbb I\left[s_n^{\text{GNN}}\right]$ and $\mathbb H\left[s_n^{\text{GNN}}\right]$ on this list of GNN scores. For instance, if an edge has been kept 47 times out of the $T=100$ passes, its average score, standard deviation and other metrics will be computed over the 47 passes. Finally, to be able to add up the intrinsic and propagated uncertainties of the GNN to obtain its total uncertainty over the total validation dataset $\mathcal D_{val}$, we group each of the uncertainties in a 100 bins histogram and then sum the histograms. In the rest of this report, we refer to $\sigma_{n}^{\text{comb}}$, the sum of the intrinsic and propagated uncertainties, as the \textbf{combined} uncertainty of the GNN.

A similar procedure is applied to obtain the relationship between $\sigma_n^{\text{GNN}}$ and $\sigma_n^{\text{Filter}}$. We start by doing a single deterministic inference (i.e. without dropout layers activated) of the Filter input dataset and we store the ids of the edges with a score $s_n^{\text{det. Filter}} \geq s_{\text{cut}}^{\text{GNN}}$. Then, $T$ MCD stochastic inferences of the Filter are done. For all of the edges that had $s_n^{\text{det. Filter}} \geq s_{\text{cut}}^{\text{GNN}}$ after the deterministic Filter inference, we compute the average stochastic Filter score $\langle s_n^{\text{Filter}}\rangle$ and the uncertainty $\sigma_n^{\text{Filter}}$. We store the obtained values. Next, $T$ MCD stochastic inferences of the GNN are applied on the deterministic output of the Filter on which the score cut has been applied, pruning the edges with $s_n^{\text{det. Filter}} < s_{\text{cut}}^{\text{GNN}}$. For all the remaining edges, we compute the average stochastic GNN score $\langle s_n^{\text{GNN}}\rangle$ and the uncertainty $\sigma_n^{\text{GNN}}$. We store the obtained values. Finally, all the edges with $s_n^{\text{det. Filter}} \geq s_{\text{cut}}^{\text{GNN}}$ have been assigned a quadruplet $\left(\langle s_n^{\text{Filter}}\rangle, \langle s_n^{\text{GNN}}\rangle, \sigma_n^{\text{Filter}}, \sigma_n^{\text{GNN}}\right)$ so that we can plot the functions $\langle s_n^{\text{GNN}}\rangle(\langle s_n^{\text{Filter}}\rangle)$ and $\sigma_n^{\text{GNN}}(\sigma_n^{\text{Filter}})$. This is useful to look for any scaling such as those presented in Section \ref{sec:uncer_prop}.

As mention in Section \ref{subsec:uncer_prop_in_acorn} we also study the uncertainty that is propagated from upstream to the rule-based track building algorithm CC\&Walk. To do so, we apply the following procedure. Starting from the input dataset of the Filter stage, each graph is passed one time through a stochastic Filter and then a stochastic GNN. The obtained scored graph dataset is passed to CC\&Walk. We measure the total track building efficiency, defined as 
\begin{equation}
    \text{Eff} = \frac{\#\text{Reconstructed particle}}{\#\text{particle}},
    \label{eq:efficiency_tracking}
\end{equation}
where the counting is done over the all inferred dataset. This value is stored and this procedure is repeat $T$ times, each one with a new instance of stochastic Filter and GNN. We thus obtain a distribution of track building efficiencies that we fit with a Gaussian. The track build efficiency uncertainty caused by upstream uncertainties is defined as the standard deviation of the best Gaussian fit achieved.

\subsection{Calibration}
The GNN is used as a binary classifier hence we would like to be able to interpret the scores as probabilities. If an edge has a score greater than 0.5 we may be inclined to conclude that it has probability greater than 50\% of being a true edge. However, in general the output of a binary classifier is not a probability. If this is the case, the classifier is said to be uncalibrated. An uncalibrated binary classifier which predictions are read as probabilities lead to misinterpretation of the results, such as leading to underconfidence or overconfidence of the score prediction. To verify the impact of the presence of a score calibration on the track building efficiency, we use a method presented in \cite{nixon2020measuringcalibrationdeeplearning}. With a deterministic GNN, we infer the total validation dataset then, we compute its reliability diagram, defined as
\begin{equation}
    \text{Rel}(s) = \frac{\# \text{True edges with scores}\ s}{\#\text{Edges with scores}\ s}.
\end{equation}
In a perfectly calibrated model we have $\text{Rel}(s) = s$. For instance, half the edges with score $s=1/2$ are true edges in a perfectly calibrated model. The calibration of the model consists then on fitting this reliability function and using this fit to recompute the scores to align them with actual probabilities. We used splines to fit the reliability curve.

As described in Section \ref{sec:ACORN} the scores predicted by the GNN are used in the track building stage. To obtain the values of the score cuts in CC\&Walk we use another type of calibration curve. We compute the rectified value of accuracy for each edge score 
\begin{equation}
    \text{Cal}(s) = \left|\text{Acc}(s) - \frac12\right|\times2    
\end{equation}
with
\begin{equation}
    \text{Acc}(s) = \frac{\#\text{Well classified edges with score}\ s}{\#\text{Edges with score}\ s}.
\end{equation}
In a perfectly calibrated model we have $\text{Cal}(s) = 2|s-\frac12|$. The accuracy function in this case is given by $\text{Acc}(s)=\max(1-s, s)$ and $\text{Acc}(1/2)=1/2$. This is because we want $\text{Acc}(0) = 1$, $\text{Acc}(1)=1$ and $\text{Acc}(1/2)=1/2$ with linear behavior in between. All edges with score 0 should be well classified (as false), all edges with score 1 should be well classified (as true) and half the edges with score $1/2$ should be well classified (as this is just a head-or-tails). Whether the scores are calibrated or not, the value $s_{opt} = \argmin_s \text{Cal}(s)$ is the score cut at which we remove half the false edges.

\section{Results}
\label{sec:results}
\begin{figure}[H]
    \centering
    \includegraphics[width=.5\textwidth]{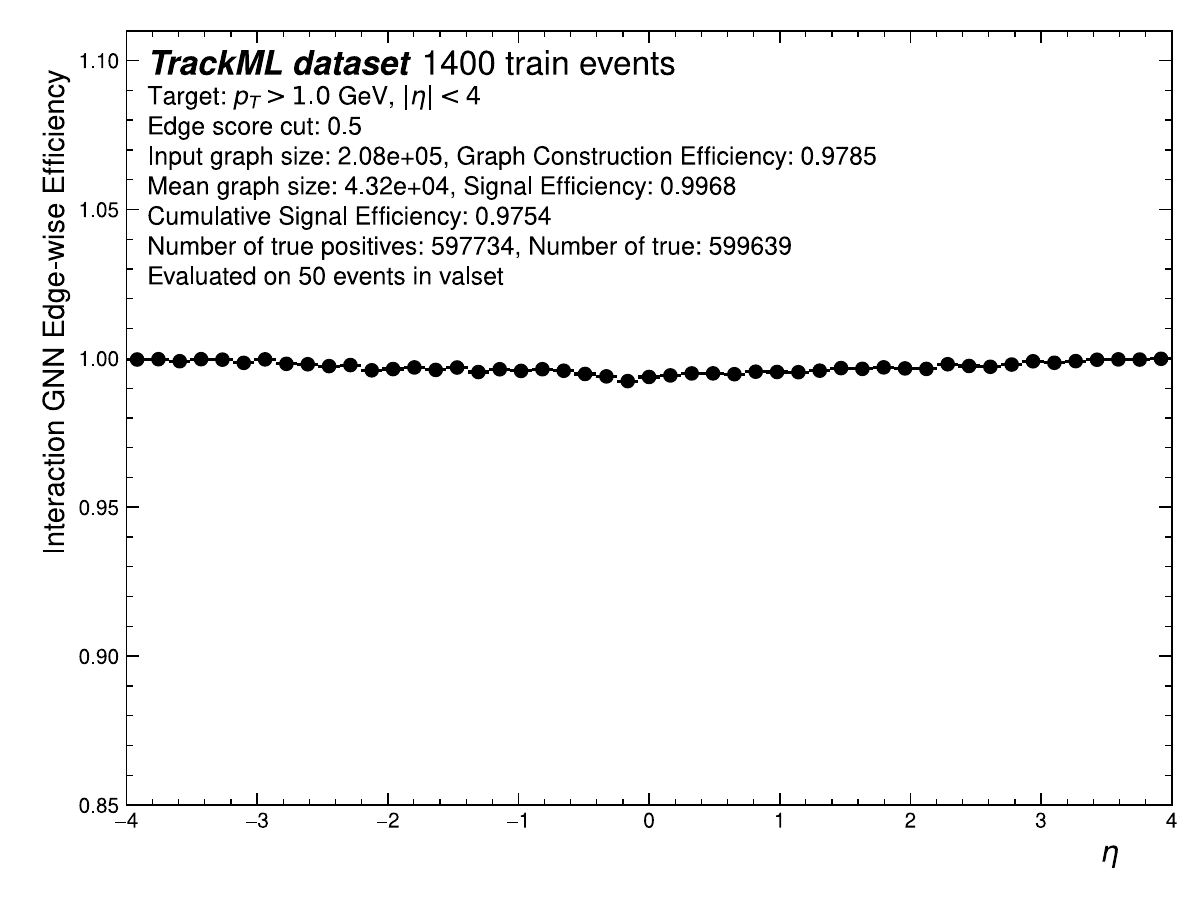}
    \caption{\centering GNN efficiency vs $\eta$.}
    \label{fig:Eff_vs_eta}
\end{figure}

\begin{figure}[H]
    \centering
    \includegraphics[width=.5\textwidth]{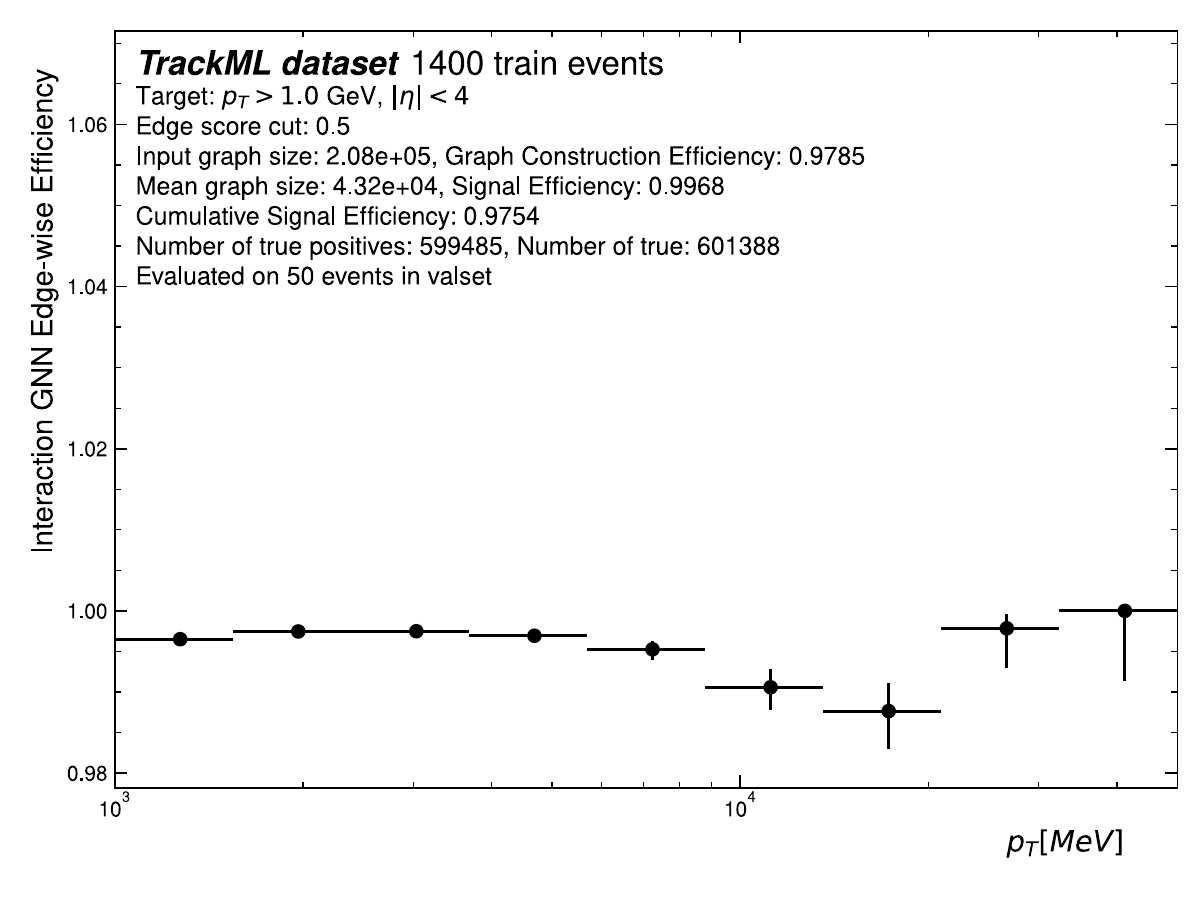}
    \caption{\centering GNN efficiency vs $p_T$.}
    \label{fig:Eff_vs_pt}
\end{figure}

\begin{figure}[H]
    \centering
    \includegraphics[width=0.5\textwidth]{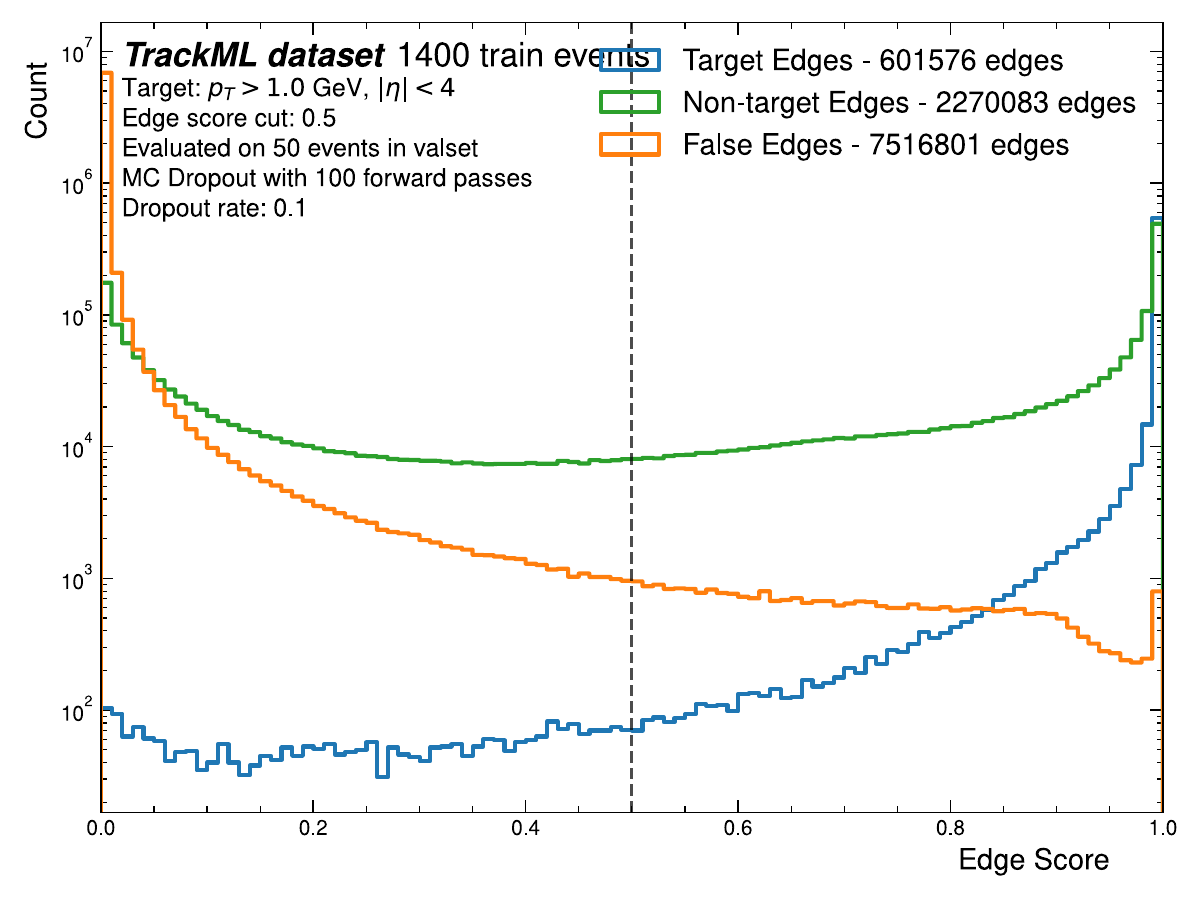}
    \caption{\centering Mean edge score $\langle s_n^{\text{GNN}}\rangle$ distribution of 100 MCD samples.}
    \label{fig:score_distribution}
\end{figure}

From now until Subsection \ref{subsec:calibration}, all the scores we discuss are not calibrated. Before diving in the quantification and propagation of uncertainty, we ensure that the pipeline is properly trained and reaches high performance by running some tests. Figures \ref{fig:Eff_vs_eta} and \ref{fig:Eff_vs_pt} show the GNN efficiency against $\eta$ and $p_T$.
 The total efficiency of the pipeline after the GNN is $97.54\%$ and the purity is $98.39\%$. The purity in the $(rz)$-plane plot can be found on Figure \ref{fig:gnn_purity} in Appendix \ref{app:other_gnn_information}. The efficiency of the GNN is defined in the same fashion as in Equation \eqref{eq:efficiency_tracking}, as the fraction of true edges that are above a score threshold of $1/2$. The purity is defined as the proportion of edges with a score above $s=1/2$ that are actually true edges. An other important measure to keep in mind of the edge score predictions distribution. This is shown on Figure \ref{fig:score_distribution}. Besides showing the ability of the GNN to correctly assign a score near 0 (resp. 1) to the majority of the false (resp. true) edges, this shows that the majority of the scores are concentrated in the extreme score regions.

\subsection{Uncertainty Quantification \& Propagation}
\label{subsec:uncertainty_comp}
\begin{figure}[H]
    \centering
    \includegraphics[width=.5\textwidth]{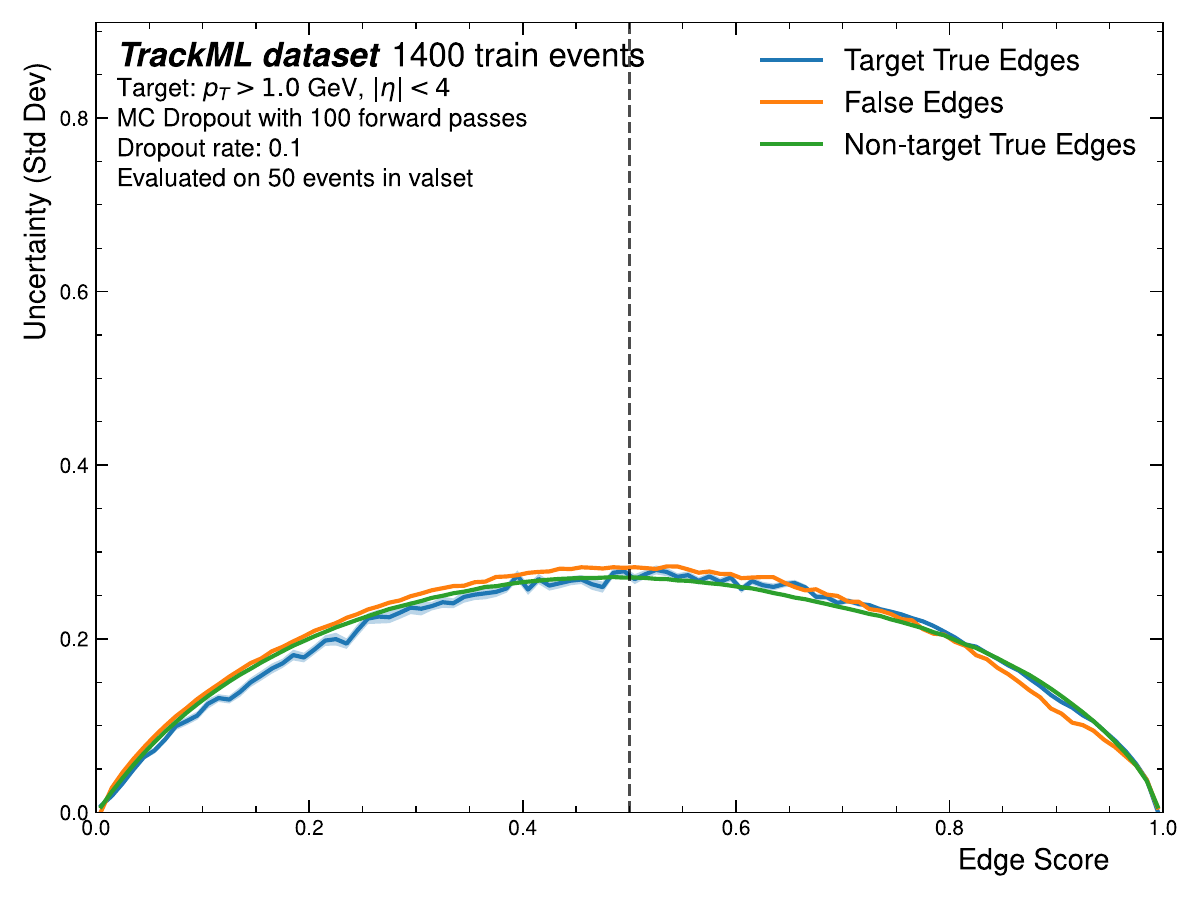}
    \caption{\centering GNN intrinsic uncertainty measured as $\sigma_n^{\text{GNN}}$ vs $\langle s_n^{\text{GNN}}\rangle$.}
    \label{fig:intrisic_uncertainty}
\end{figure}

\begin{figure}[H]
    \centering
    \includegraphics[width=.5\textwidth]{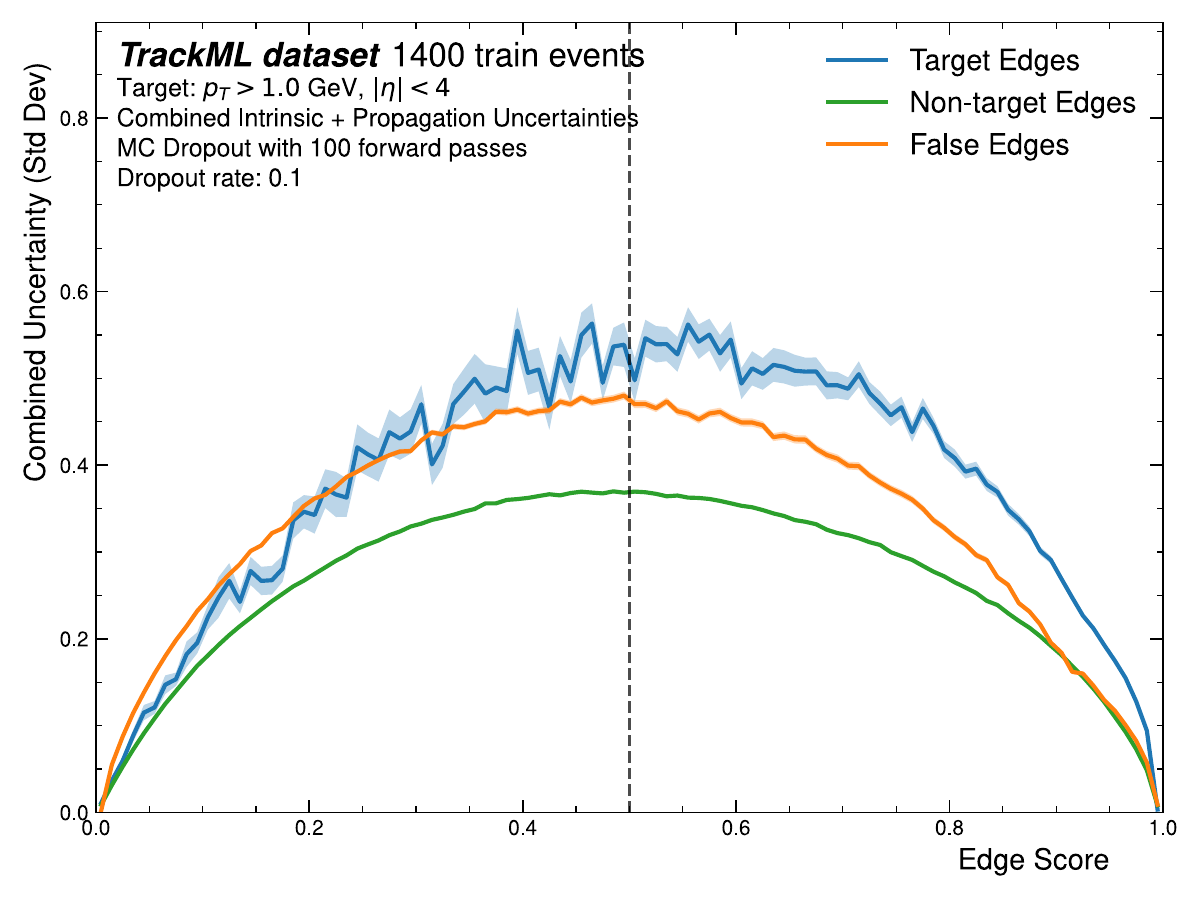}
    \caption{\centering GNN combined uncertainty $\sigma_{n}^{\text{comb}}$.}
    \label{fig:combined_uncertainty}
\end{figure}

As described in Sections \ref{subsec:source_of_uncertainty} and \ref{sec:methods} there are different ways to measure the uncertainty of the GNN and to separate various sources of uncertainty. Figure \ref{fig:intrisic_uncertainty} shows the intrinsic uncertainty of the GNN measured as the standard deviation of the edge score prediction $\sigma_n^{\text{GNN}}$. The combined GNN uncertainty $\sigma_{n}^{\text{comb}}$ is shown on Figure \ref{fig:combined_uncertainty}.
One can observe that the combined uncertainty is roughly equally distributed between the intrinsic and propagated uncertainties. We also see that $\sigma_n^{\text{comb}}$ is skewed for target and false edges. This is due to the topological uncertainty. True target tracks contain at least three edges and have, by physical constraints, certain topology properties. Therefore, if a true target edge is pruned by the Filter, it is likely that all the track edges will be assigned lower scores. The high score true target edges get a higher uncertainties, skewing the distribution to the right. The same reasoning applied to for false edges lead to a skewing to the left.

\begin{figure}[H]
    \centering
    \includegraphics[width=.5\textwidth]{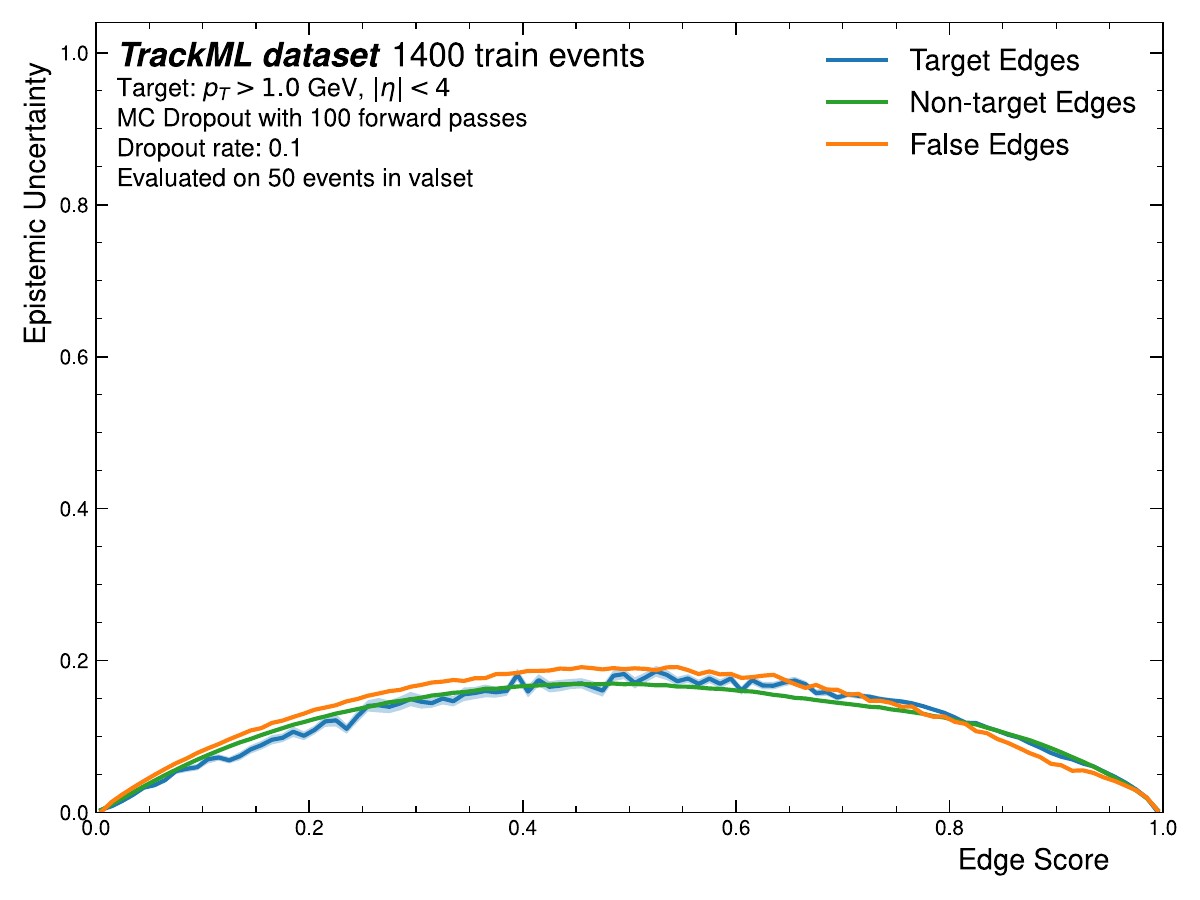}
    \caption{\centering GNN intrinsic epistemic uncertainty $\mathbb I\left[s_n^{\text{GNN}}\right]$ vs $\langle s_n^{\text{GNN}}\rangle$.}
    \label{fig:intrinsic_epistemic_uncertainty}
\end{figure}

\begin{figure}[H]
    \centering
    \includegraphics[width=.5\textwidth]{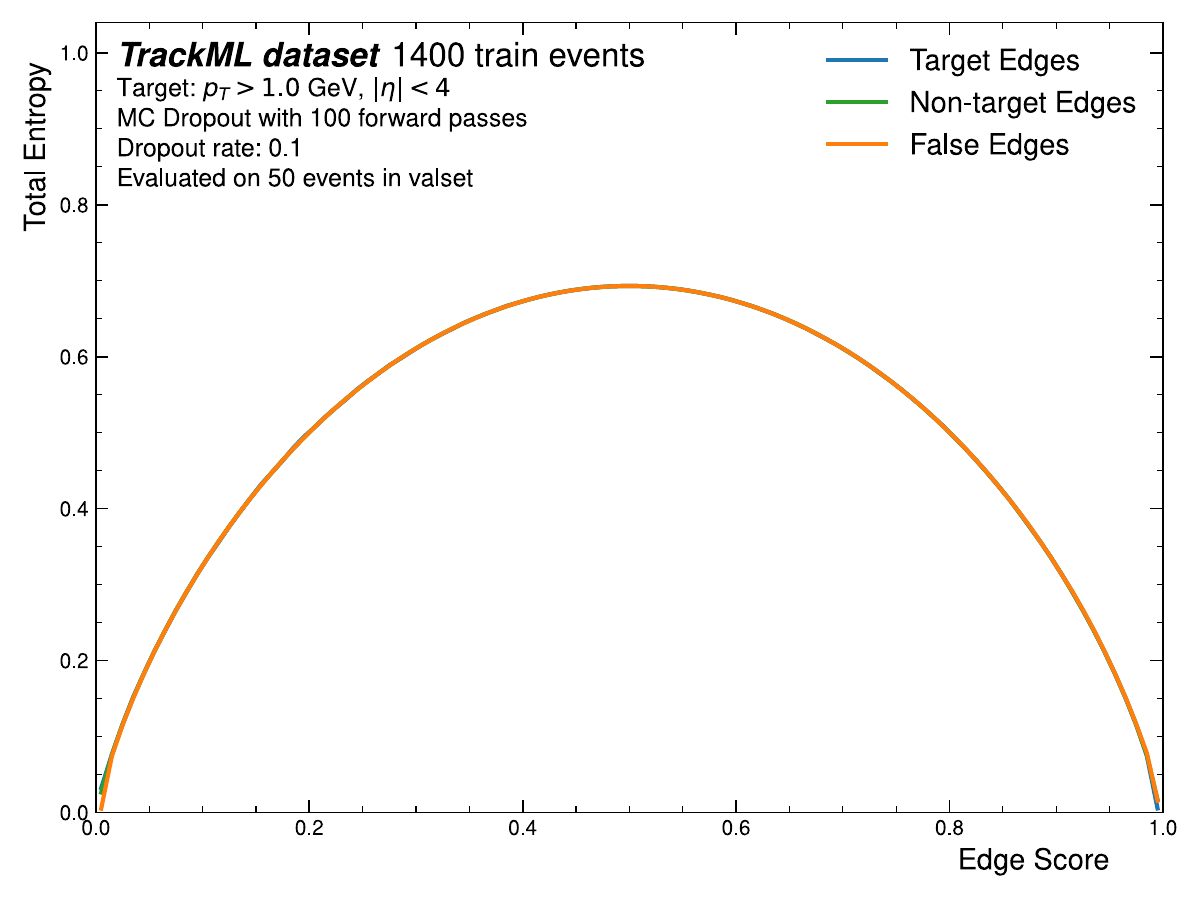}
    \caption{\centering GNN intrinsic total uncertainty $\mathbb H\left[\langle s_n^{\text{GNN}}\rangle\right]$ vs $\langle s_n^{\text{GNN}}\rangle$. The curves overlap almost fully.}
    \label{fig:intrinsic_total_uncertainty}
\end{figure}

\begin{figure}[H]
    \centering
    \includegraphics[width=.5\textwidth]{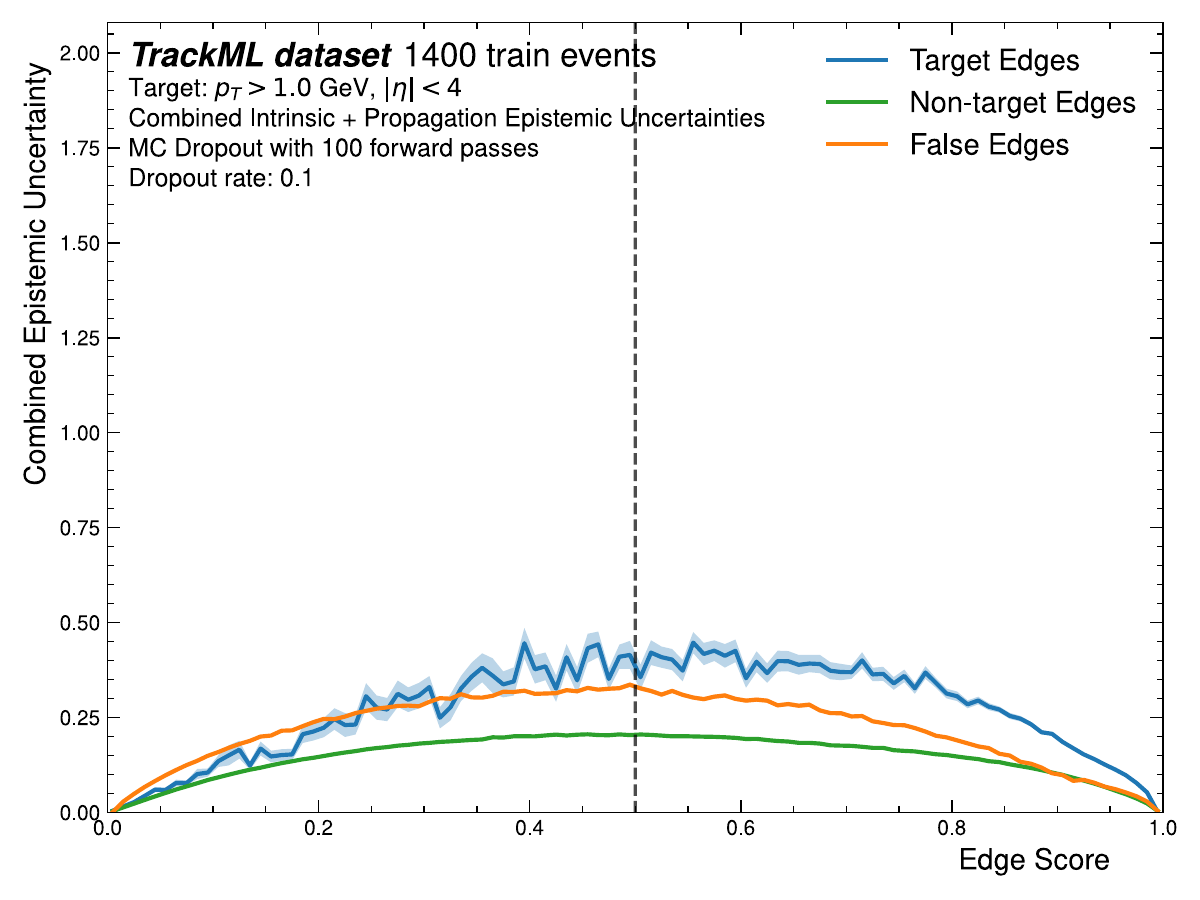}
    \caption{\centering GNN combined epistemic uncertainty.}
    \label{fig:combined_epistemic_uncertainty}
\end{figure}

\begin{figure}[H]
    \centering
    \includegraphics[width=.5\textwidth]{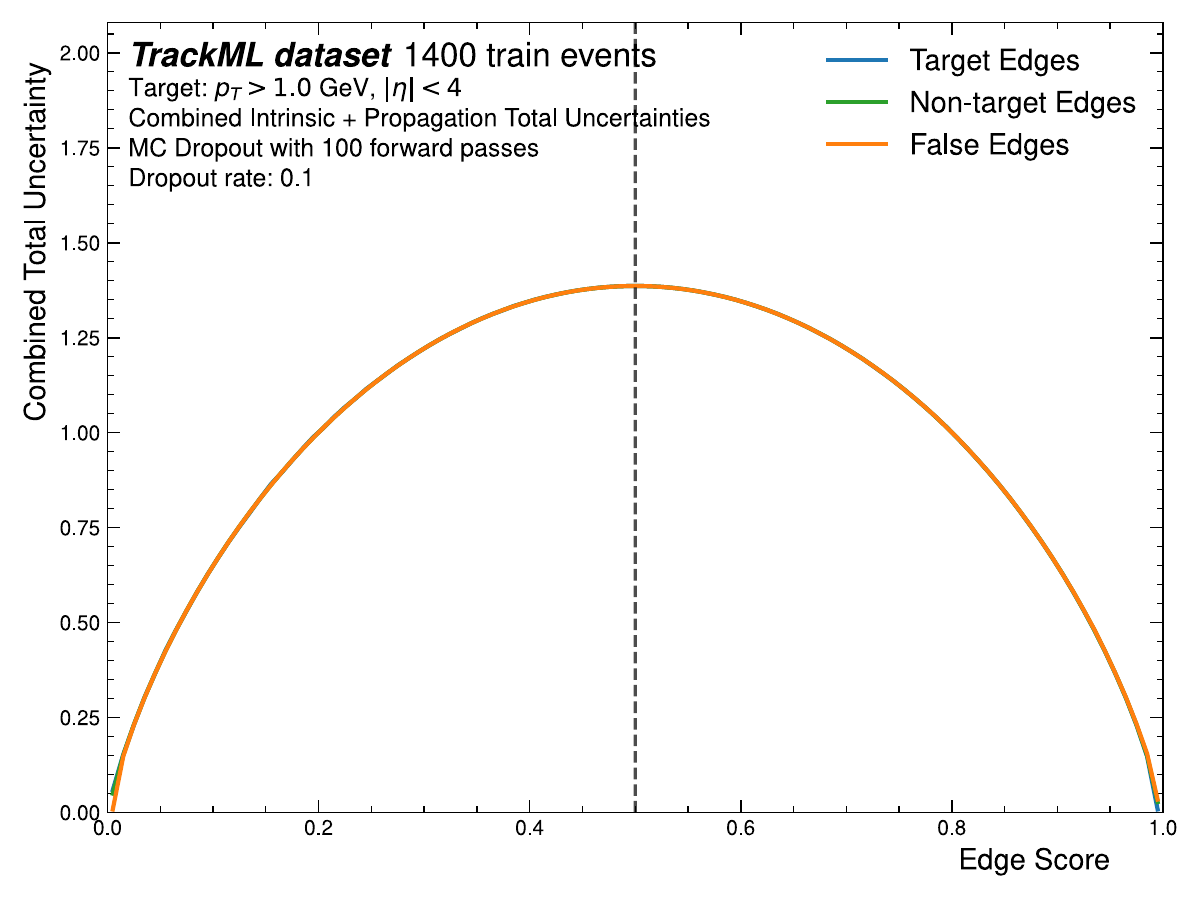}
    \caption{\centering GNN combined total uncertainty. The curves overlap almost fully.}
    \label{fig:combined_total_uncertainty}
\end{figure}

\begin{figure}[H]
    \centering
    \includegraphics[width=.5\textwidth]{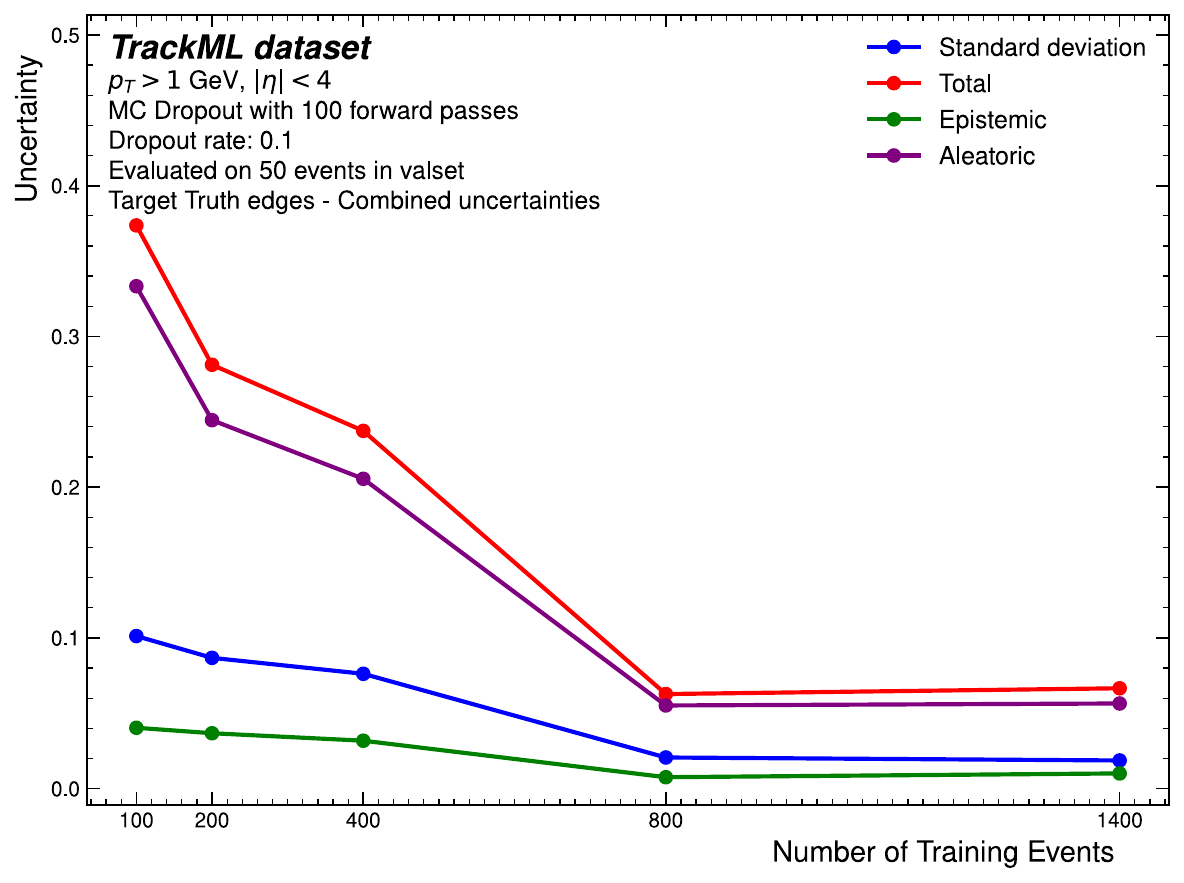}
    \caption{\centering Combined uncertainties measurements vs $|\mathcal D^{\text{train}}|$ for target edges.}
    \label{fig:uncertainty_vs_dataset_size}
\end{figure}

The same analysis is made for the epistemic uncertainty $\mathbb I\left[s_n^{\text{GNN}}\right]$ and total uncertainty $\mathbb H\left[\langle s_n^{\text{GNN}}\rangle\right]$. Figure \ref{fig:intrinsic_epistemic_uncertainty} (resp. \ref{fig:intrinsic_total_uncertainty}) shows the intrinsic epistemic (resp. total) uncertainty.
Figure \ref{fig:combined_epistemic_uncertainty} shows the combined epistemic uncertainty and Figure \ref{fig:combined_total_uncertainty} shows the combined total uncertainty.
We recover the same features observed with the standard deviation $\sigma_n^{\text{GNN}}$ measurement and especially that the uncertainties on the target and false edges are skewed when calculated with its propagated part. This is also due to the topological effect previously described. These figures allow us to see that the epistemic uncertainty is not dominant, i.e. the total GNN uncertainty, whether is it intrinsic or combined, is dominated by the aleatoric uncertainty. To verify this result, we computed the different uncertainties obtained with models trained on different dataset sizes. Figure \ref{fig:uncertainty_vs_dataset_size} shows the various combined uncertainty measurements obtained with training dataset sizes ranging from 100 to 1,400 events for target edges only.
We observe that the uncertainty is decreasing as the training dataset size increases, as expected. This plot also confirms the dominant behavior of the aleatoric uncertainty for all training dataset sizes. 

\begin{figure}[H]
    \centering
    \includegraphics[width=0.5\textwidth]{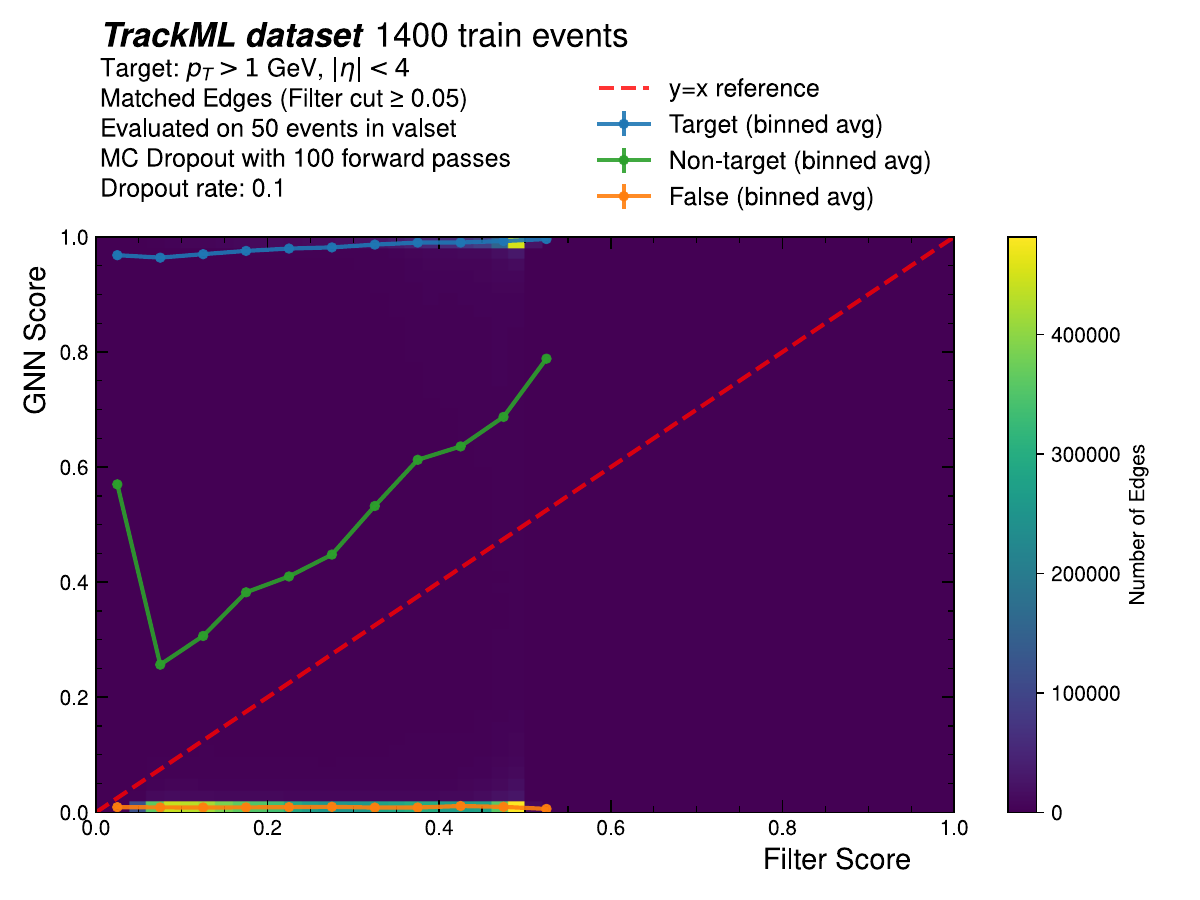}
    \caption{\centering Mean prediction evolution function $\langle s_n^{\text{GNN}}\rangle(\langle s_n^{\text{Filter}}\rangle)$.}
    \label{fig:mean_pred_function}
\end{figure}

\begin{figure}[H]
    \centering
    \includegraphics[width=0.5\textwidth]{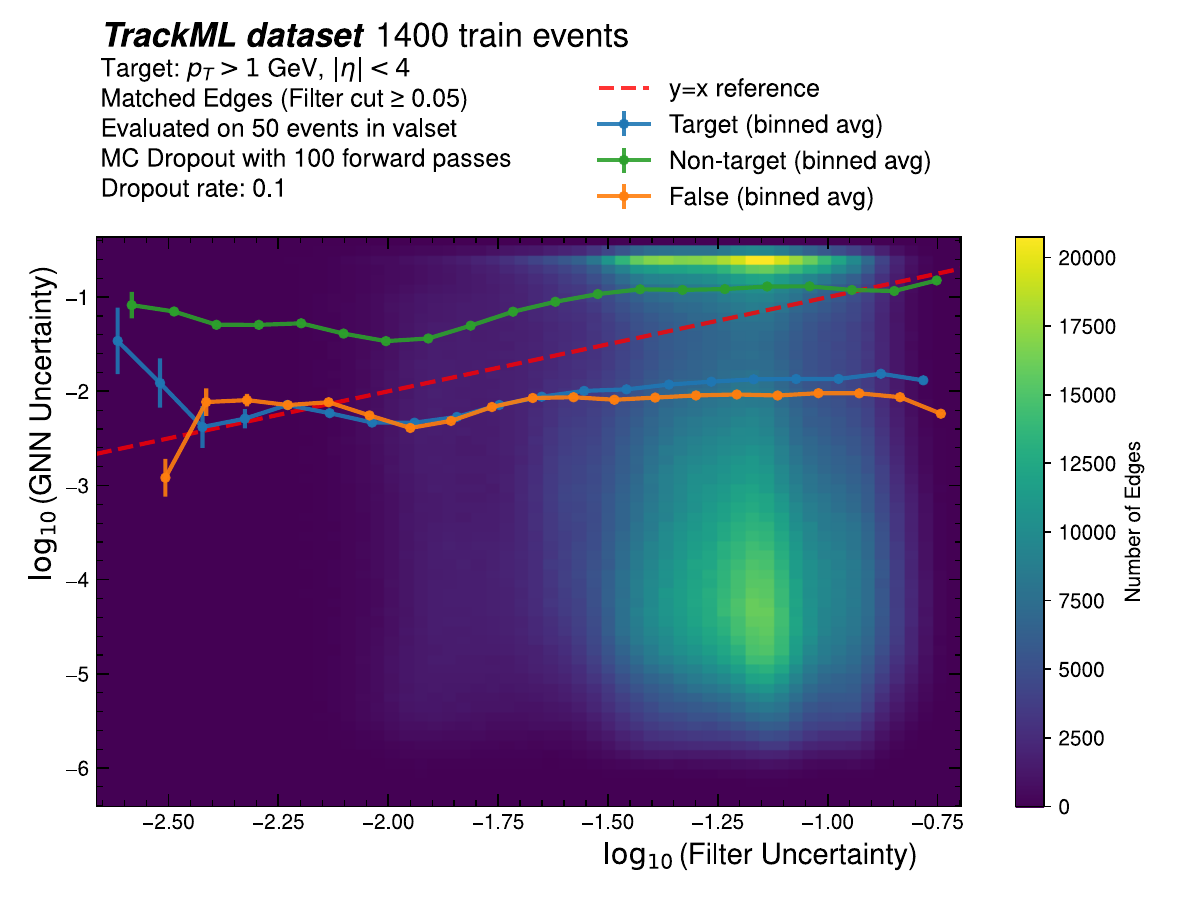}
    \caption{\centering Standard deviation evolution function $\sigma^{\text{GNN}}_n(\sigma^{\text{Filter}}_n)$.}
    \label{fig:sigma_gnn_vs_sigma_filter}
\end{figure}

\begin{figure}[H]
    \centering
    \includegraphics[width=.5\textwidth]{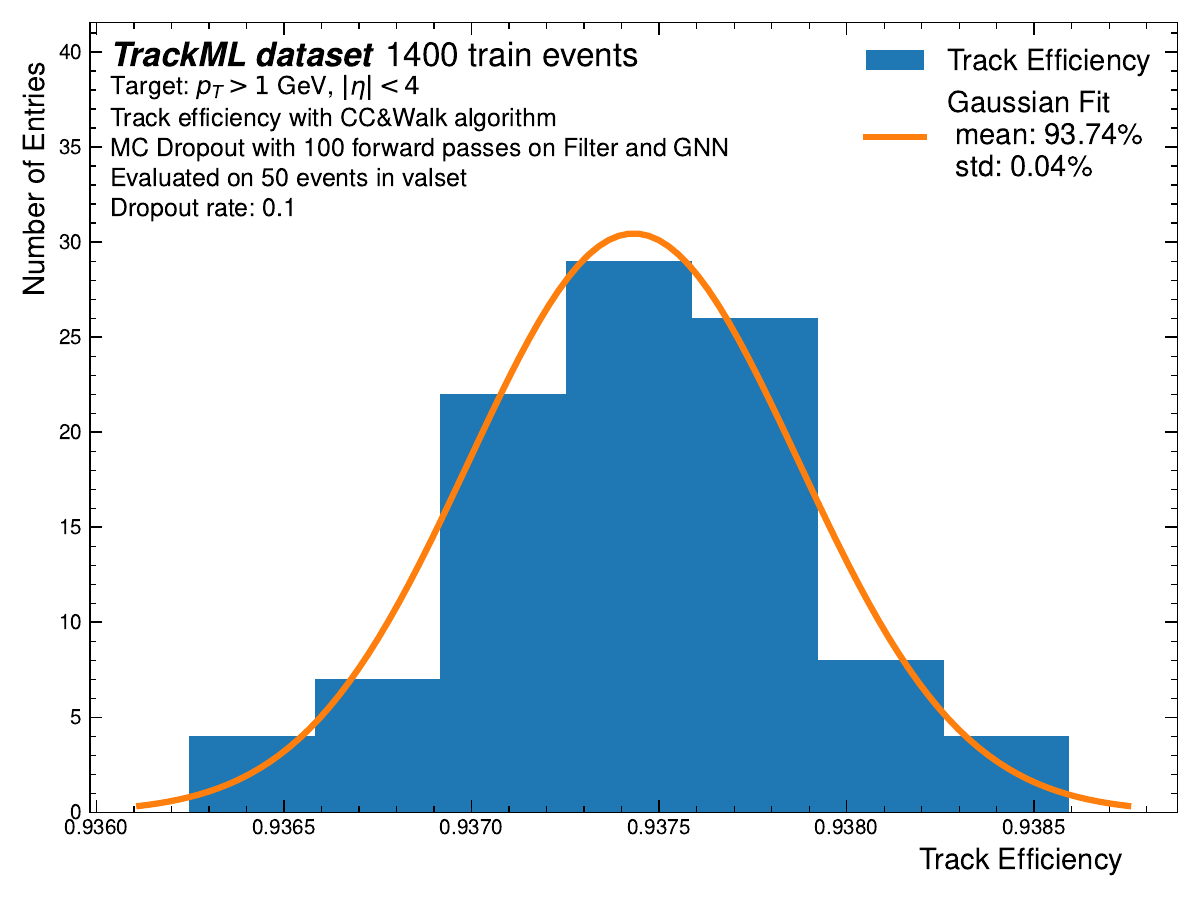}
    \caption{\centering Track building efficiencies.}
    \label{fig:track_eff_histo}
\end{figure}

As discussed in Section \ref{sec:uncer_prop}, it is likely impossible to provide an analytical treatment of the UP for a realistic case such as our. Figure \ref{fig:mean_pred_function} shows the average behavior of the GNN scores predictions as function of the Filter scores (equivalent of $\mu_n$ function in Eq. \eqref{eq:law_total_var}).
We observe that, for target and false edges, the score predictions of the GNN is largely independent of the Filter score. Figure \ref{fig:sigma_gnn_vs_sigma_filter} shows the average behavior of the standard deviation of the GNN as a function of the standard deviation of the Filter (equivalent of $\varsigma_n$ function in Eq. \eqref{eq:law_total_var}). No power-law scaling is observed.
So far, we have considered the uncertainty propagated from the Filter to the GNN and we now evaluate the impact of these uncertainties on the final stage: the track building. We proceed as described in Subsection \ref{subsec:methods_uncer_prop}. Figure \ref{fig:track_eff_histo} shows this histogram.
We observe that the uncertainty on the track building efficiency is approximately $0.05$\%. This shows that the ACORN pipeline is robust against its components uncertainties or at least that the upstream uncertainties are small enough to not affect the track building efficiency significantly. 
\subsection{$\sigma_n^{\text{GNN}}$ dependency on $\eta$ and $p_T$}
\begin{figure}[H]
    \centering
    \includegraphics[width=.5\textwidth]{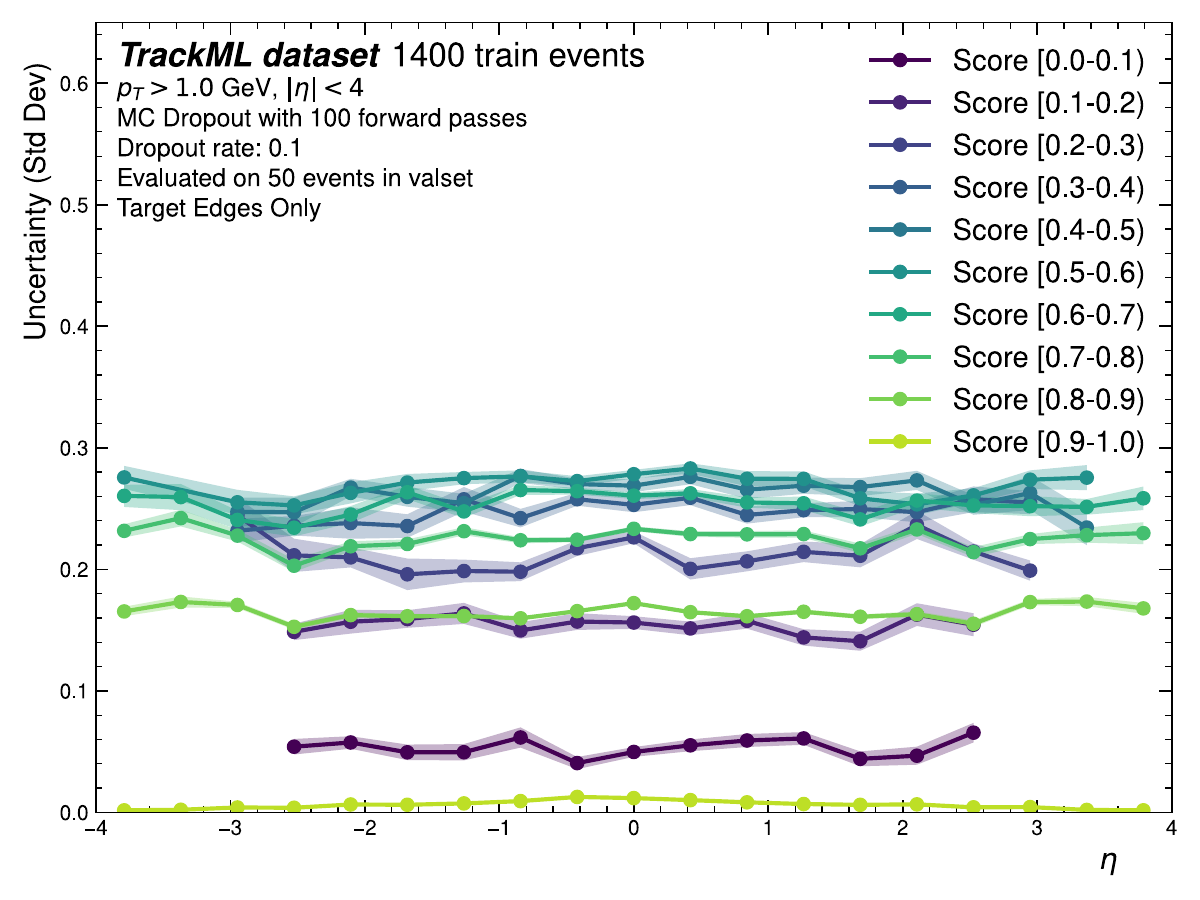}
    \caption{\centering $\sigma_n^{\text{GNN}}$ vs $\eta$ for target edges.}
    \label{fig:sigma_vs_eta}
\end{figure}

\begin{figure}[H]
    \centering
    \includegraphics[width=.5\textwidth]{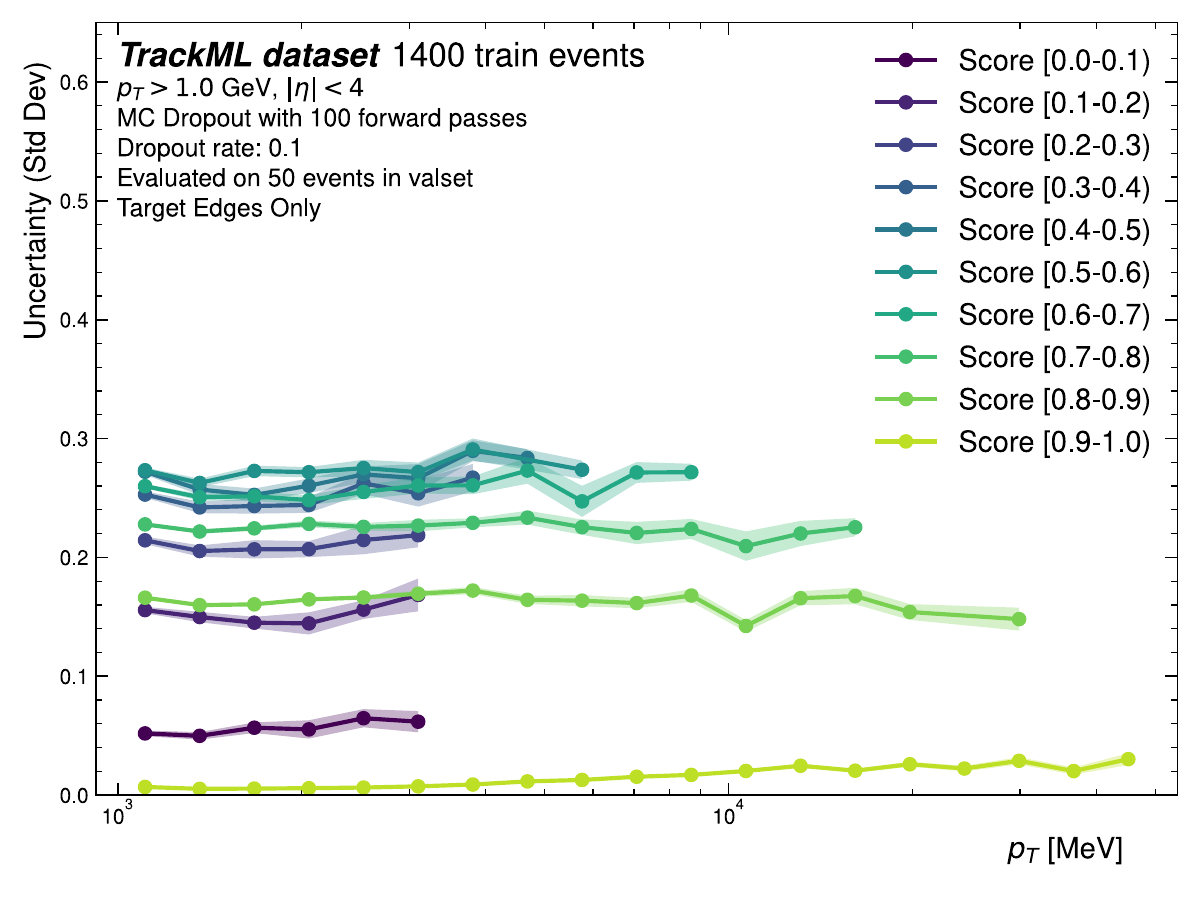}
    \caption{\centering $\sigma_n^{\text{GNN}}$ vs $p_T$ for target edges.}
    \label{fig:sigma_vs_pt}
\end{figure}

\begin{figure}[H]
    \centering
    \includegraphics[width=.5\textwidth]{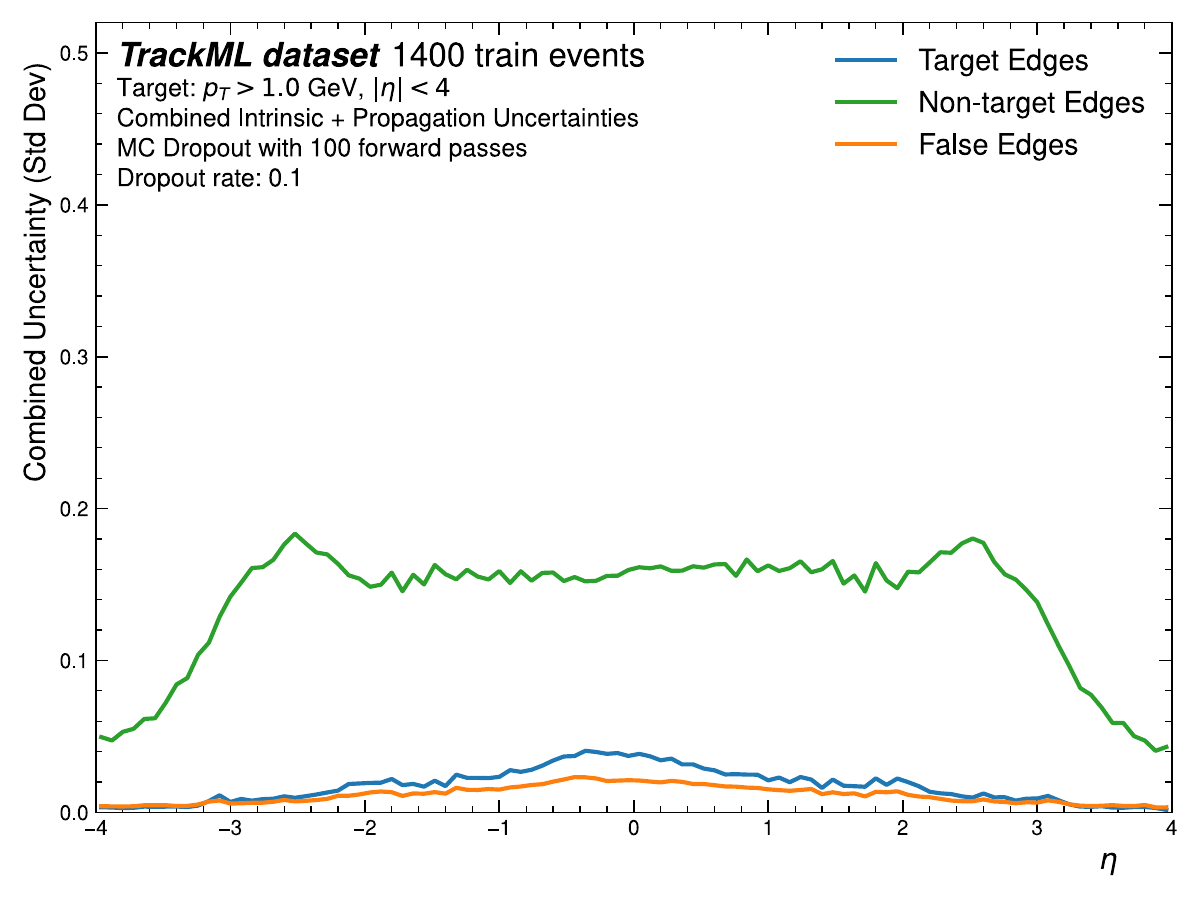}
    \caption{\centering Combined uncertainty $\sigma_n^{\text{comb}}$ vs $\eta$.}
    \label{fig:total_std_uncertainty_vs_eta}
\end{figure}

\begin{figure}[H]
    \centering
    \includegraphics[width=.5\textwidth]{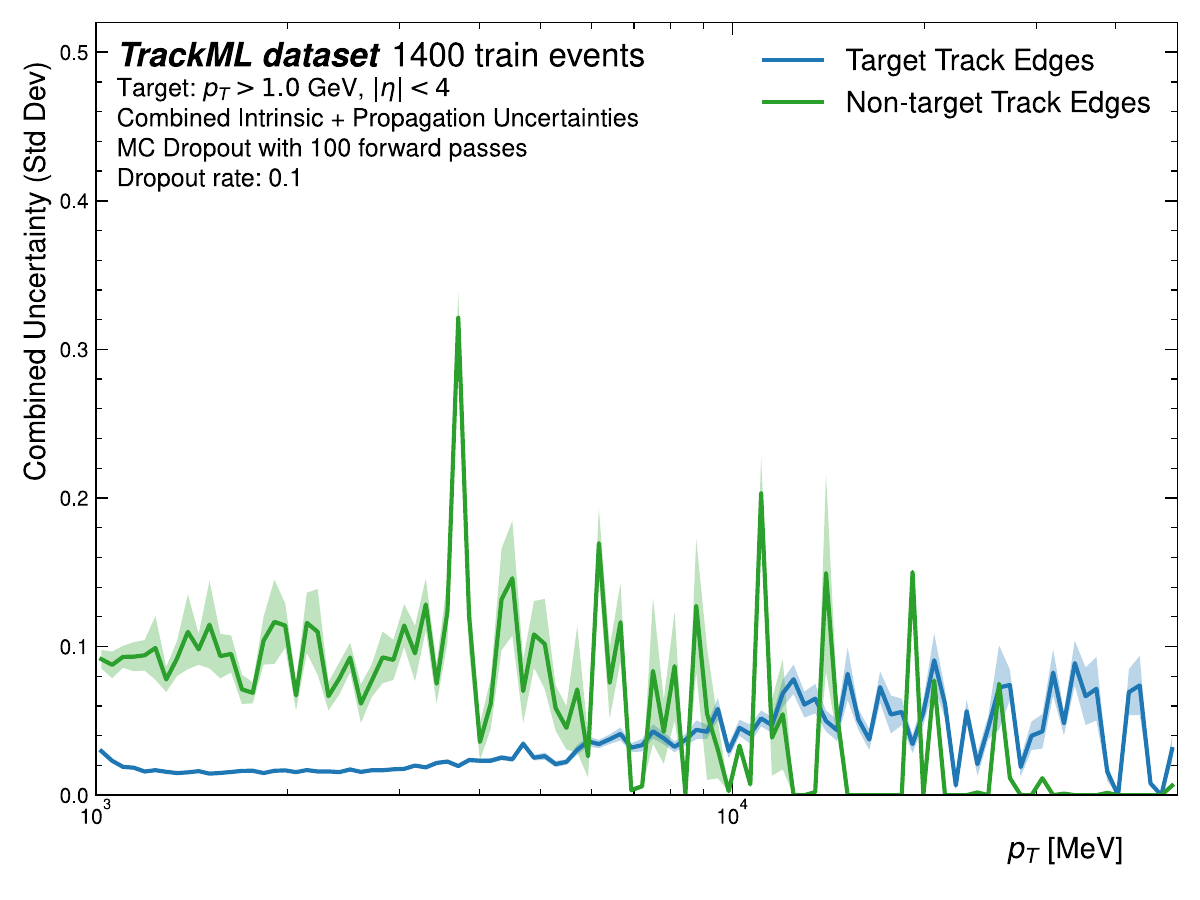}
    \caption{\centering Combined uncertainty $\sigma_n^{\text{comb}}$ vs $p_T$.}
    \label{fig:total_std_uncertainty_vs_pt}
\end{figure}

We test how detector geometry and transverse momentum $p_T$ affect uncertainty to verify pipeline consistency. Figure \ref{fig:sigma_vs_eta} shows that the GNN uncertainty $\sigma_n^{\text{GNN}}$ is mostly independent of $\eta$, with a slight rise in the central region due to the concentration of false edges near $\eta \sim 0$ (see Figure \ref{fig:app_number_of_edges_eta} in Appendix \ref{app:other_gnn_information}).
Figure \ref{fig:sigma_vs_pt} shows a slight increase of $\sigma_n^{\text{GNN}}$ with $p_T$. This increase is also explained by the low amount of high $p_T$ tracks in the dataset (see Figure \ref{fig:app_pt_spectrum} in Appendix \ref{app:other_gnn_information}).
We perform the same analysis for the combined uncertainty $\sigma_n^{\text{comb}}$. Figure \ref{fig:total_std_uncertainty_vs_eta} shows the behavior of the combined uncertainty against the pseudo rapidity $\eta$. One can observe a slight rise of the uncertainty (for target and false edges) in the central region $\eta\simeq0$ as previously with the intrinsic uncertainty. The two effects have the same origin. 
Figure \ref{fig:total_std_uncertainty_vs_pt} shows the behavior of the combined uncertainty with $p_T$. The spikes in the non target edges are due to low statistics. Here again, a slight increase of the uncertainty with $p_T$ can be observed for target edges. This is once more due to a low training dataset size for high $p_T$ tracks.

\subsection{Properties of the probability distribution of the score prediction}
\begin{figure}[H]
    \centering
    \includegraphics[width=.5\textwidth]{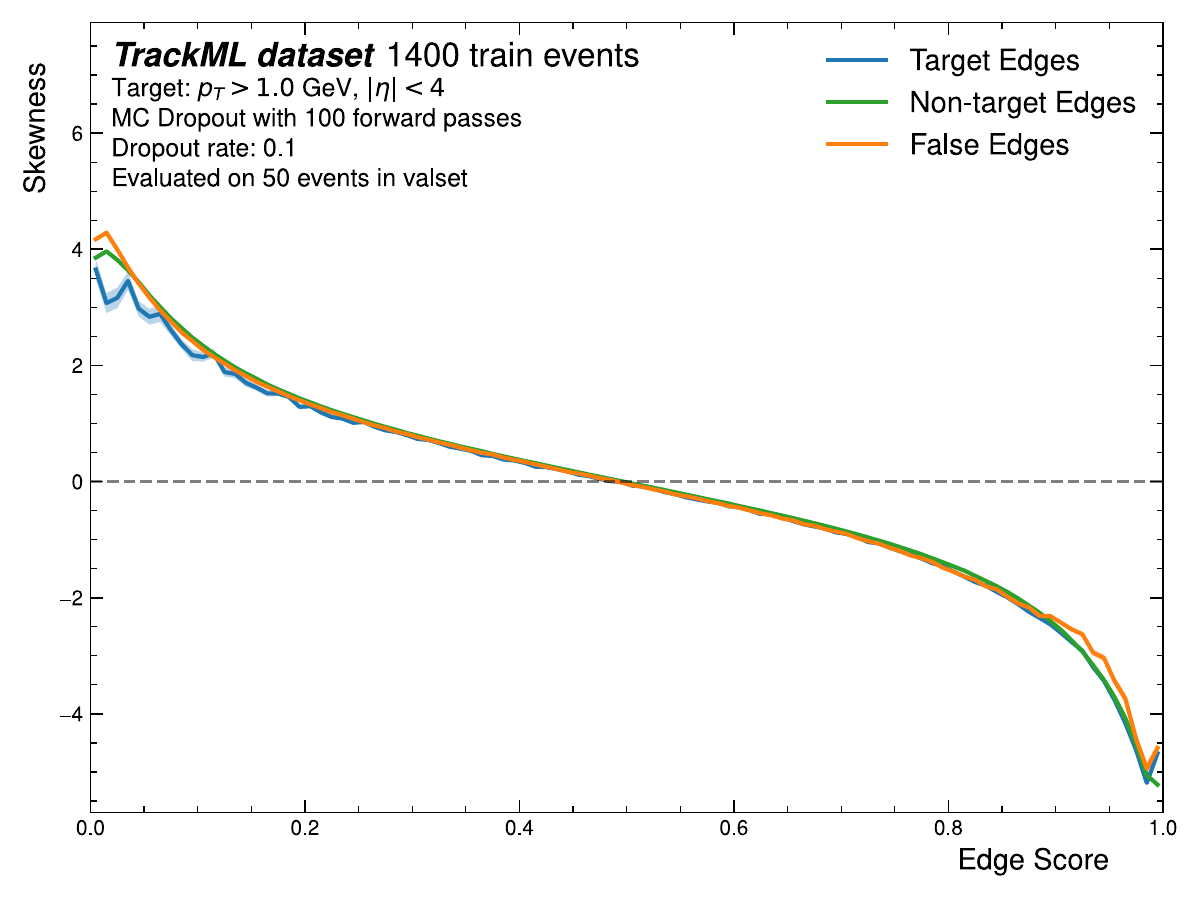}
    \caption{\centering Skewness of the MCD samples.}
    \label{fig:skewness}
\end{figure}

\begin{figure}[H]
    \centering
    \includegraphics[width=.5\textwidth]{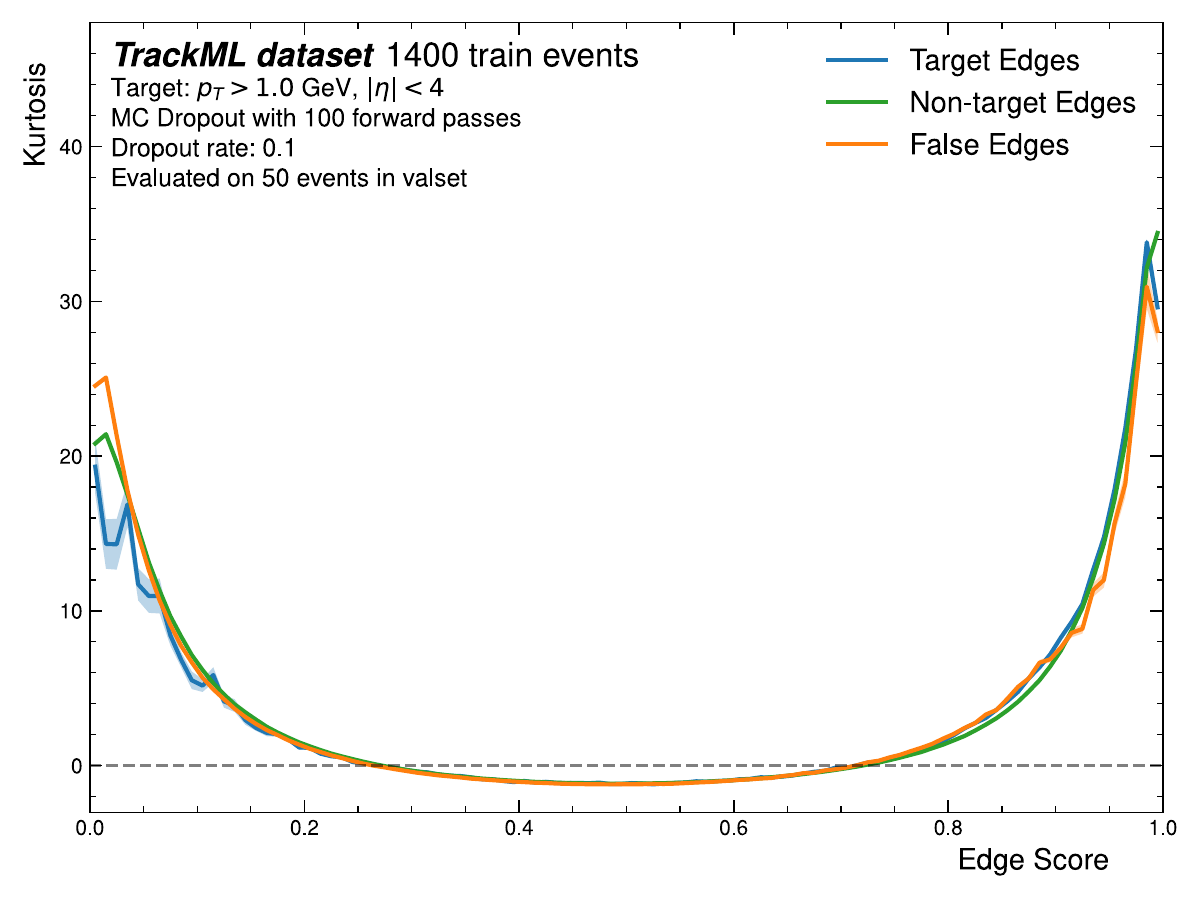}
    \caption{\centering Kurtosis of the MCD samples.}
    \label{fig:kurtosis}
\end{figure}

\begin{figure}[H]
    \centering
    \includegraphics[width=.5\textwidth]{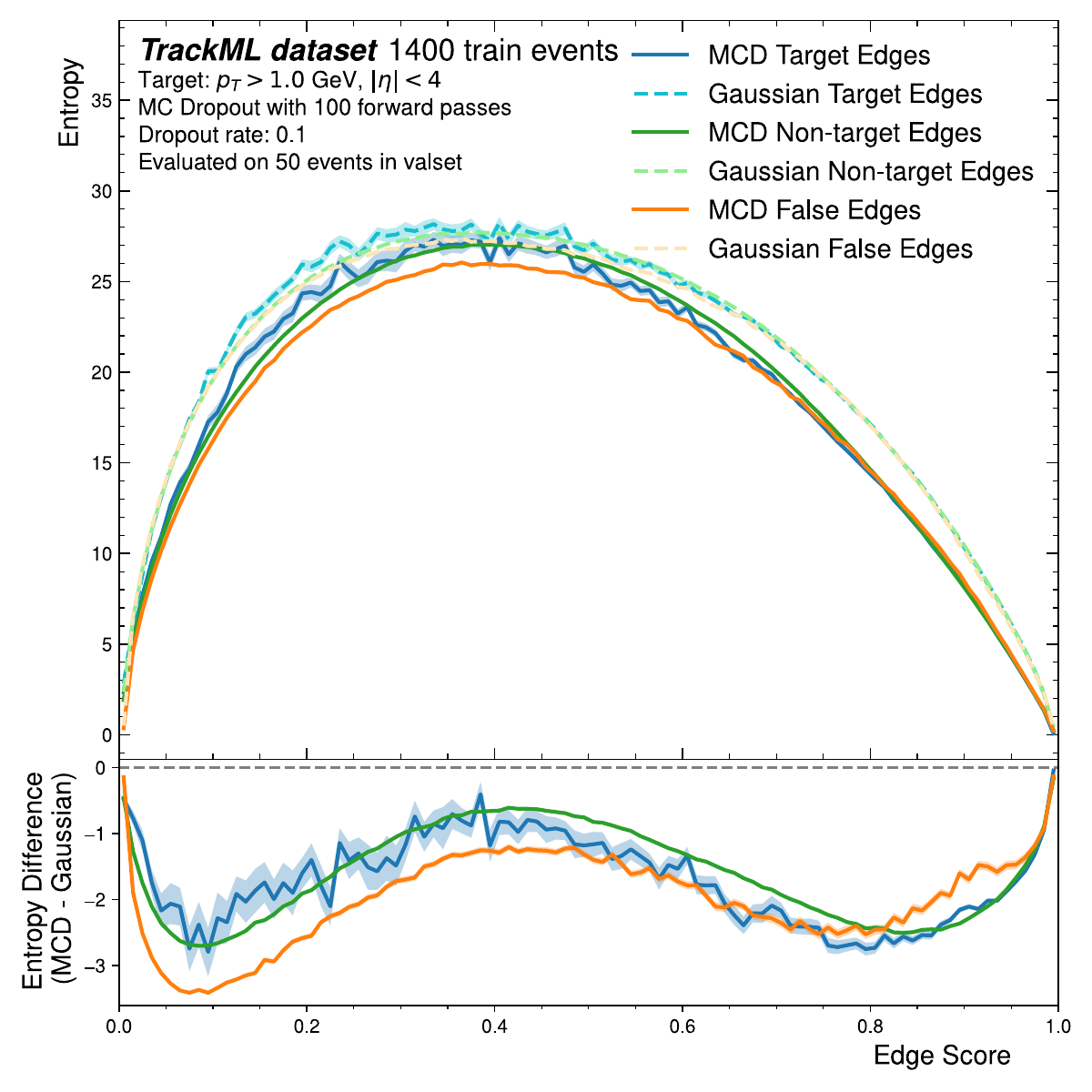}
    \caption{\centering Shanon's entropy comparison between MCD and Gaussian score distributions.}
    \label{fig:compare_entropy}
\end{figure}

As mentioned in Sections \ref{sec:prior_works_review} and \ref{sec:methods}, the goal of MCD is to model the GNN as a posterior $p(s_n|\mathcal G_n, \mathcal D^{\text{train}}, \theta)$. Since the scores are predicted in a compact set of values (see Fig. \ref{fig:score_distribution}), this posterior cannot be Gaussian for every score prediction. To show that, we propose two methods. The first one consists on computing the skewness and kurtosis of the stochastic score samples yielded by the MCD procedure. Figures \ref{fig:skewness} and \ref{fig:kurtosis} show the skewness and kurtosis of the score distribution obtained and compare them to the theoretical values for a Gaussian distribution\footnote{We use the Fisher's definition of kurtosis so that it is 0 for a Gaussian random variable.}. We observe that the samples are distributed with non-Gaussian behavior for every mean score $\langle s_n^{\text{GNN}} \rangle$. The skewness of these samples is symmetric around $s=1/2$ which is explained by the fact that the score distribution is bounded by 0 and 1 which is symmetrical around $1/2$. One could achieve this distribution by having a Gaussian error profile for each value of $\langle s_n^{\text{GNN}} \rangle$ clipped between 0 and 1. The kurtosis however is non-Gaussian in this central region, with larger tails than the Gaussian case and thinner tails in the low and high score regions. Hence, we can conclude on the non gaussianity of the empirical posterior obtained by the MCD method.
We propose a second method to obtain this non gaussianity that consists on the following approach. For a fixed edge, we gather the $T$ MCD score predictions and compute their mean and standard deviation. We then generate 100 Gaussian distributed samples (clipped in $(0,1)$) with the same mean and standard deviation. We then compare the Shanon's entropy of both distributions. Figure \ref{fig:compare_entropy} shows the results of this procedure for the complete validation dataset.
A Gaussian MCD posterior would lead to an average null difference of entropy. However, we observe a non-zero difference everywhere and a higher entropy for the Gaussian samples. This is expected in the case of non-Gaussian MCD posterior since it is known that the Gaussian distribution is the one of maximum entropy for given first and second moments.

\subsubsection{Choice of the dropout rate}
\begin{figure}[H]
    \centering
    \includegraphics[width=.5\textwidth]{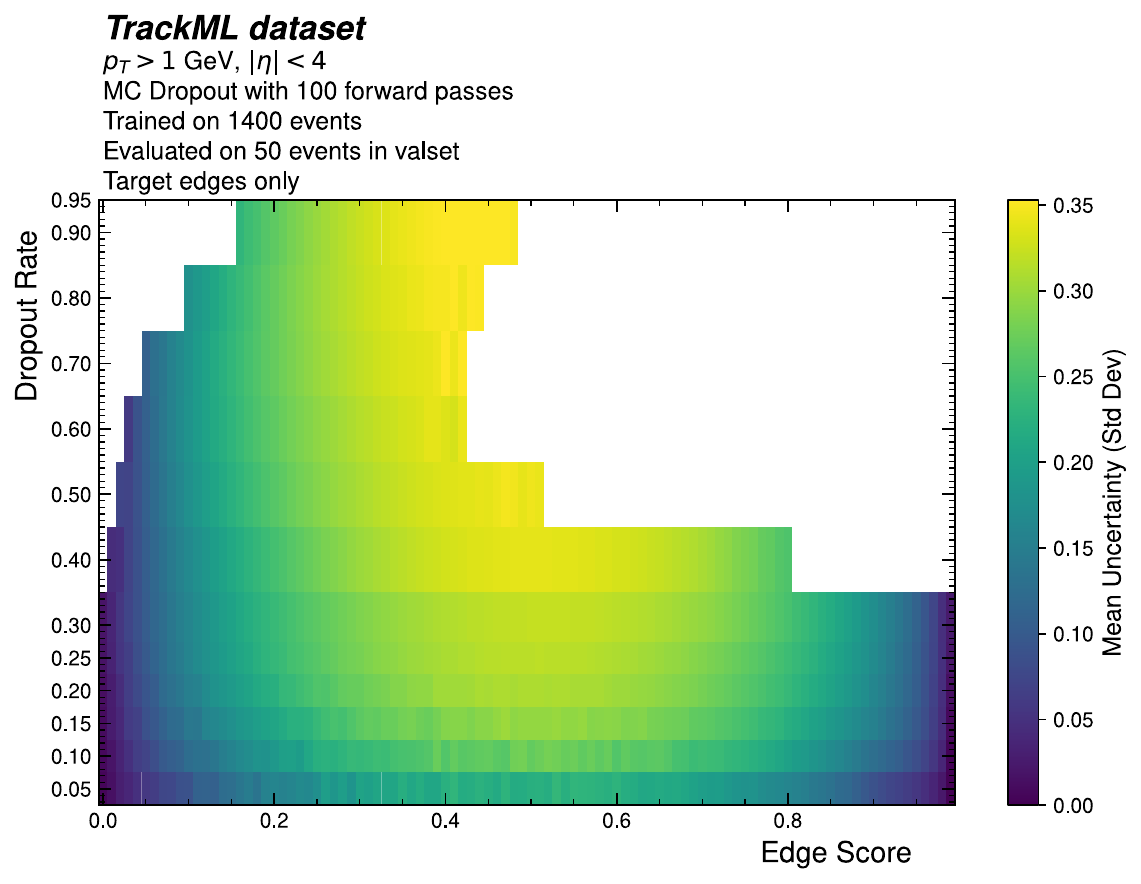}
    \caption{\centering Uncertainty $\sigma_n^{\text{GNN}}$ heatmap for varying dropout rate and $\langle s_n^{\text{GNN}} \rangle$.}
    \label{fig:sigma_vs_dropout_target}
\end{figure}

The complete analysis presented in this report has been conducted with a 10\% dropout rate. One could argue that the choice of this value, although if classical within deep learning studies, is arbitrary and therefore fails to provide an objective evaluation of the pipeline uncertainty. To overcome this potential bias, we apply the MCD procedure on the GNN with different dropout rates. For computational reasons, we utilize a single model, trained with a 10\% dropout rate, but employ dropout rates ranging from 5\% to 95\% during the stochastic inferences. Figure \ref{fig:sigma_vs_dropout_target} shows the obtained uncertainty profiles for target edges. 
One observes that the order of magnitude of uncertainty for a fixed mean edge score remains the same for dropout rates lower than 30\%. This justifies that the choice of the dropout rate does not matter and therefore a low rate can be used. The same effect is observed for false and non-target edges, the plots can be found in Appendix \ref{app:other_gnn_information}.

\subsection{Calibration of edge scores}
\label{subsec:calibration}
\begin{figure}[H]
    \centering
    \includegraphics[width=.5\textwidth]{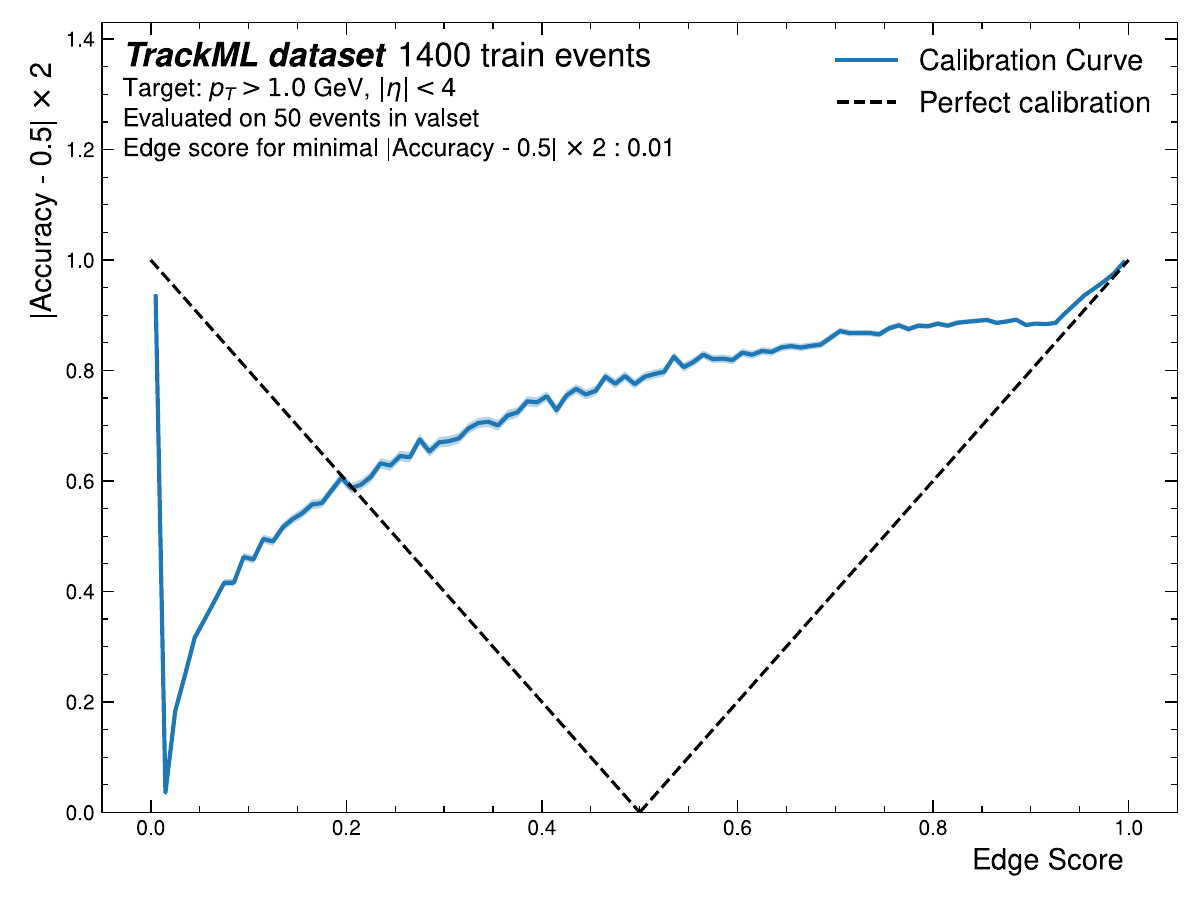}
    \caption{\centering Calibration curve for non calibrated GNN.}
    \label{fig:calibration_curve_uncal}
\end{figure}

\begin{figure}[H]
    \centering
    \includegraphics[width=.5\textwidth]{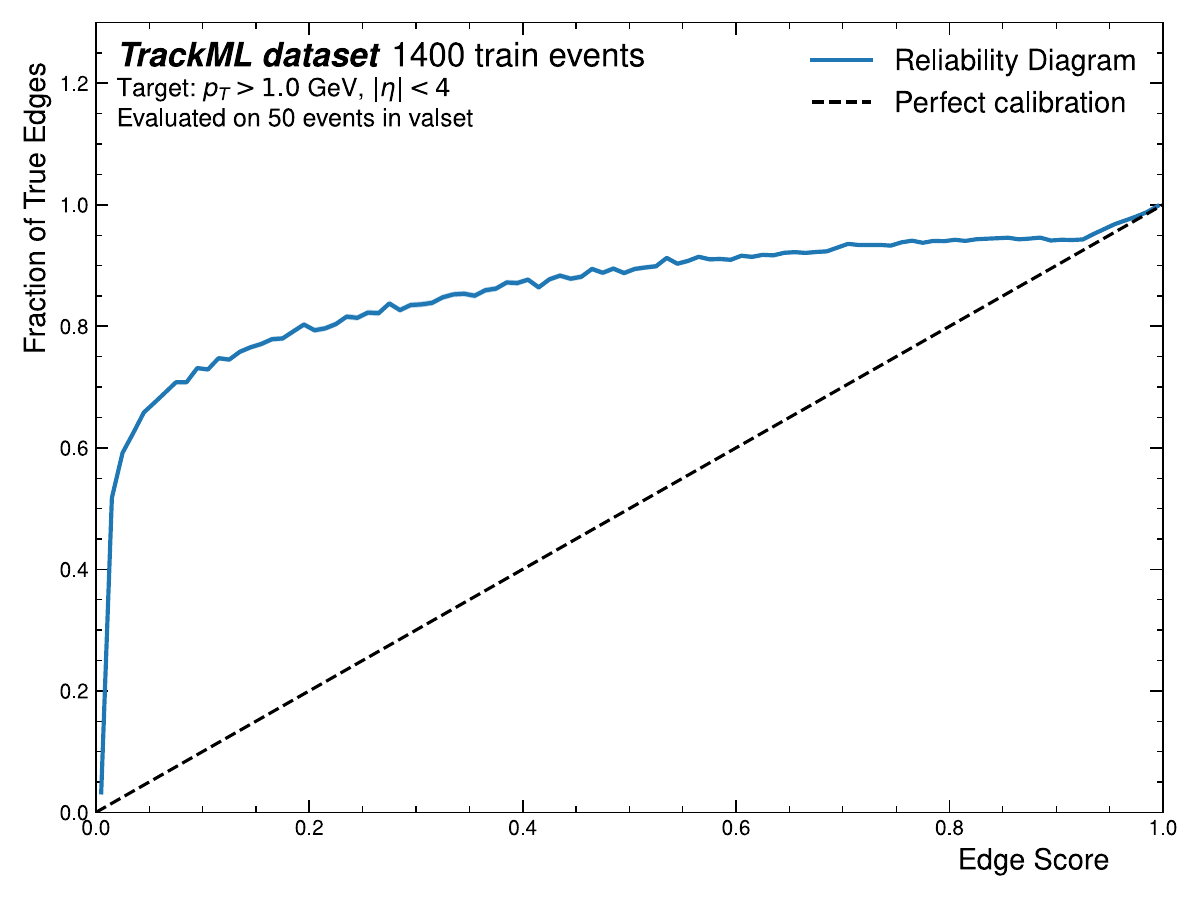}
    \caption{\centering Reliability curve for non calibrated GNN.}
    \label{fig:reliability_curve_uncal}
\end{figure}

\begin{figure}[H]
    \centering
    \includegraphics[width=.5\textwidth]{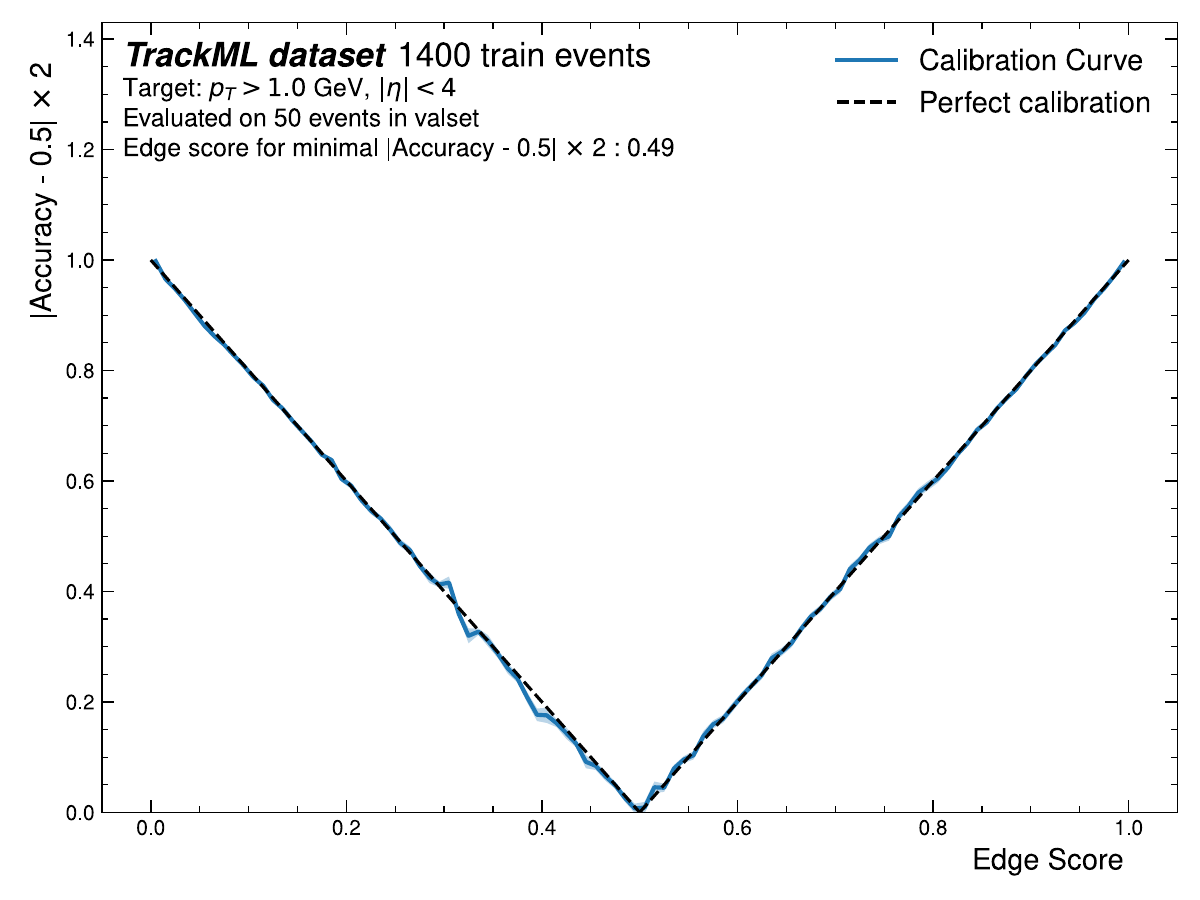}
    \caption{\centering Calibration curve for calibrated GNN.}
    \label{fig:calibration_curve_cal}
\end{figure}

\begin{figure}[H]
    \centering
    \includegraphics[width=.5\textwidth]{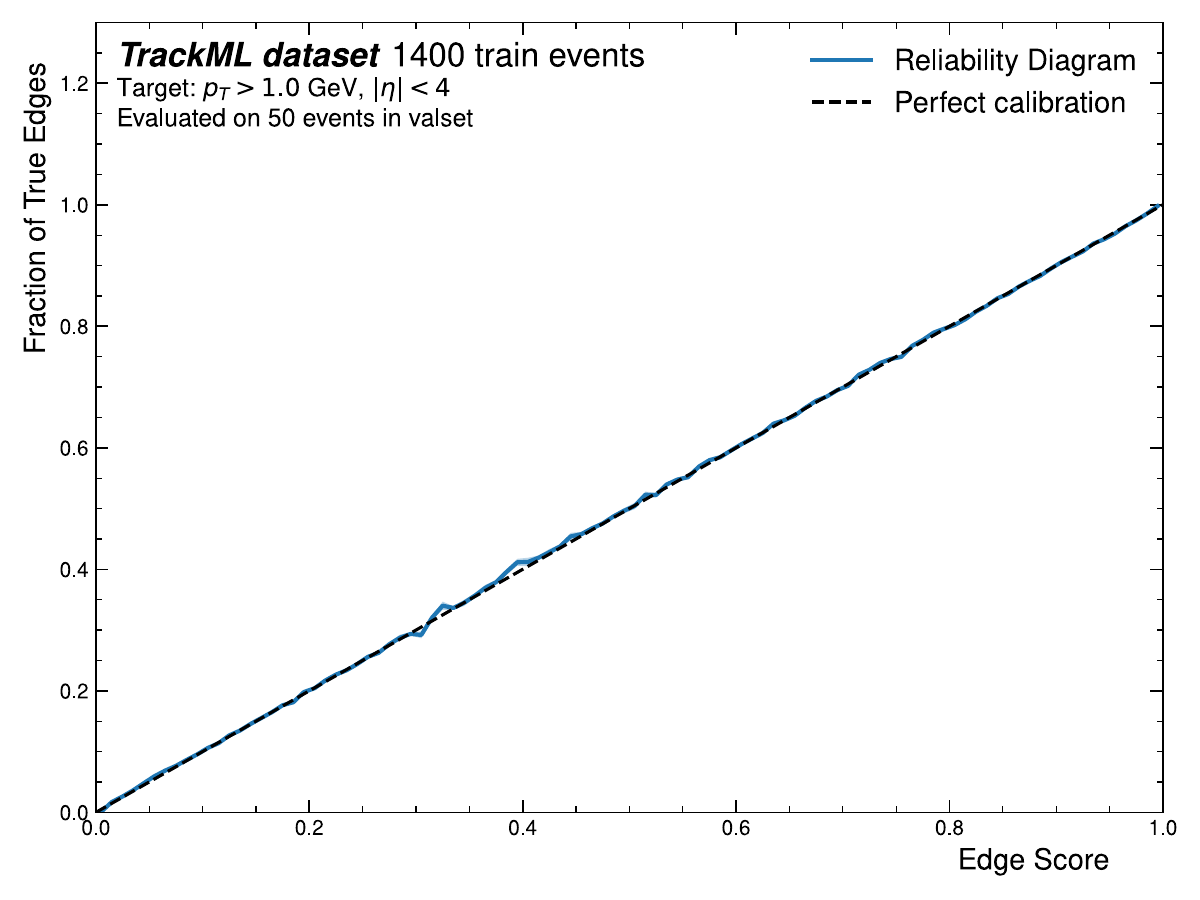}
    \caption{\centering Reliability curve for calibrated GNN.}
    \label{fig:reliability_curve_cal}
\end{figure}

\begin{figure}[H]
    \centering
    \includegraphics[width=.5\textwidth]{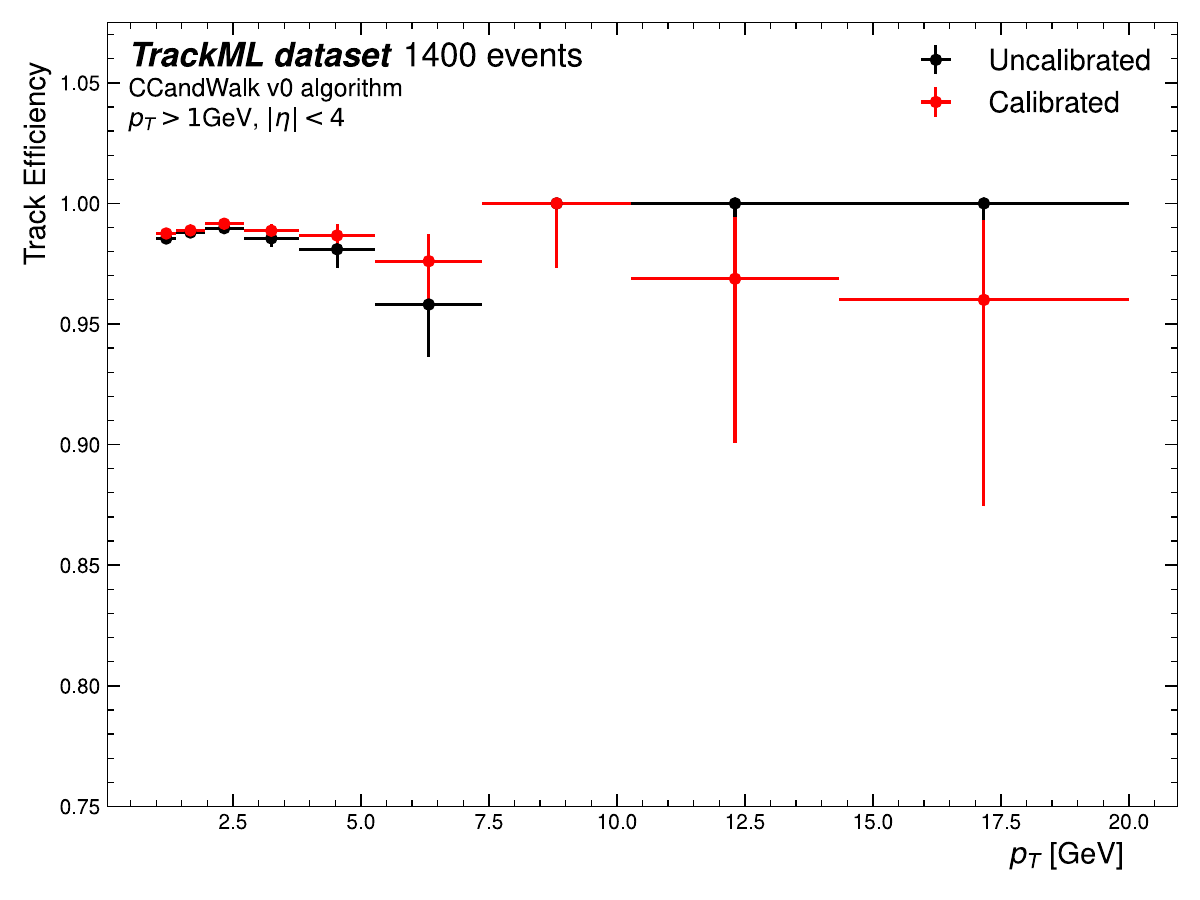}
    \caption{\centering Track building efficiency vs $p_T$ for calibrated and uncalibrated GNN.}
    \label{fig:track_calibration}
\end{figure}

To study the impact of the GNN score predictions on the track building efficiency, we use the method presented in Section \ref{sec:methods}. Figure \ref{fig:calibration_curve_uncal} and Figure \ref{fig:reliability_curve_uncal} show the calibration curve and reliability diagram for the GNN before the score calibration.
We observe a strong miscalibration of the score predictions which leads to an optimal score cut values for the connected components part of CC\&Walk of $s_{opt}=0.01$. The interpretation of the excess observed in the reliability diagram is that the GNN is underconfident, giving low scores to actual true edges. A naive choice of connected components score cut would lead to a dramatic loss of true edges that could significantly affect the track building performances. We chose the two other cuts, in the walkthrough part of CC\&Walk, to be $0.1$ for the minimal branching cut and $0.6$ for the additive branching cut. We apply the calibration procedure using splines and Figure \ref{fig:calibration_curve_cal} and \ref{fig:reliability_curve_cal} show the calibration curve and reliability diagram after the score calibration.
Once calibrated, the optimal score cut for the connected components part is now of almost $1/2$ and overconfidence or underconfidence is detected in the reliability diagram plot. We chose the other two cuts in the walkthrough part to be $0.9$ for the minimal branching cut and $1.0$ for the additive branching cut. We now compare the results of tracking efficiencies for different values of $p_T$. The result is shown on Figure \ref{fig:track_calibration}.
We observe no statistically significant difference between calibrated and uncalibrated GNN. Overall, the average tracking efficiency without calibration is 94.4\% and 94.5\% with calibration. Table \ref{tab:recap_calib} shows all the differences between the calibrated and uncalibrated GNN scores on track construction.
\begin{table}[H]
    \centering
    \resizebox{0.48\textwidth}{!}{\begin{tabular}{c c c}
        \hline
        \hline
        & Uncalibrated & Calibrated \\
        \hline
        Reco. particle & 55254 & \textbf{55318} \\
        Tracks proposed & \textbf{247643} & 240591 \\
        Matched tracks & \textbf{245791} & 239573 \\
        Efficiency & 0.944 & \textbf{0.945} \\
        Fake rate & 0.007 & \textbf{0.004} \\
        Duplication rate & \textbf{0.039} & \textbf{0.039}\\
        \hline
        \hline
    \end{tabular}}
    \caption{\centering CC\&Walk performances for uncalibrated and calibrated GNN scores.}
    \label{tab:recap_calib}
\end{table}

\section{Conclusion}
We have studied Uncertainty Quantification and Propagation (UQ\&P) in the ACORN pipeline with the MCD method on the TrackML dataset. We have shown that the GNN uncertainty is low for the vast majority of edges and reaches its maximum for edges with predicted scores near $1/2$. We found that the uncertainty is equally distributed between the intrinsic GNN uncertainty and the uncertainty propagated from the upstream Filter. This combined GNN uncertainty is dominated by the aleatoric uncertainty and results in low uncertainty in the track reconstruction efficiency. The uncertainty also exhibits non-Gaussian properties and is independent of the chosen dropout rate. We have also demonstrated that the ACORN pipeline is robust against miscalibration of the GNN edge scores by tuning the track reconstruction score cuts.

The logical continuation of this work would be to apply the UQ\&P scheme we developed to actual ATLAS data. It would also be of great interest to study the UQ\&P of the metric learning graph construction stage. It is non-trivial how this could be achieved as it is not clear which metric should be used to quantify the uncertainty in this case. A possible idea would be to modify the metric learning in order to infer, rather than a list of edges, a real valued adjacency matrix. The uncertainty in this case would be the standard deviation of the samples for each coefficient in the matrix. However, it is unclear then how to study the propagation of this uncertainty. Tools from manifold learning may be needed. The code used for this work is available at \url{https://github.com/LukasPeron/ACORN\_UQ\_UP\_with\_MCD}.
\section*{Acknowledgments}
This research was supported in part by the Institute Philippe Meyer, french Ministry of Higher Education and Research, U.S. Department of Energy's Office of Science, Office of High Energy Physics, under Contracts No. DE-AC02 05CH11231 (CompHEP Exa.TrkX) and
DE-SC0024364 (CompHEP FAIR). This research used resources of the National Energy Research Scientific Computing Center (NERSC), a U.S. Department of Energy Office of Science User Facility located at Lawrence Berkeley National Laboratory, operated under Contract No. DE-AC02-05CH11231.

\bibliographystyle{unsrt}
\bibliography{mabiblio}

\appendix

\section{Filter UQ procedure}
\label{app:Filter_uq}
\begin{figure}[H]
    \centering
    \includegraphics[width=.5\textwidth]{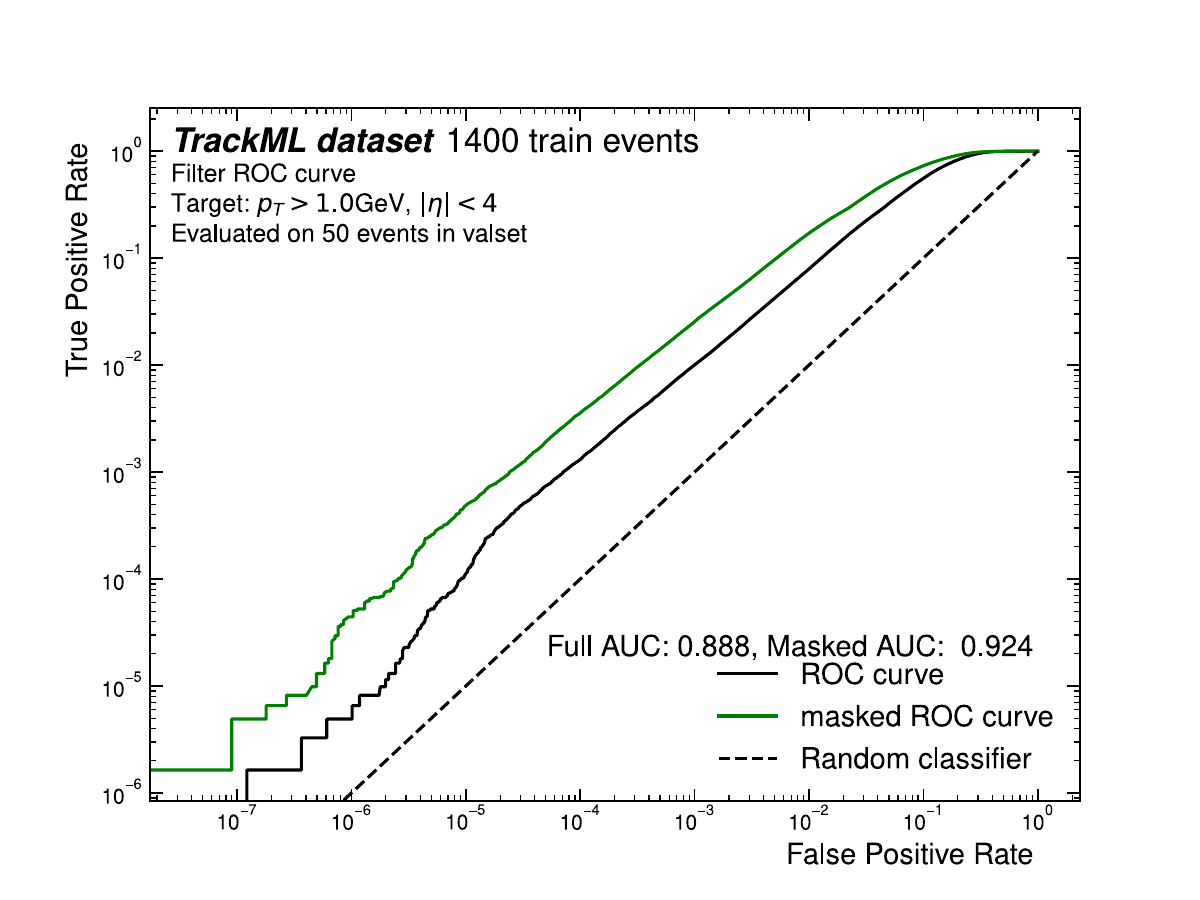}
    \caption{\centering Filter ROC curve.}
    \label{fig:Filter_roc}
\end{figure}
\begin{figure}[H]
    \centering
    \includegraphics[width=.5\textwidth]{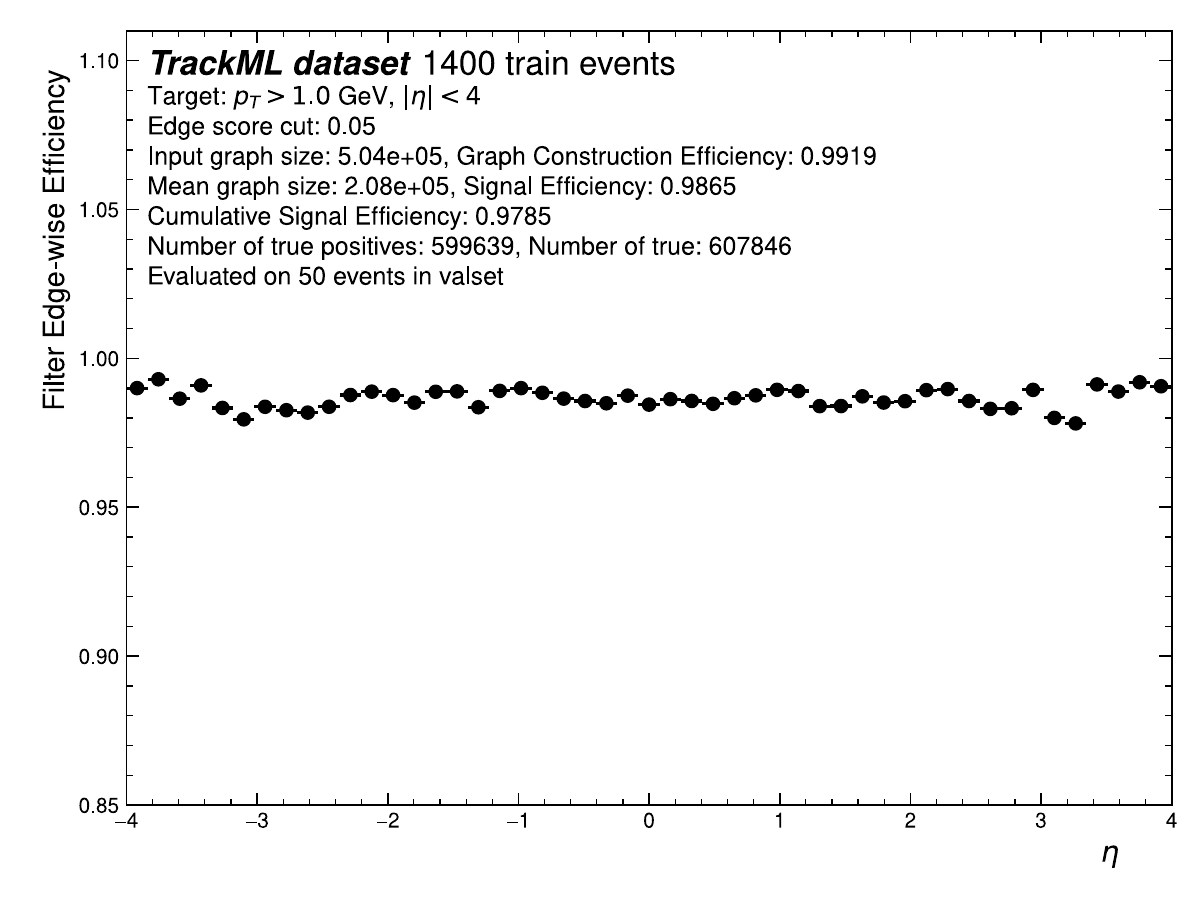}
    \caption{\centering Filter efficiency vs $\eta$.}
    \label{fig:Filter_eff_eta}
\end{figure}
\begin{figure}[H]
    \centering
    \includegraphics[width=.5\textwidth]{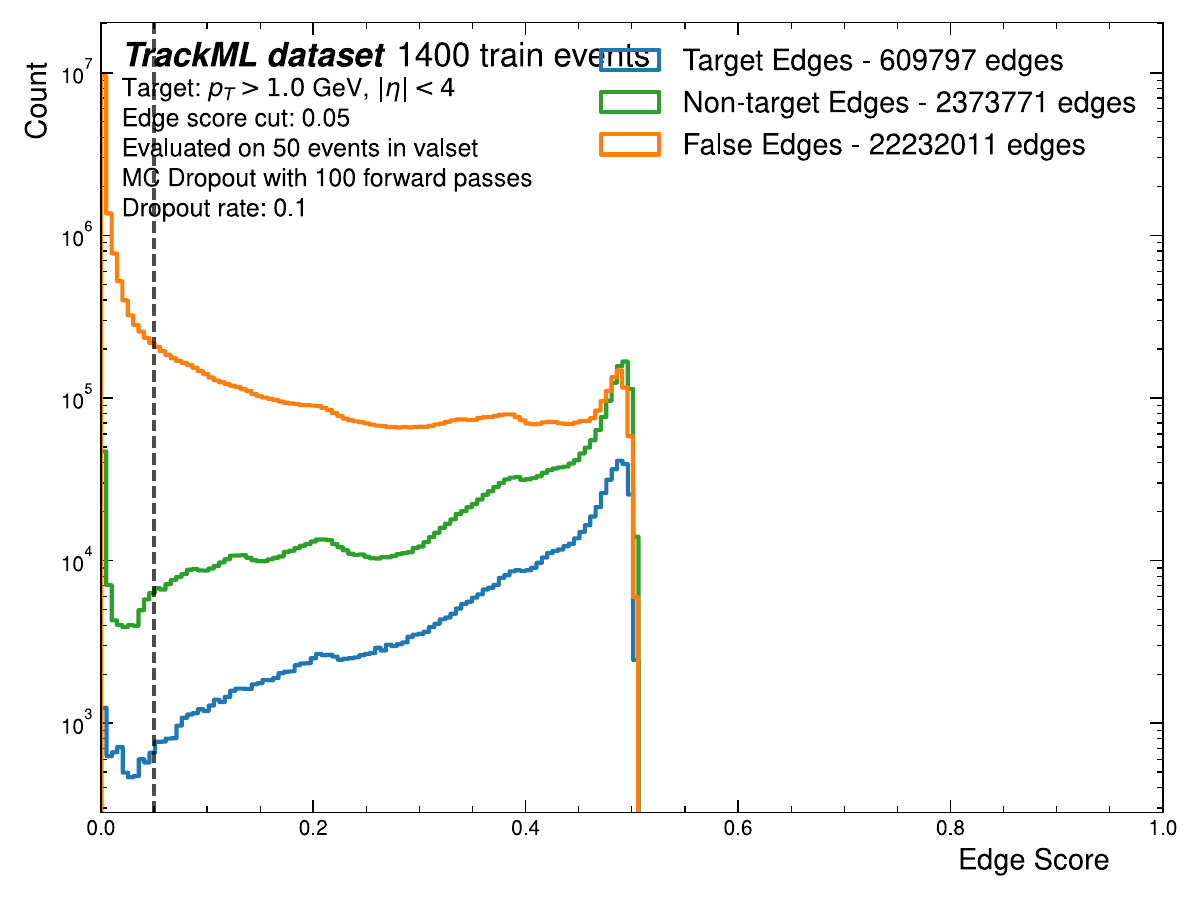}
    \caption{\centering Mean edge score $\langle s_n^{\text{Filter}} \rangle$.}
    \label{fig:Filter_score_distrib}
\end{figure}
\begin{figure}[H]
    \centering
    \includegraphics[width=.5\textwidth]{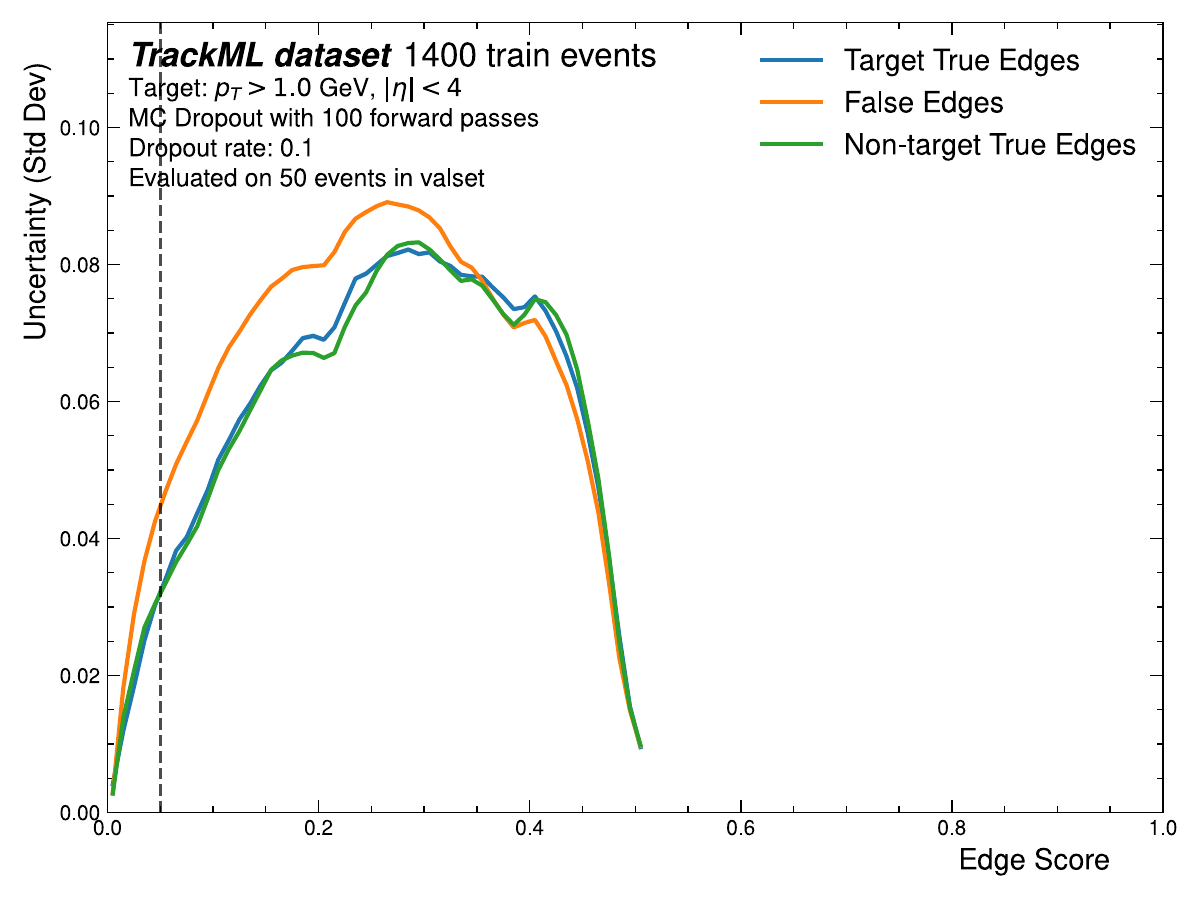}
    \caption{\centering Filter uncertainty $\sigma_n^{\text{Filter}}$ vs $\langle s_n^{\text{Filter}} \rangle$.}
    \label{fig:Filter_sigma}
\end{figure}
\begin{figure}[H]
    \centering
    \includegraphics[width=.5\textwidth]{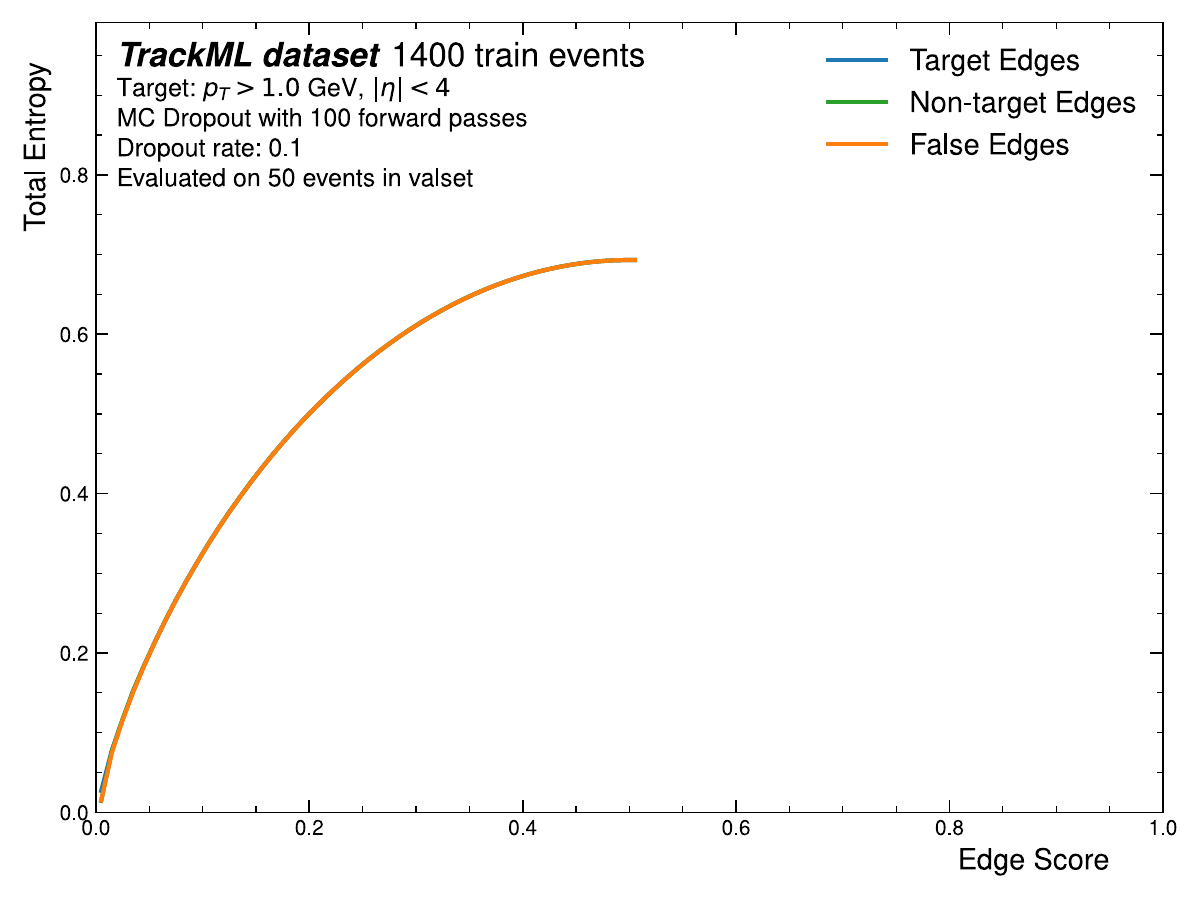}
    \caption{\centering Filter total uncertainty $\mathbb H\left[\langle s_n^{\text{Filter}} \rangle\right]$ vs $\langle s_n^{\text{Filter}} \rangle$.}
    \label{fig:Filter_total_uncertainty}
\end{figure}
\begin{figure}[H]
    \centering
    \includegraphics[width=.5\textwidth]{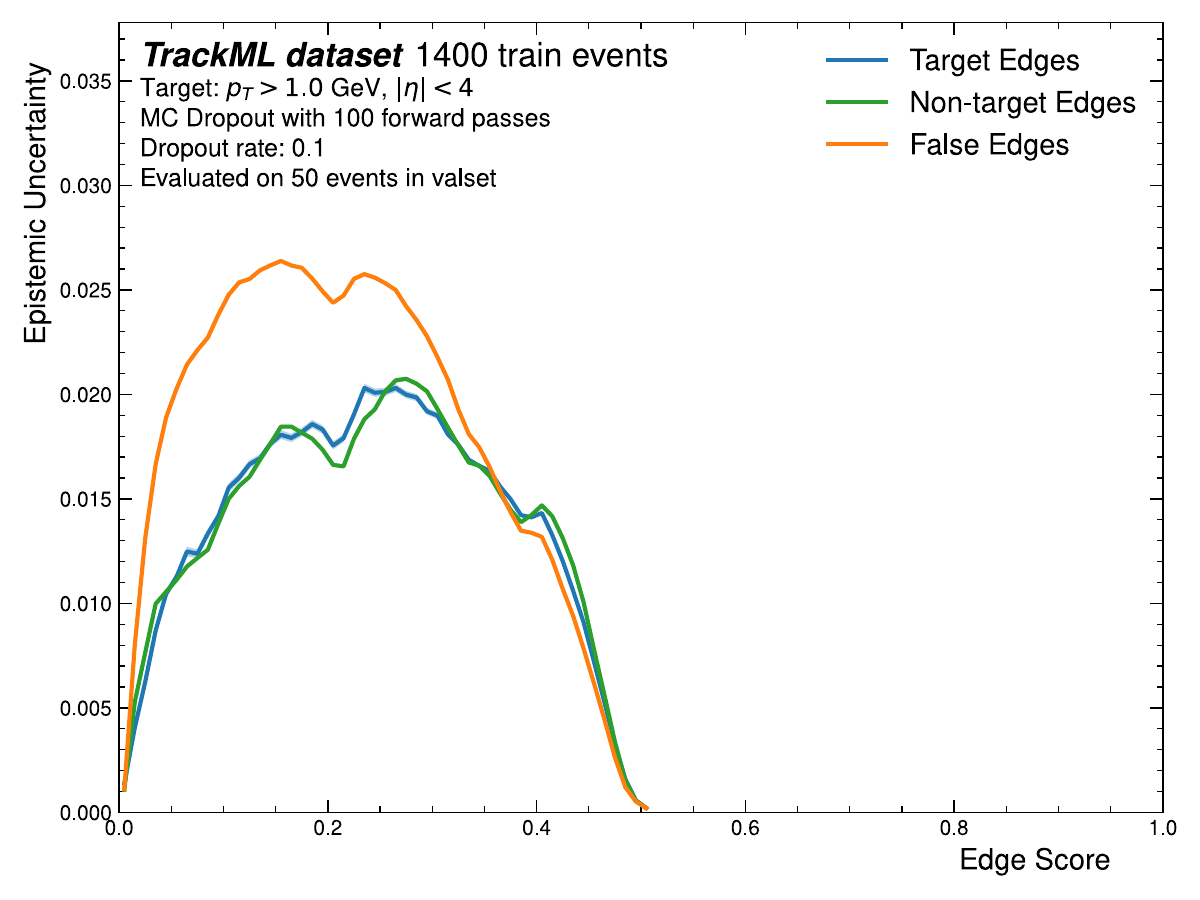}
    \caption{\centering Filter epistemic uncertainty $\mathbb I\left[s_n^{\text{Filter}}\right]$ vs $\langle s_n^{\text{Filter}} \rangle$.}
    \label{fig:Filter_epistemic_uncertainty}
\end{figure}
\begin{figure}[H]
    \centering
    \includegraphics[width=.5\textwidth]{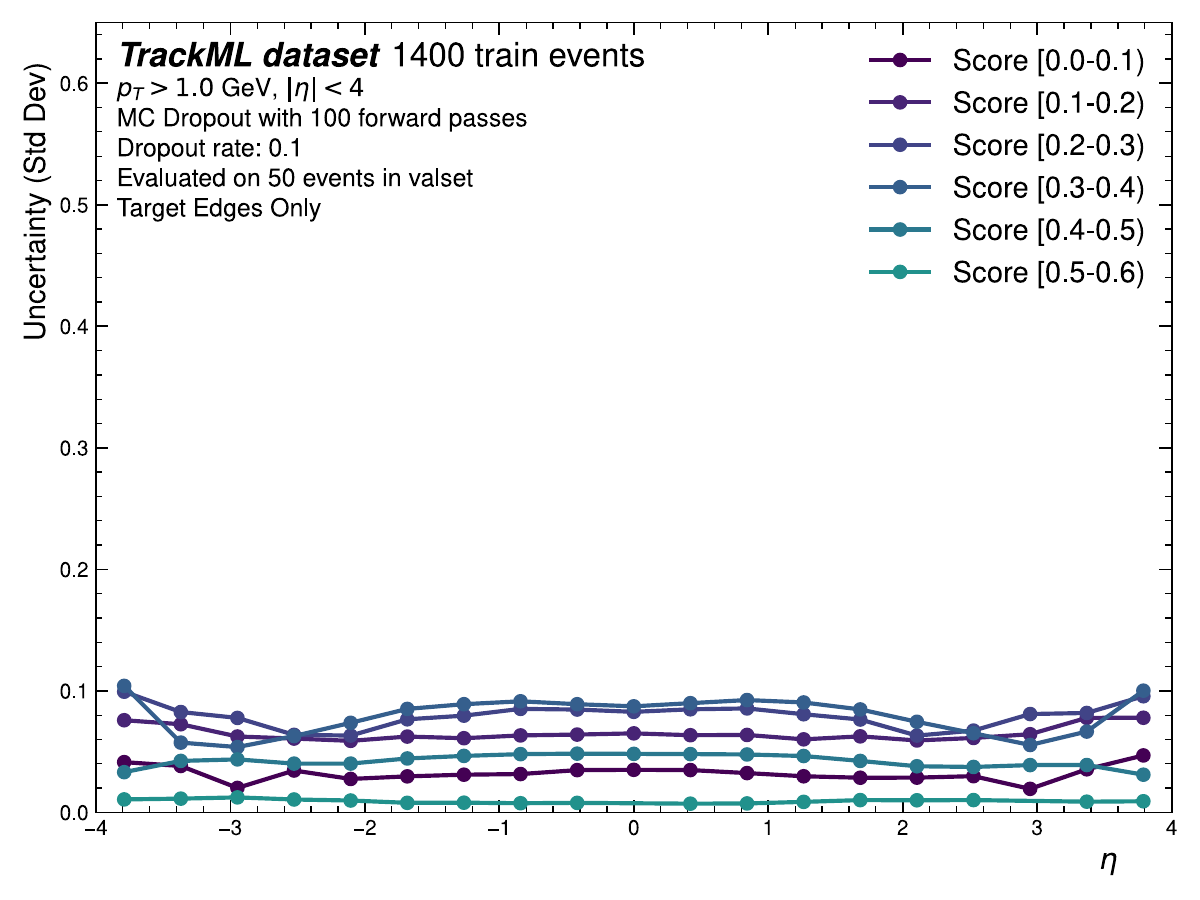}
    \caption{\centering Filter uncertainty $\sigma_n^{\text{Filter}}$ vs $\eta$ for target edges.}
    \label{fig:Filter_sigma_vs_eta}
\end{figure}
\begin{figure}[H]
    \centering
    \includegraphics[width=.5\textwidth]{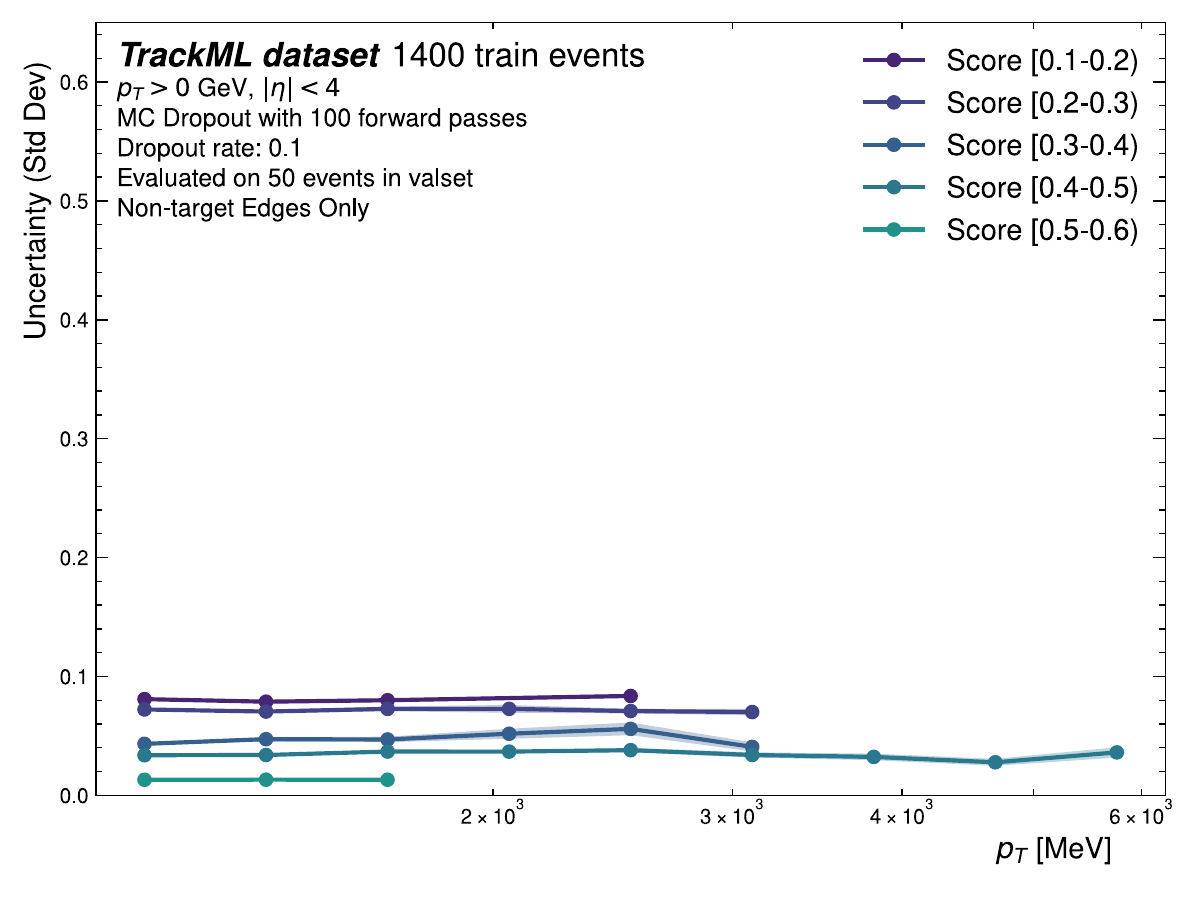}
    \caption{\centering Filter uncertainty $\sigma_n^{\text{Filter}}$ vs $p_T$.}
    \label{fig:Filter_sigma_vs_pt}
\end{figure}
\begin{figure}[H]
    \centering
    \includegraphics[width=.5\textwidth]{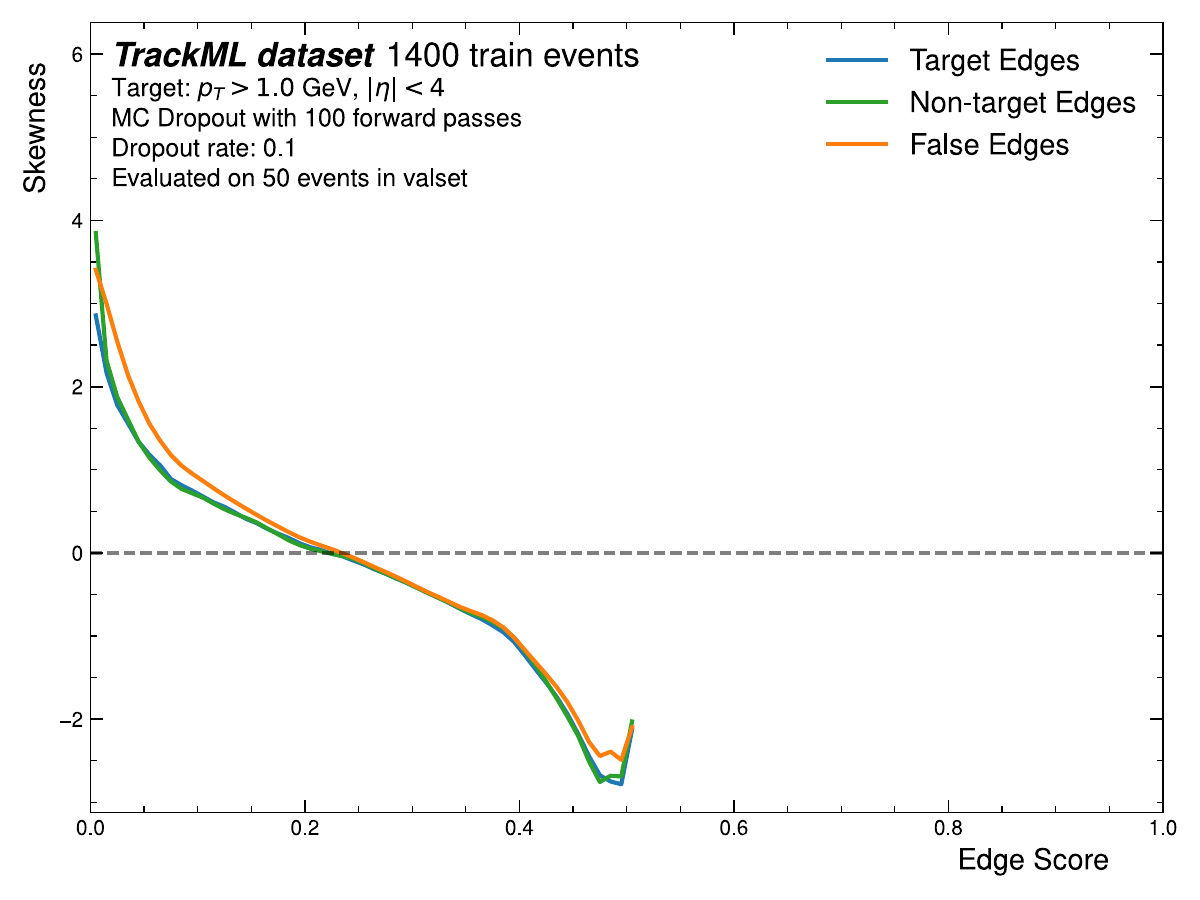}
    \caption{\centering Skewness of the empirical Filter MCD samples}
    \label{fig:Filter_skewness}
\end{figure}
\begin{figure}[H]
    \centering
    \includegraphics[width=.5\textwidth]{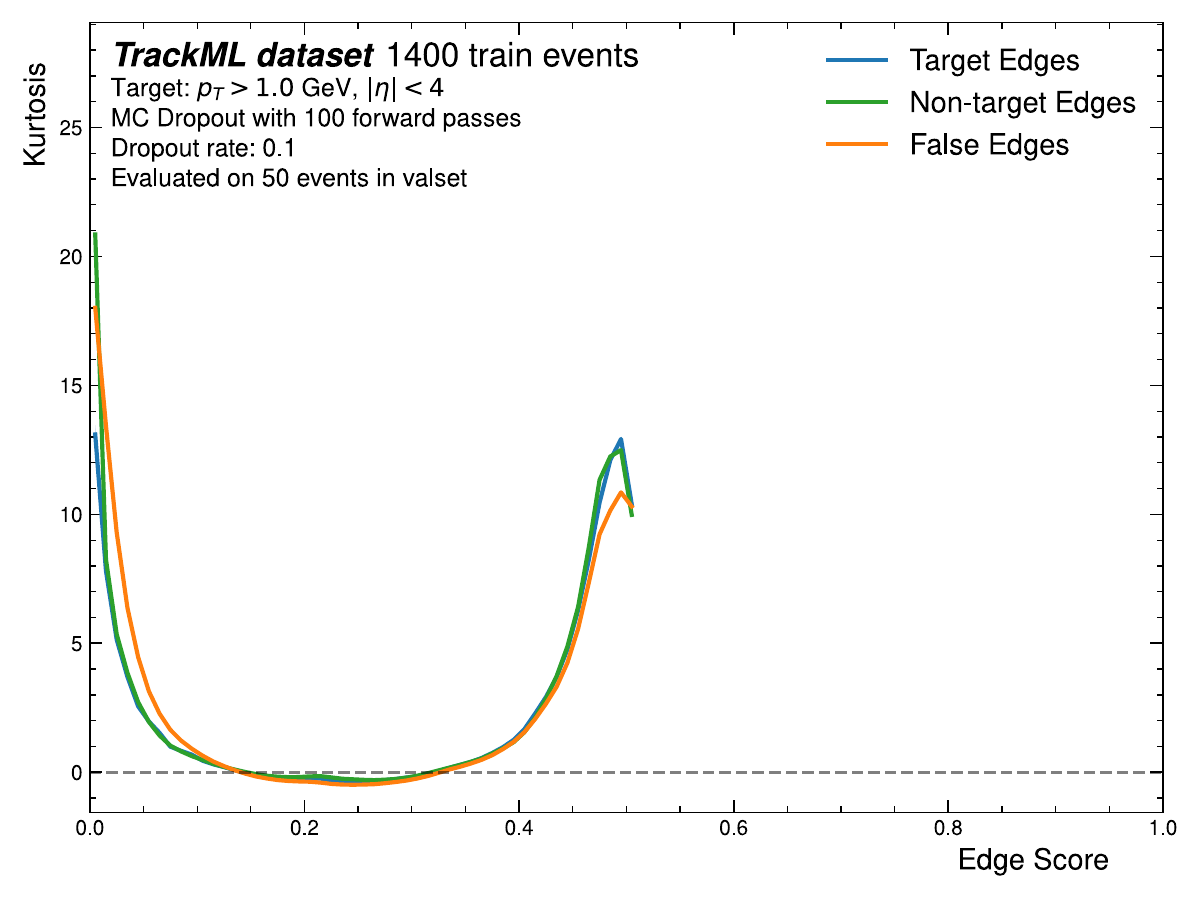}
    \caption{\centering Kurtosis (Fisher's definition) of the empirical Filter MCD samples}
    \label{fig:Filter_kurtosis}
\end{figure}
\begin{figure}[H]
    \centering
    \includegraphics[width=.5\textwidth]{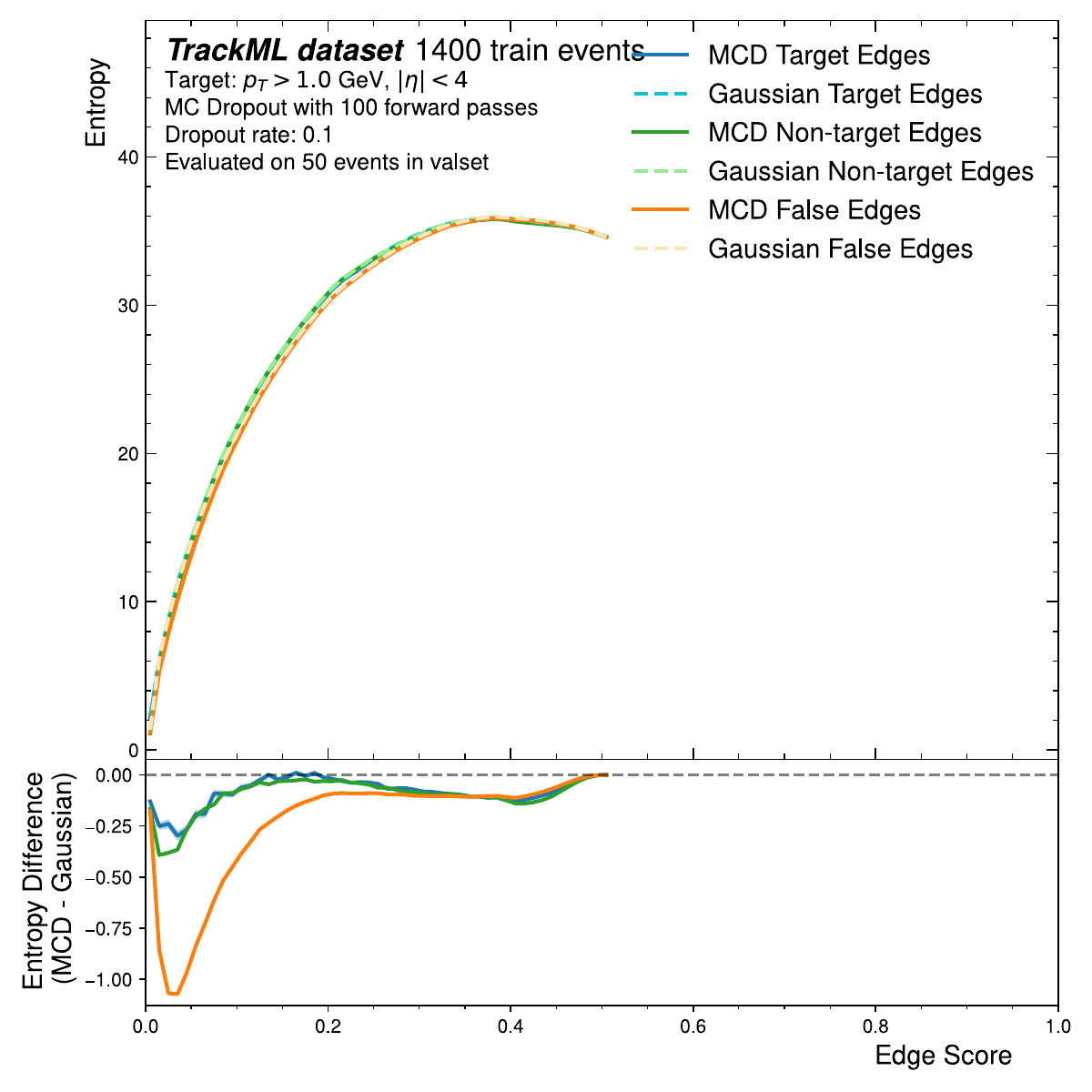}
    \caption{\centering Shannon's entropy comparison between MCD and Gaussian Filter score distribution}
    \label{fig:Filter_entropy_comparison}
\end{figure}
\begin{figure}[H]
    \centering
    \includegraphics[width=.5\textwidth]{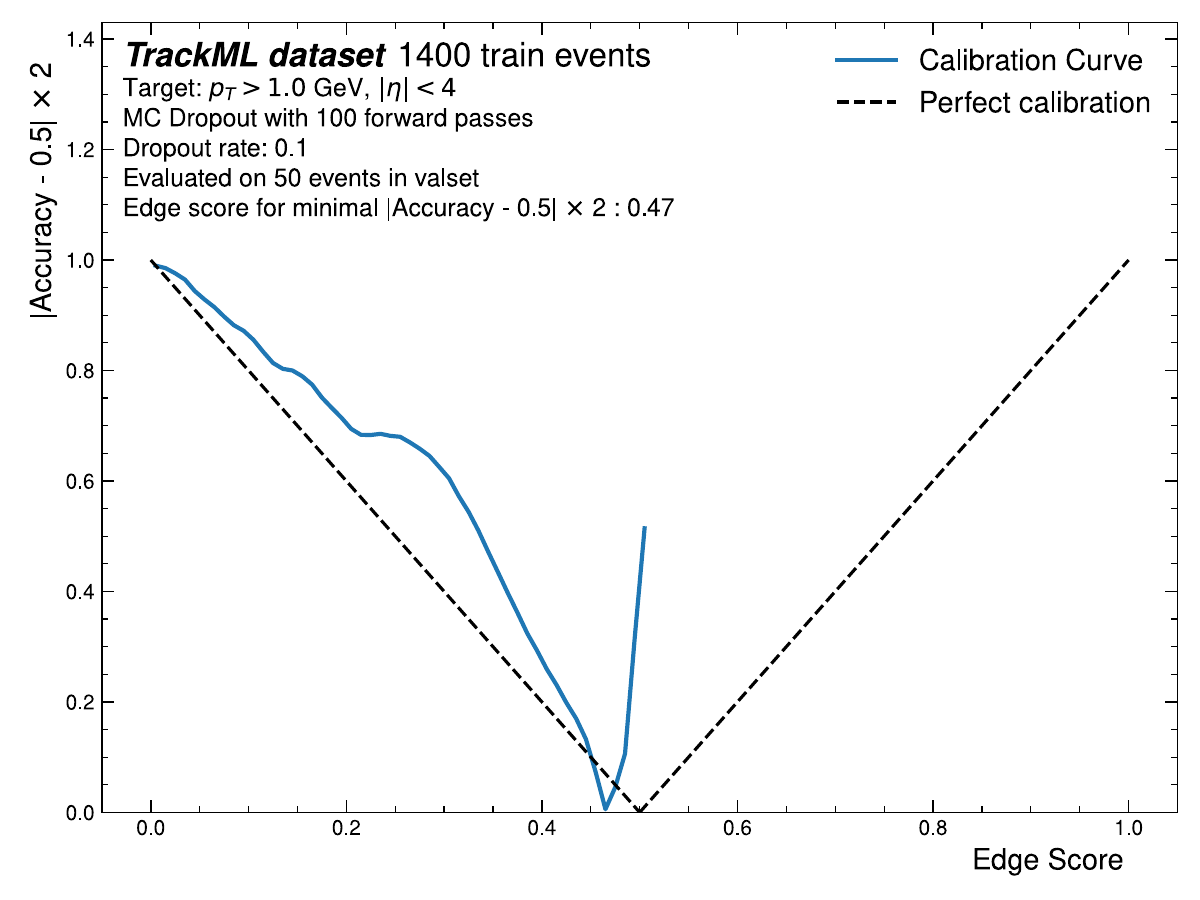}
    \caption{\centering Calibration curve for non calibrated Filter.}
    \label{fig:Filter_calibration}
\end{figure}
\begin{figure}[H]
    \centering
    \includegraphics[width=.5\textwidth]{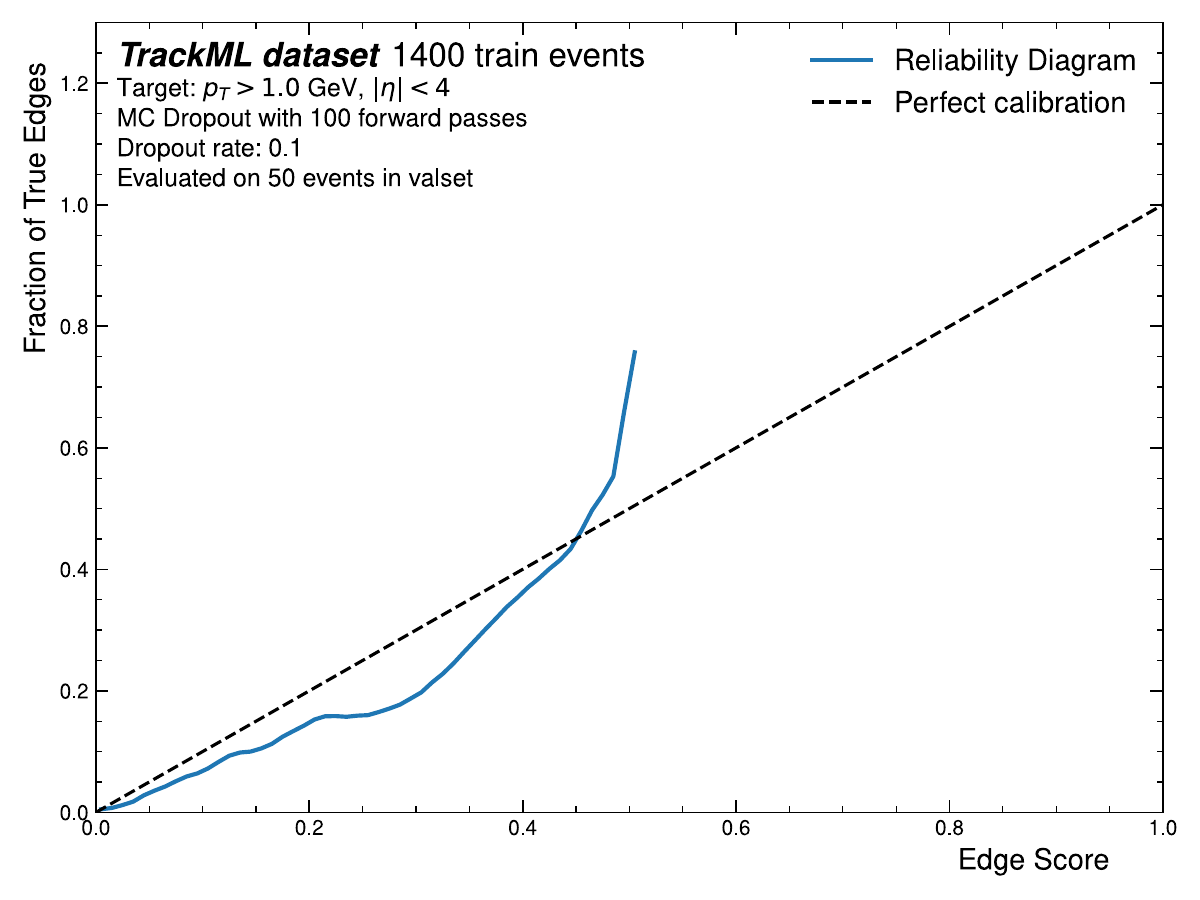}
    \caption{\centering Reliability diagram for non calibrated Filter.}
    \label{fig:Filter_reliability}
\end{figure}
\newpage
\section{Other plots}
\label{app:other_gnn_information}
\begin{figure}[H]
    \centering
    \includegraphics[width=.5\textwidth]{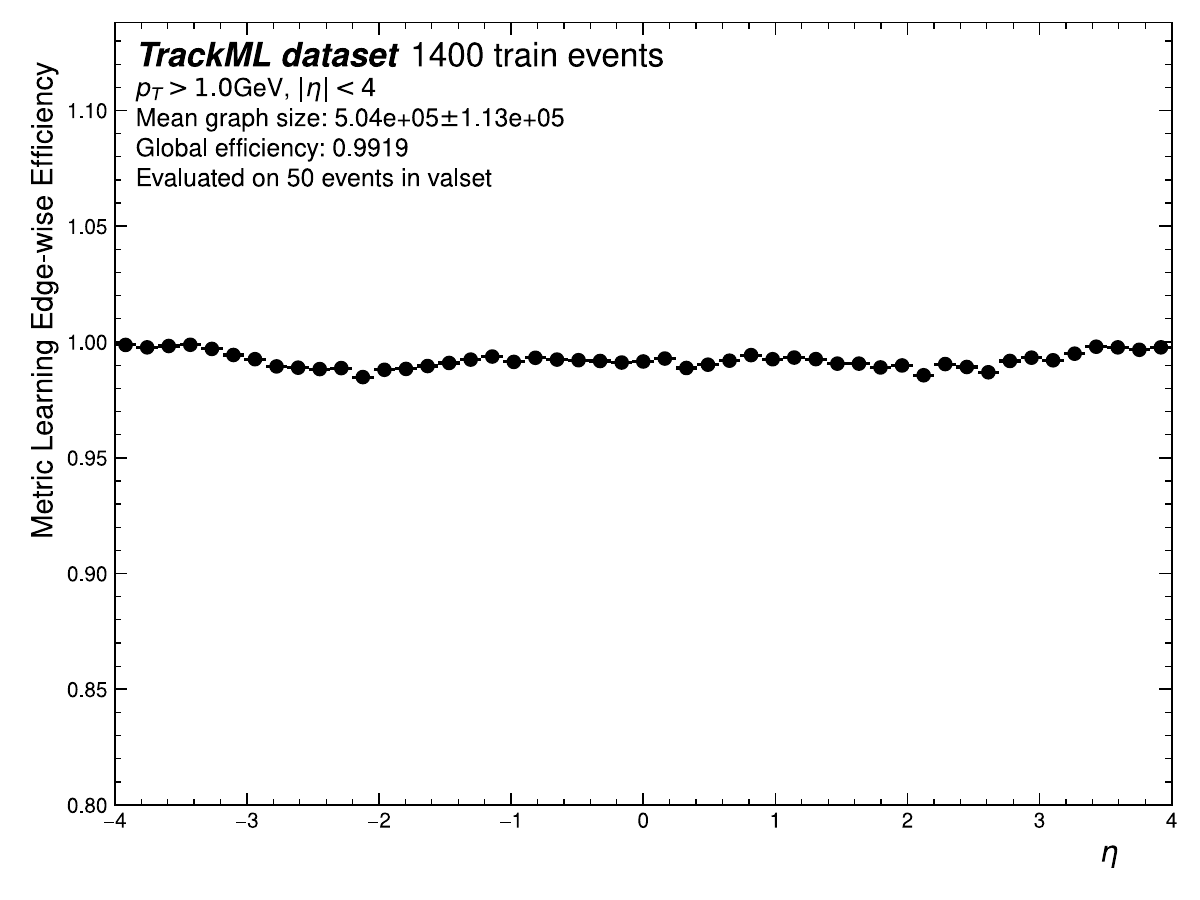}
    \caption{\centering Metric learning efficiency vs $\eta$.}
    \label{fig:metric_eff_eta}
\end{figure}
\begin{figure}[H]
    \centering
    \includegraphics[width=.5\textwidth]{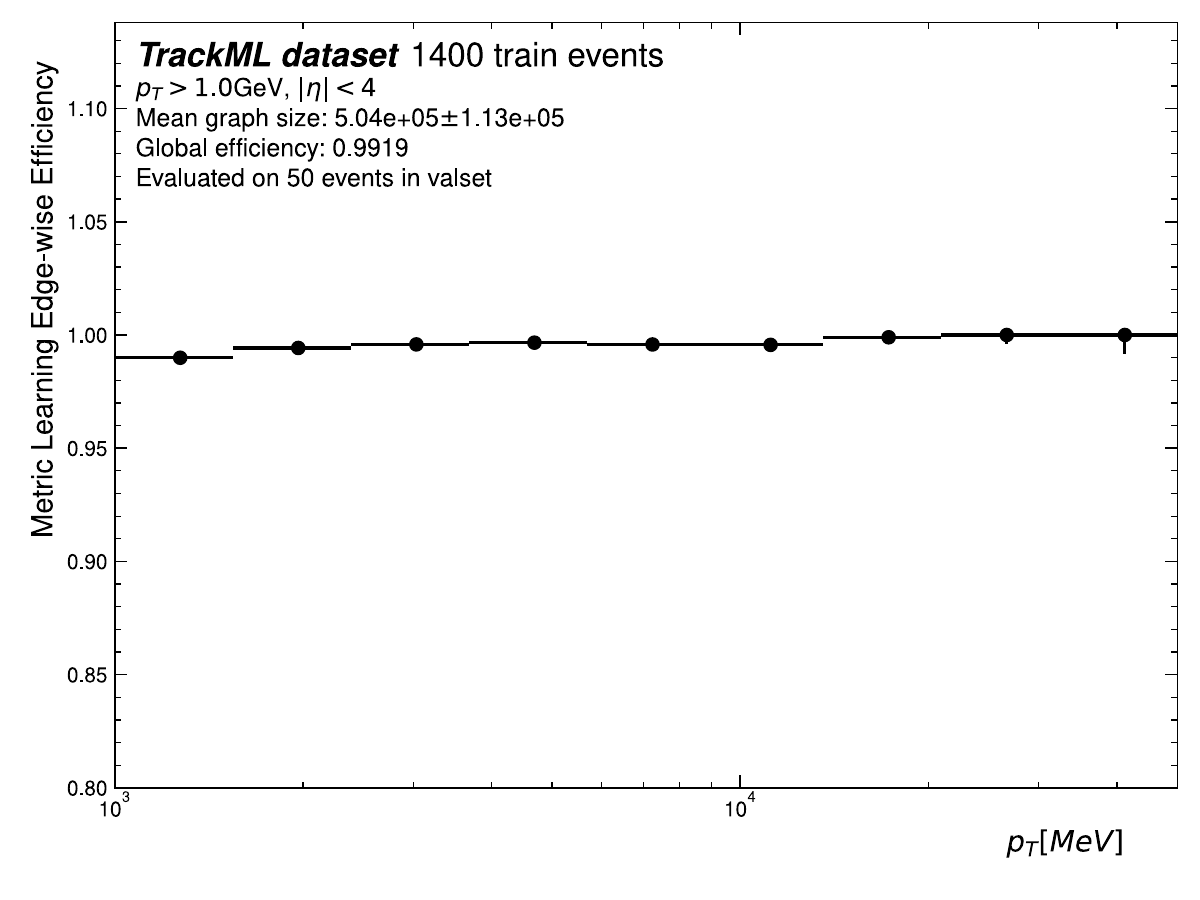}
    \caption{\centering Metric learning efficiency vs $p_T$.}
    \label{fig:metric_eff_pt}
\end{figure}
\begin{figure}[H]
    \centering
    \includegraphics[width=.5\textwidth]{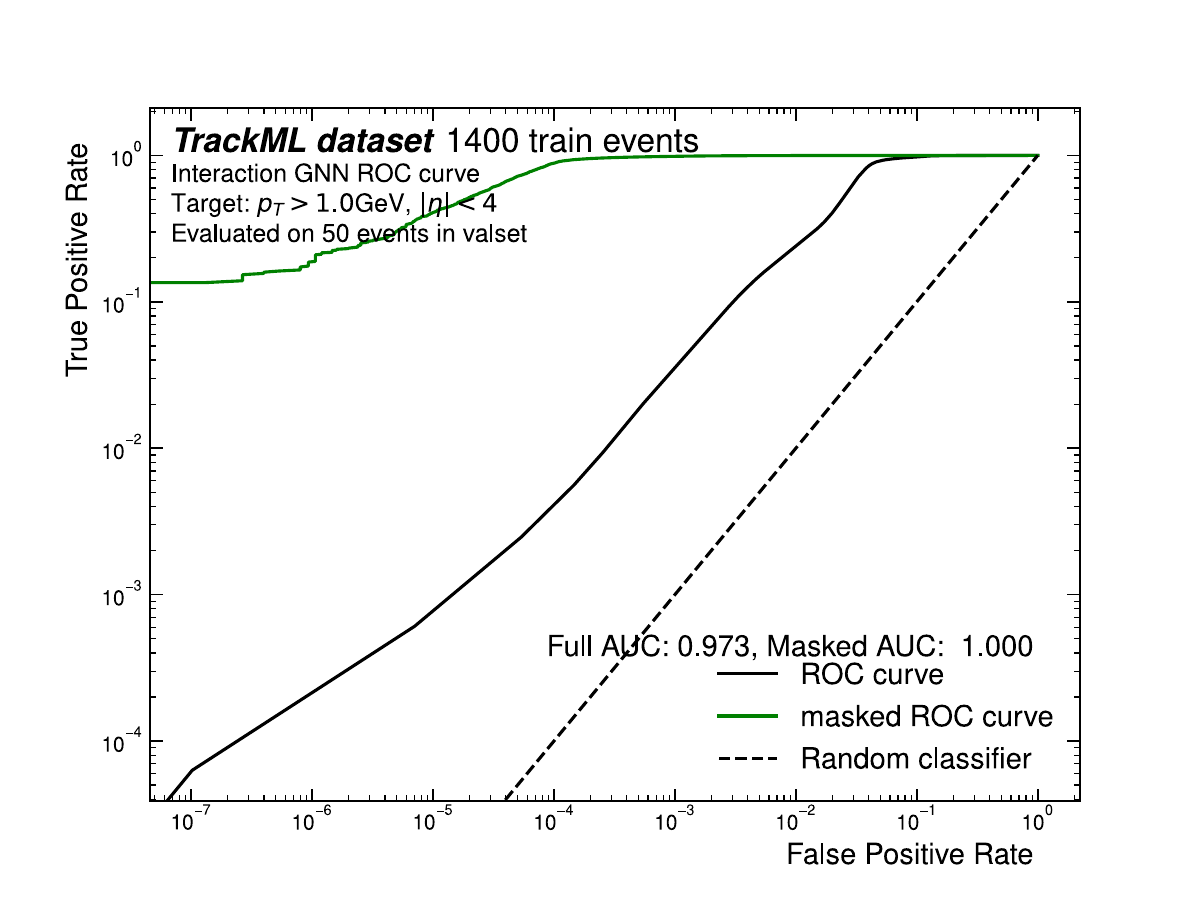}
    \caption{\centering GNN ROC curve.}
    \label{fig:gnn_roc}
\end{figure}
\begin{figure}[H]
    \centering
    \includegraphics[width=.5\textwidth]{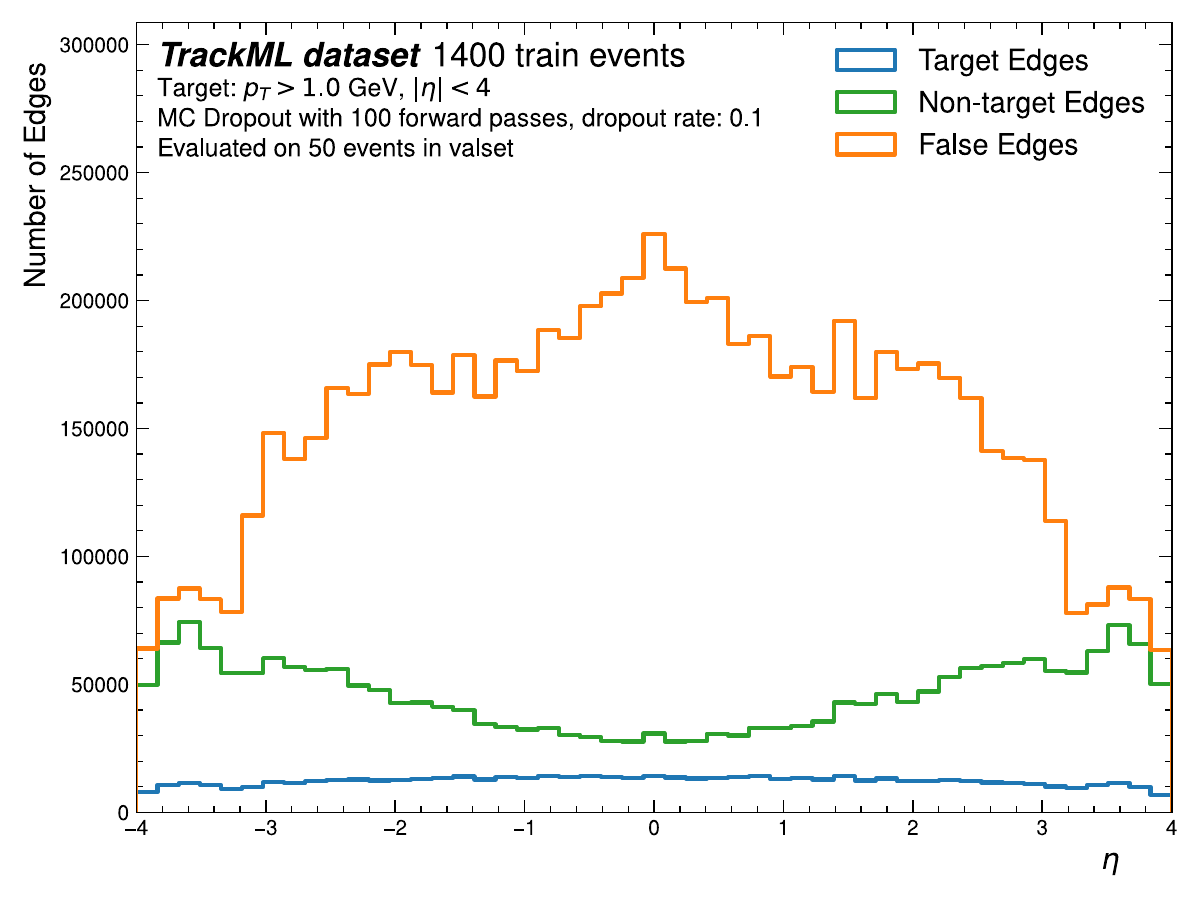}
    \caption{\centering Number of edges in the validation dataset vs $\eta$.}
    \label{fig:app_number_of_edges_eta}
\end{figure}
\begin{figure}[H]
    \centering
    \includegraphics[width=.5\textwidth]{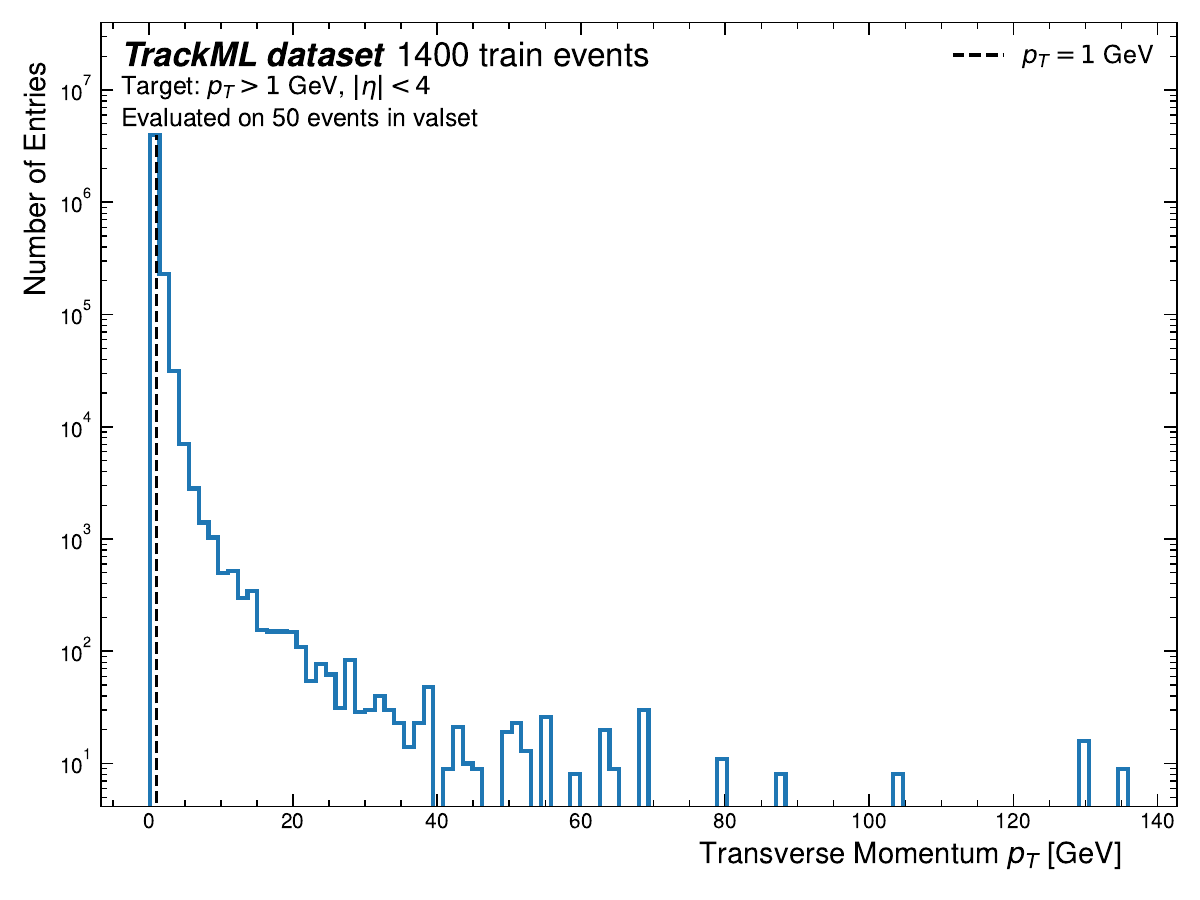}
    \caption{\centering $p_T$ spectrum of the validation dataset.}
    \label{fig:app_pt_spectrum}
\end{figure}
\begin{figure}[H]
    \centering
    \includegraphics[width=.5\textwidth]{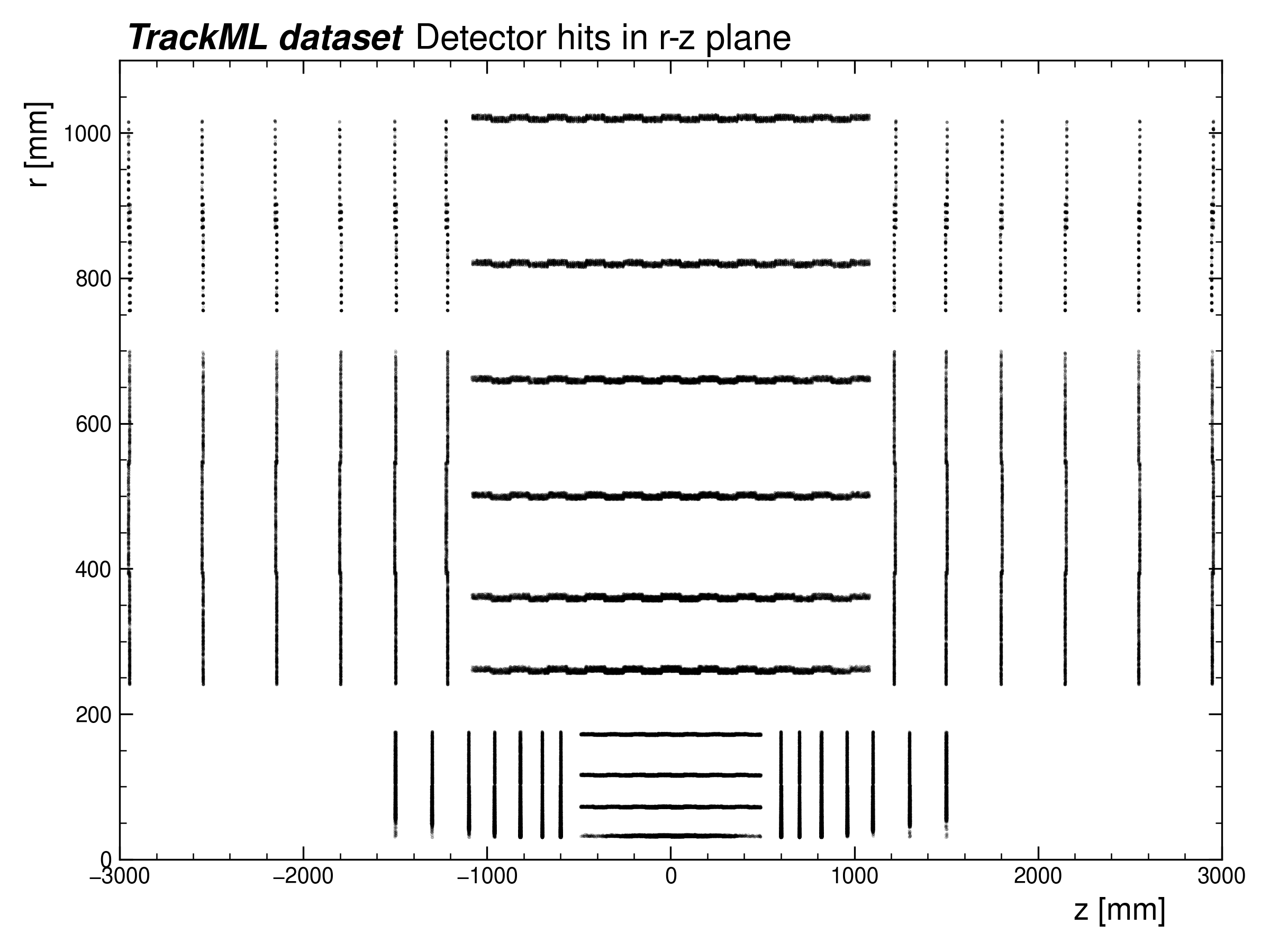}
    \caption{\centering Example of detector hits from the TrackML dataset in the $(rz)$-plane.}
    \label{fig:hits_rz}
\end{figure}
\begin{figure*}
    \centering
    \includegraphics[width=\textwidth]{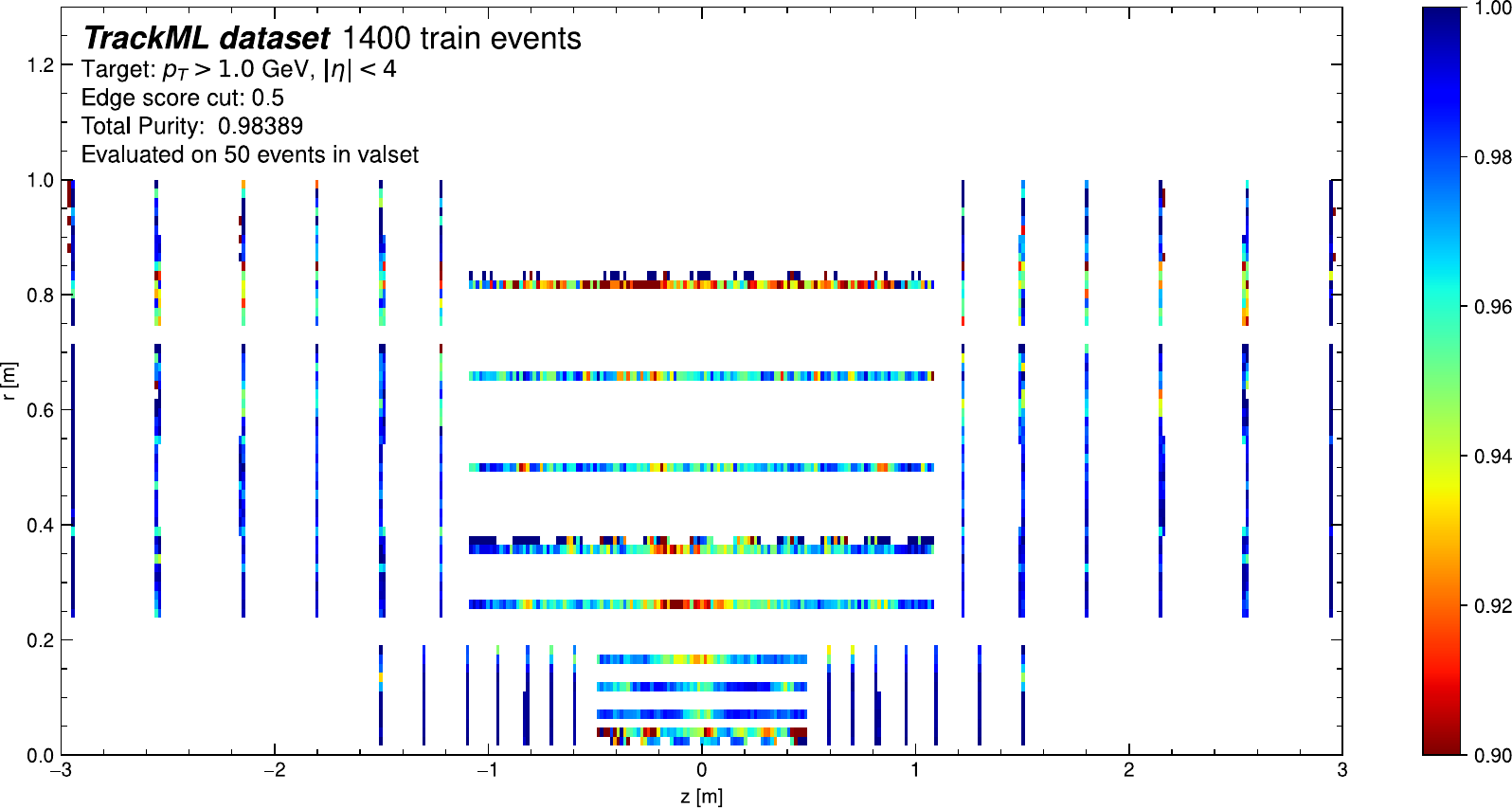}
    \caption{\centering GNN purity in the $(rz)$-plane.}
    \label{fig:gnn_purity}
\end{figure*}
\begin{figure*}
    \centering
    \includegraphics[width=\textwidth]{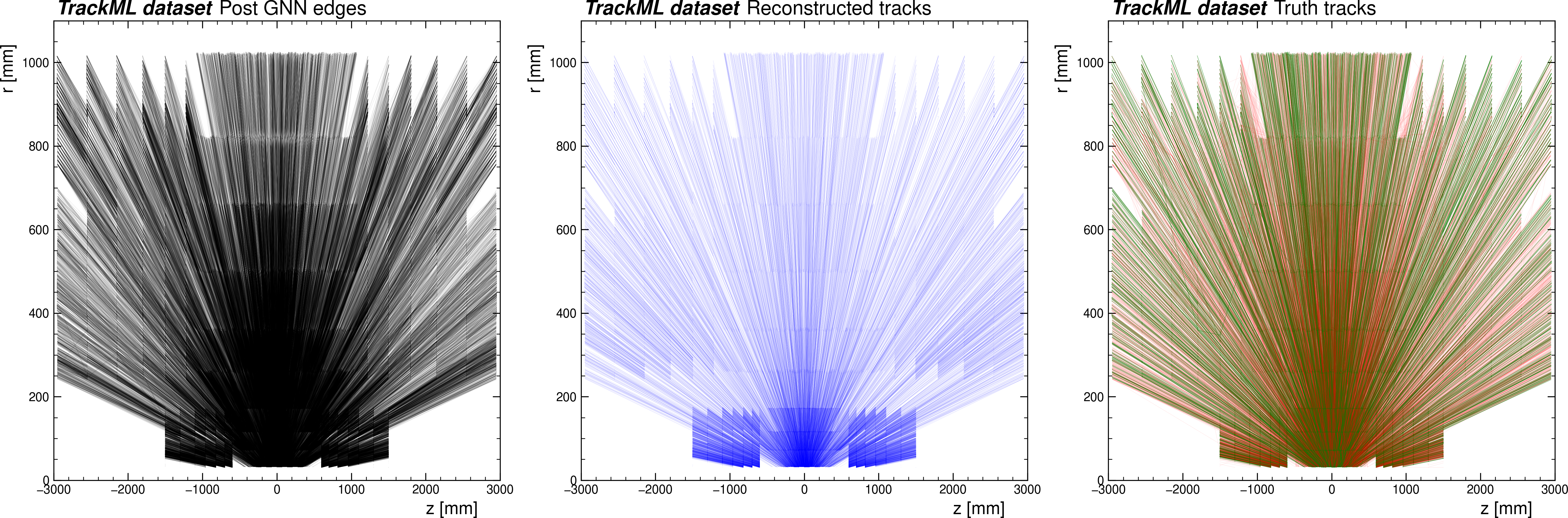}
    \caption{\centering Example of GNN output graph, true tracks objective and reconstructed tracks by CC\&Walk in the $(rz)$-plane. Green (red) tracks are tracks with $p_T \geq (<) 1$GeV.}
    \label{fig:tracks_rz}
\end{figure*}
\begin{figure*}
    \centering
    \includegraphics[width=.9\textwidth]{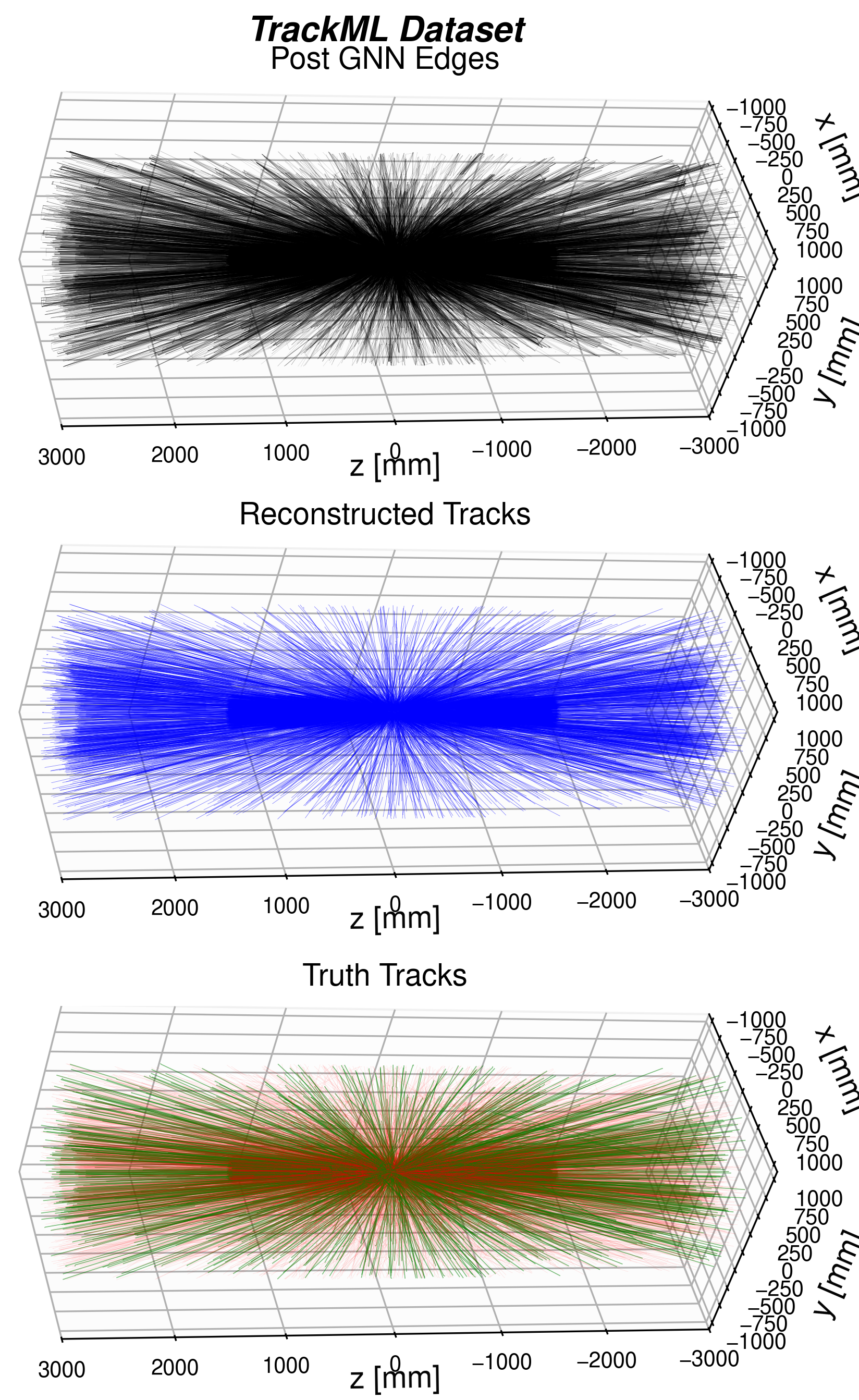}
    \caption{\centering 3D visualization of tracks from TrackML dataset. See caption of Fig. \ref{fig:tracks_rz} for details.}
    \label{fig:tracks_3d}
\end{figure*}

\end{multicols*}
\end{document}